\documentclass[12pt]{book}
\usepackage[square, numbers]{natbib}
\usepackage[british]{babel}
\usepackage{epsf}
\usepackage{epsfig}
\usepackage{psfrag}
\usepackage{changebar}
\usepackage{indentfirst}
\usepackage{amsmath}
\usepackage{amssymb}
\usepackage{theorem}
\usepackage{default}
\usepackage{longtable}

\usepackage{fancyhdr}
\usepackage{geometry}
\usepackage[small,bf]{caption}

\fancypagestyle{plain}{
			\fancyhead{}
			
}

\begin{document}
   \frontmatter
      \title{\LARGE \bf Spin-Torsion, Braneworlds and Changing Symmetry in
the Universe \\
       From the Beginning to the End}
\author{Leong Chung Wei Bernard \\ Darwin College \\ Cavendish Laboratory
}
\date{}

\vspace{3 in}

      \maketitle
\chapter*{Preface}

This dissertation is the result of work carried out in the
Astrophysics Group (formerly Radio Astronomy Group) of the Cavendish
Laboratory, Cambridge between October 1999 and September 2002. Except
where explicit reference is made to the work of others, the work
contained in this dissertation is my own and is not the outcome of
work done in collaboration. No part of this dissertation has been
submitted for a degree or diploma or other qualification at this or
any other university. The total length of this dissertation does not
exceed sixty thousand words.

\vspace{1in}
\par \hspace{4.5in} Bernard Leong
\par \hspace{4.5in} October 2002

      \clearpage{\pagestyle{empty}\cleardoublepage} 
\chapter*{Acknowledgements}
\begin{quote}
{\it The subject is a very much difficult and intricate one than at first
one  is inclined to  think and  I feel  that I  have not  succeeded in
catching the  keynote. When  that is found,  the various  results here
given will no doubt appear  in their real connection with one another,
perhaps  even  as  immediate  consequences of  a  thoroughly  adequate
conception of the question.} 
\par{\bf - Tait}
\end{quote}

\begin{quote} 
{\it Our whole problem (in theoretical physics) is to make the
mistakes as fast as possible.} 
\par{\bf- J.A. Wheeler}
\end{quote}

There are  many people who I like to thank in the  process of
making this   thesis a  reality.    First and   foremost, I  thank  my
supervisor,  Professor  Anthony  Lasenby   for   the   guidance  and
encouragement  which he has  given me over the  past three years of my
Ph.D. Of equal importance, I express my thanks to Dr Anthony
Challinor, who I have benefitted from his guidance and ideas over the
course of the past three years. I am also indebted to Dr Peter
Dunsby, Professor Roy  Maartens  and Professor  William Saslaw whom  I
have  benefitted for their time and  guidance with the exposure to the
many facets of the subject in cosmology. The results of their guidance
have led to the writing of this thesis. I thank D. Baumann, C. Doran,
S. Gull, M. Hobson, Y. L. Loh, C. Martins, C. Van de Bruck and
A. Sinha and my many colleagues in the Cavendish Astrophysics Group
for many illuminating discussions and assistance. \\

Finally, I thank my parents, relatives, friends and teachers for their support
and encouragement. \\

I gratefully acknowledge financial support from Overseas Research
Studentship (ORS), Cambridge Commonwealth Trust (CCT) and Lee
Foundation, Singapore. \\

      \clearpage{\pagestyle{empty}\cleardoublepage} 
      \chapter*{Abstract}
      
In this thesis, we explore three alternatives to the current paradigm of the
standard inflationary big bang scenario. The three alternative themes are spin
torsion (or Einstein-Cartan-Kibble-Sciama) theories, extra dimensions
(braneworld cosmology) and changing global symmetry. In the spin
torsion theories, we found new cosmological solutions with a
cosmological constant as alternative to the standard scalar field
driven inflationary scenario and we conclude that these toy models do
not exhibit an inflationary phase. In the theme of extra dimensions,
we discuss the dynamics of linearized scalar and tensor perturbations
in an almost Friedmann-Robertson-Walker braneworld cosmology of
Randall-Sundrum type II using the 1+3 covariant approach.  We derive a
complete set of frame-independent equations for the total matter
variables, and a partial set of equations for the non-local variables,
which arise from the projection of the Weyl tensor in the bulk. The
latter equations are incomplete since there is no propagation equation
for the non-local anisotropic stress. For the scalar perturbations, we
found solutions that reveal the existence of new modes arising from
the two additional non-local degrees of freedom (extra dimensions)
under the assumption that the non-local anisotropic stress
vanishes. Our solutions should prove useful in setting up initial
conditions for numerical codes aimed at exploring the effect of
braneworld corrections on the cosmic microwave background (CMB) power
spectrum. For the tensor perturbations, we set out the framework of a program to
compute the tensor anisotropies in the CMB that are generated in
braneworld models. In the simplest approximation, we show the
braneworld imprint as a correction to the power spectra for standard
temperature and polarization anisotropies and similarly show that the
tensor anisotropies are also insensitive to the high energy
effects. Finally in the theme of changing global symmetry, we
constructed a bounded isothermal solution embedded in an expanding
Einstein de Sitter universe and showed that there is a possible phase
transition in the far future.

      \clearpage{\pagestyle{empty}\cleardoublepage} 
      \tableofcontents
      \clearpage{\pagestyle{empty}\cleardoublepage} 
      \listoffigures
      \clearpage{\pagestyle{empty}\cleardoublepage} 
    \mainmatter
\chapter{Prologue} \label{prologue}
\begin{quote}
{\it I want to know how God created this world. I am not interested in
this or that phenomenon, in the spectrum of this or that element. I
want to know His thoughts; the rest are details." }
\par{\bf - Albert Einstein}
\end{quote}

\begin{quote}
{\it ``The future of the empirical  world (or of the phenomenal world)
is completely predetermined by its present state, down to its smallest
detail.''}
\par{\bf - Karl Popper, ``Conjectures and Refutations''}
\end{quote}

\begin{quote}
{\it ``There is a coherent plan  in the universe,  though I don't know
what it's a plan for.''}
\par{\bf - Fred Hoyle}
\end{quote}  

\section{Introduction}

{\it How did our Universe begin? Where did it come from and where is it
heading to? How will our Universe end?} These questions about the
origin and the end of our observable Universe have fascinated many
great thinkers from the ancient Greeks to the dawn of the 21st
century. There is, perhaps, no question other than that of
the origin of our Universe, which so transcends cultural and
temporal divides, inspiring the ancient and modern philosophers, and
the modern cosmologists. At a fundamental level, we yearn to find an
explanation of why there is a Universe, how it has come to assume the form
which we observe it to take, and what are the fundamental laws and
principles that govern its evolution. Although some of the above questions
are philosophical and metaphysical, there are still others which are
within the boundaries of scientific enquiry. Of course, we are 
interested here in the latter rather than the former. 

{\it How do we describe our physical universe?} The global description of the
observable Universe rests on two important attributes, namely the
the {\bf cosmological parameters} from a physical theory and the
{\bf irregularities or inhomogeneities} which we observe. The
cosmological parameters describe quantitatively the various features
of the Universe, for example, the baryon density, $\Omega_b$ and the
age of the Universe, $t_0$. The irregularities
or {\bf density perturbations} describe the physics of the evolution
of the matter components and how they are distributed irregularly
throughout the Universe.  
Einstein's theory of general relativity is a theory of gravity which
is used to study cosmology, and is thought to be an effective
low energy theory from a more fundamental theory that we have no grasp
of. The dynamics of the expanding Universe are characterized by the
Hubble parameter that describes the expansion rate, $H_0$, and the spatial curvature,
$\Omega_K$. The cosmological parameters are obtained from experiments,
for example, the measurement of the cosmic microwave background (CMB)
radiation, which can be characterized by a thermal distribution 
at a temperature of about $2.728$ K. However, not all the
cosmological parameters are measured accurately by one type of
experiment. For example, there are experiments such as large galactic surveys and
measurement of type I supernova that can give us other accurate
cosmological parameters to compare to the CMB experiments.  
For the matter content of our Universe, we
observe the visible baryonic matter. However, there is circumstantial evidence
which suggest that there exists a large and dominant amount of
nonbaryonic matter (or dark matter) in our Universe. To explain the
issue of how the large scale structure of clusters and
galaxies has emerged, we assume the existence of cold dark matter
which comprises particles with negligible velocity. There could be
other possibilities such as hot dark matter (particles that decouple
when their velocities are relativistic) or something exotic. The final
possibility with increasing support is that our Universe possesses a
nonzero positive cosmological constant, $\Lambda$. 

With these features, we have a reasonable description of our
Universe. It remains a question how these features come to
be the way they are. The answer lies in the search for an underlying
theory of gravity which works from the Planck scale to the present
size of our observable Universe. At present, there exists no theory of
gravity which could work at the Planck scale although there are many
candidates for these theories, namely {\bf M/string-theory} and {\bf loop quantum
gravity}. On the other hand, an effective description beyond the Planck
scale is expressed elegantly with Einstein's general
relativity. General relativity provides an reasonable idea of how the 
cosmological parameters and the density perturbations come about. The
resulting cosmology from general relativity is the standard {\bf
inflationary big bang cosmology} which assumes that 
there is a beginning of our Universe. In the next section,
we will summarize the central tenets of the standard inflationary big
bang cosmology. 

\section{The Standard Inflationary Big Bang Cosmology}

In this section, we briefly sketch the theory of the standard hot big bang
cosmology which is presented in many standard texts \cite{kolb,
liddle, longair, padmanabhan, padmanabhan2, peacock}. The central
premises of the theory are: 
\begin{enumerate}
\item The Universe is isotropic and homogeneous on large scales based
on the {\bf Copernican} or {\bf cosmological principle} which states
that {\it there are no priviledged observers in the Universe}. 
In general relativity, the observable Universe is described
by the Friedmann-Robertson-Walker (FRW) metric\footnote{In this
thesis, we adopt the choice of natural units and set $\hbar=c=G=1$.} :
\begin{equation} \label{e:frw}
ds^2 = - dt^2 + a(t)^2 \l[\frac{dr^2}{1 - Kr^2} + r^2 (d \theta^2 +
\sin^2 \theta d \phi^2 ) \r] \;,
\end{equation}
where $K$ is the curvature constant which is $+1$
for a closed Universe, $-1$ for an open Universe and $0$ for
a flat Universe.   
\item The Universe is expanding and obeys an empirical relation
known as the Hubble law. An example of the Hubble law is the 
relationship between the ``distance'' to a galaxy\footnote{An example
of a distance indicator is the absolute luminosity $L$, 
 which is defined as the energy per unit time produced by the source
in its rest frame.}, $d_L$ and the observed redshift of a galaxy, $z$, 
which can be expressed by the following power series: 
\begin{equation} \label{e:hubblelaw}
z = H_0 d_L + \frac{1}{2} (q_0 -1) (H_0 d_L)^2 + \dots \;,
\end{equation}   
where $H_0$ is the present value of the Hubble constant ($H=(\dot{a}/a)$) which
characterizes the age of the Universe, and $q_0$ is the deceleration
parameter which measures the rate of slowing of the
expansion\footnote{
$q_0$ is defined in terms of the scale factor of the
FRW metric to be:
\begin{equation}
q_0 = -\frac{\ddot{a}(t_0) }{a(t_0) H_0^2}
\end{equation}
where $t_0$ denotes the present age of the Universe.
}.
\item The Universe began from a state of infinite (or near infinite)
density and temperature and then cooled with the expansion of the
Universe.  
\end{enumerate}

The big bang model makes accurate and scientifically testable
hypotheses. The agreement with data from astronomical
observations has given us considerable confidence in the model. The
four key observational successes of the standard Hot Big Bang model
are the following:   
\begin{itemize}
\item {\bf Expansion of the Universe}: Our Universe began about fifteen billion
years ago in violent ``explosions'' at every point of spacetime. 
In an early super-dense phase, every particle started rushing apart
from every other particle. At lower redshifts $z \lesssim 1$, the linear relationship
between $d_L$ and $z$ is clear and convincing. One may determine $H_0$
from this linear relationship by making use of galaxies at relatively
low redshifts. This linear relation at low redshifts was first
discovered by Hubble~\cite{hubble}.    
\item {\bf Origin of the cosmic background radiation}: About $3 \times
10^{5}$ years after the Big Bang, the temperature of the Universe
had dropped sufficiently for electrons and protons to combine into
hydrogen atoms. This is known as the {\bf recombination era}. After
the recombination era, the cosmic microwave background photons just
stream toward the observer at the present. Hence the temperature   
differences on this surface of last scattering become the anisotropies
in the CMB temperature we see today. Since the recombination era, the
radiation has propagated freely as it is unable to interact with the
electrically neutral background matter (hydrogen atoms). The radiation
constantly loses energy because its wavelength is stretched by the
expansion of the Universe. The radiation temperature has fallen to approximately
2.728 K in the present day. This background temperature was discovered by Penzias and
Wilson~\cite{penzias} in 1965. Subsequently, the temperature
anisotropies ($(\Delta T/T) \sim 10^{-5}$) were discovered by the
COBE satellite in 1992~\cite{smoot}. The large angle CMB anisotropies
probe the fluctuations in the early Universe, possibly due to the
result of quantum mechanical processes during an epoch of
inflation. The large scale structures of our present Universe are
generated by these primordial fluctuations due to gravitational
instability. A detailed description of the physics of these CMB
anisotropies is given in \cite{hu1, hu2}.    
\item {\bf Nucleosynthesis of the light elements}: After the big bang
occurred, the matter which consisted of free neutrons and protons, was
very dense and hot. The temperature of the matter fell and some of the
nucleons combined to form the light elements, for example,
deuterium\footnote{The formation of deuterium depends crucially on the
baryon/photon ratio.}, helium-3, and helium-4. Theoretical
calculations predicted that 
25 per cent of the matter should be in the form of
helium-4, a result which is in good agreement with current stellar
observations. The heavier elements, of which we are partly made, were
created later in the interiors of stars and spread widely in supernova
explosions.   
\item {\bf Formation of galaxies and large-scale structure}: The formation
of galaxies and large-scale structure of our observable Universe, can
be understood in the framework of the hot big bang model. At about
10,000 years after the big bang, the temperature had fallen to the
point at which massive particles dominate the energy density of the
Universe. The matter content of the Universe can affect structure
formation by different processes which modify any primordial
perturbations, for example, growth under self gravitation and effects
of pressure and dissipative processes. As a result,
we observe the large-scale structure of our Universe today.
\end{itemize}

Although the Big Bang model is successful, it is also plagued by various
problems. Since there is no quantum theory of gravity, there exist the
problem of initial conditions for the standard big bang cosmology. It 
does not explain why the Universe we observe is nearly flat (the
flatness problem). Another interesting question, is that the comoving distance over
which causal interactions can occur before the release of the cosmic
microwave background is considerably less than the comoving distance
that the radiation traverses after decoupling. It would imply that the
microwaves arriving from regions seperated by more than the
horizon scale at last scattering which typically subtends about a
degree, cannot have interacted before decoupling. The big bang model
does not offer a satisfactory explanation for the nearly homogeneous
temperature of the cosmic microwave background in the Universe. It
requires that the homogeneity must be part of the initial
conditions and this is known as the horizon problem. Finally, for the
big bang to begin at a very high
temperature, there are possible unwanted relics forbidden by
observation which would survive till today, for example, the gravitino
in supergravity theories or the moduli fields in superstring theory.

To deal with these inherent problems, an inflationary phase is
incorporated into the standard big bang cosmology. The inflationary
cosmology was proposed by Guth~\cite{guth}, Albrecht and
Steinhardt~\cite{albrecht} and Linde~\cite{linde, linde2}. It assumes 
that there is an epoch during which the scale factor of the
Universe is accelerating. It is often described as a rapid
superluminal expansion driven by a source, for example, a scalar field
(or the {\bf inflaton field}). A physical characterization of inflation is
given by the following expression
\begin{equation} \label{e:inflation}
\dd{}{t} \frac{1}{a H} < 0 \;,
\end{equation}  
where $1/(aH)$ is the comoving Hubble length which is the most
important characteristic scale of the expanding Universe. The condition for
inflation is that the comoving Hubble length is decreasing with
time. With inflation, the above mentioned problems of the big bang
cosmology are more or less cured (see \cite{kolb, liddle,
padmanabhan, peacock}). In the case of the
flatness problem, inflation requires the condition that the critical
density $\Omega$ is driven towards 1 rather than away from it. It
would explain why the Universe is nearly flat. As for the horizon
problem, we can allow our present observable Universe to emerge from a
tiny region which was well inside the Hubble radius early on during
inflation because of the large reduction in the comoving Hubble length
during the inflationary phase. Finally, the relic
abundances can be reduced by the expansion during inflation if they
are produced before the inflationary epoch. 

Another example of the many successes of inflation is that it provides the mechanism
for setting the scale-invariant spectrum of density perturbations
present in the CMB anisotropies. The discovery of the accoustic peaks in CMB
anisotropies came from the recent balloon-based experiments such as BOOMERANG
\cite{boomerang1} and MAXIMA \cite{maxima1}. These accoustic peaks
probe the information of the early Universe, since they depend mainly
on the spectrum of initial fluctuations and fundamental cosmological
parameters. Although inflation is a successful phenomenological theory that
accounts for the large scale structure and various other features of
the observable Universe, the inflaton field (or the scalar field) is
not related to the Standard Model which has unified electromagnetism,
strong and weak nuclear interactions in high energy theories. As
a result, there are many different models of inflation (see
\cite{kolb, liddle, lyth3}). The future experiments in the CMB and
gravitational waves will be able to constrain the
possible type of inflationary models 
that give us the present description of our observable Universe. 

\begin{center}
\begin{figure}[!bth]
\psfrag{a}{$t=15$ billion years, $T=3$ K}
\psfrag{b}{$t=10^{5}$ years, $T=3000$ K}
\psfrag{c}{$t=3$ minutes}
\psfrag{d}{$t=1$ s, $T=1$ MeV}
\psfrag{e}{$t=10^{-6}$ s, $T=1$ GeV}
\psfrag{f}{$t=10^{-11}$ s, $T=10^{3}$ GeV}
\psfrag{g}{$t=10^{-35}$ s, $T=10^{15}$ GeV}
\psfrag{h}{$t=10^{-43}$ s, $T=10^{19}$ GeV}
\psfrag{i}{$t=0$ s}
\includegraphics[scale=0.8]{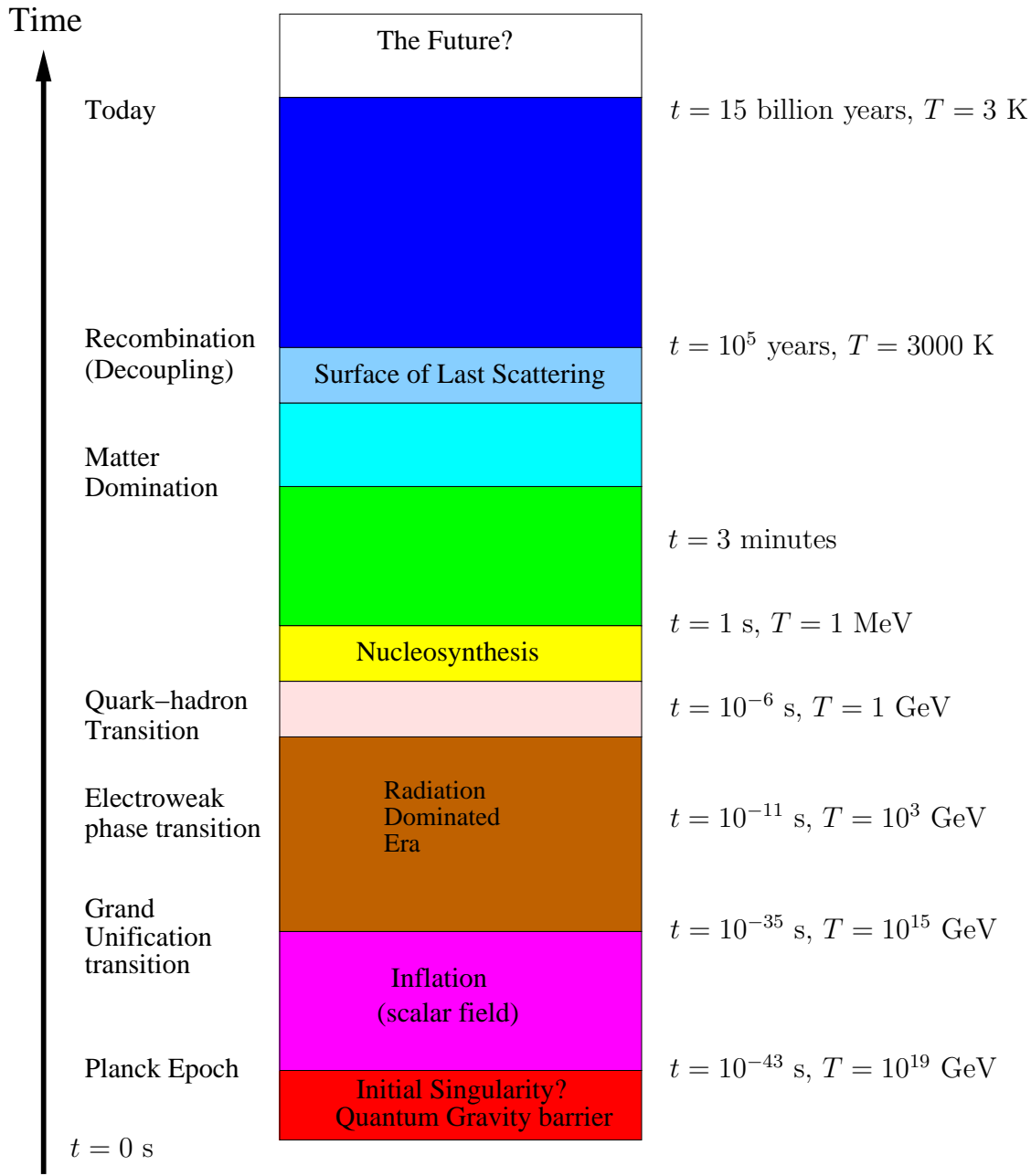}
\caption{A diagram (not drawn to scale) illustrating the history of the standard
inflationary big bang cosmology.}
\label{chap1-0a}
\end{figure}
\end{center}

Of course, the above summary does not do enough justice to the
development of cosmology for the last few decades. However, we can
formulate an updated picture of the history of our Universe (see
fig. \ref{chap1-0a}). In the picture, we could see that considering the period
after inflation, we can account for most of the features of our
Universe with the present standard inflationary big bang
cosmology. 

\section{Three Alternatives to Standard Cosmology}

\begin{figure}[!bth]
\psfrag{a}{$t=15$ billion years, $T=3$ K}
\psfrag{b}{$t=10^{5}$ years, $T=1$ eV $=3000$ K}
\psfrag{c}{$t=3$ minutes}
\psfrag{d}{$t=1$ s, $T=1$ MeV}
\psfrag{e}{$t=10^{-6}$ s, $T=1$ GeV}
\psfrag{f}{$t=10^{-11}$ s, $T=10^{3}$ GeV}
\psfrag{g}{$t=10^{-35}$ s, $T=10^{15}$ GeV}
\psfrag{h}{$t=10^{-43}$ s, $T=10^{19}$ GeV}
\psfrag{i}{$t=0$ s}
\includegraphics[scale=0.8]{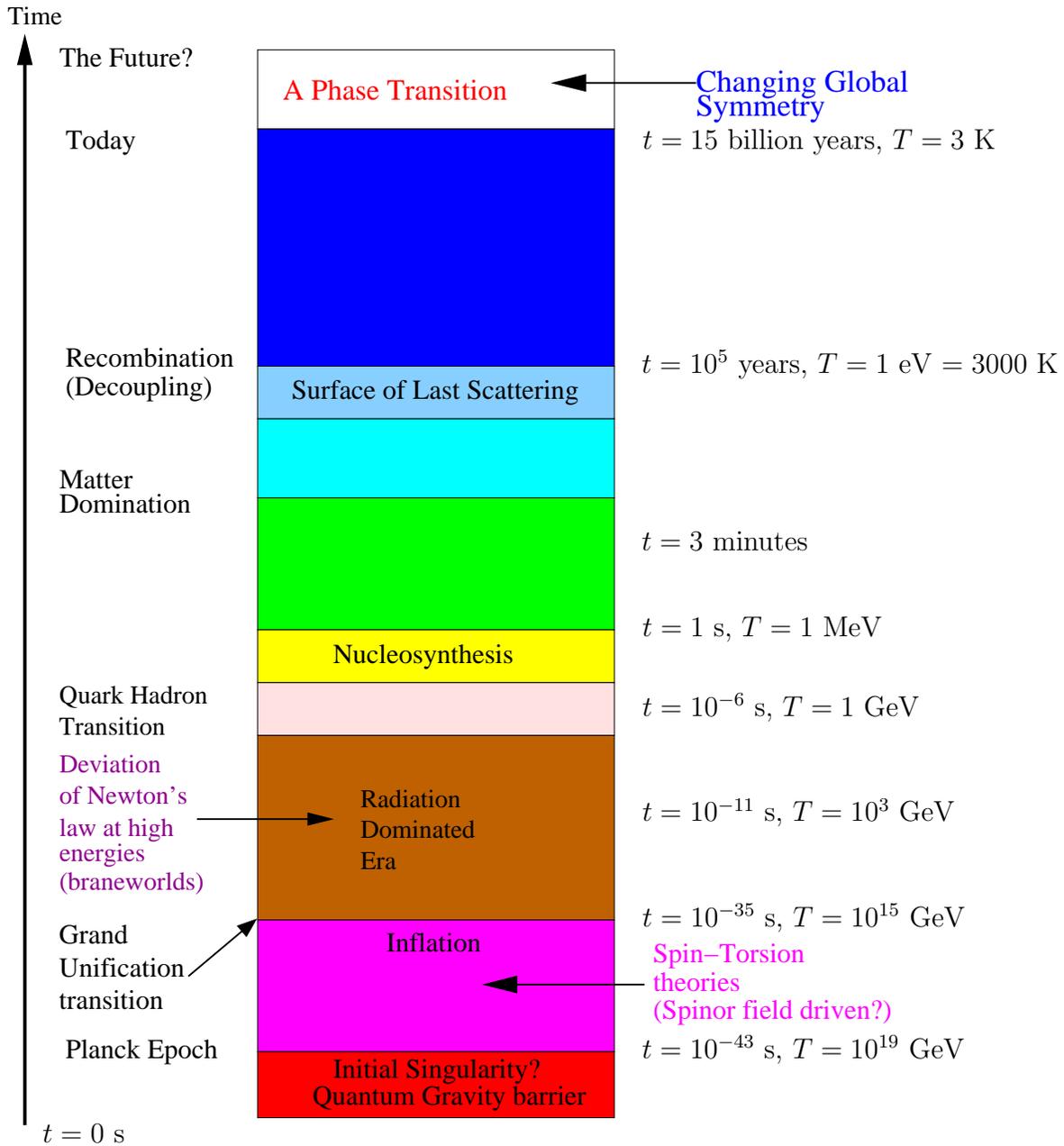}
\caption{A diagram (not drawn to scale) illustrating how the three alternative themes to
the standard inflationary big-bang cosmology is altered with respect
to fig.\ref{chap1-0a}.}
\label{chap1-0b}
\end{figure}

The central aim of this thesis is to look at three alternative themes
to the present cosmology. Although there exist recent alternative
models of inflation \cite{banks, bojowald, brandenberger, gasperini, hollands,
khoury1} which are constructed in a top-down approach from a plausible 
theory of quantum gravity, the focus of this thesis is taking
the themes in alternative cosmologies from the bottom-up approach (or
phenomenological theories) after the Planck scale. We adopt a
modest approach in the subject and would like examine not all but some
of these alternative models and their implications. We hope to explore
phenomenological possibilities in these alternative models if
possible. There are reasons why these 
 alternative themes would be an interest of study (see
fig. \ref{chap1-0b} to see how they fit into the big picture). The first reason is
to explore alternative mechanisms to the standard inflationary big
bang cosmology. One could ask whether we could substitute other
possible inflationary driven fields that might give us an alternative
to the standard scalar fields, for example, this theme will be explored in
the {\bf spin-torsion} or {\bf Einstein-Cartain-Kibble-Sciama} (ECKS)
theories. The second reason is that some of these alternative themes
are required as basic premises to be integrated into theories which
are plausible candidates for quantum gravity theories. For example, the
premise of {\bf extra-dimensions} is an essential feature in M/String
theory. Finally, we explore the topic of {\bf changing global
symmetry} where we look at the possibility of our Universe undergoing
a phase transition. We will discuss these topics in detail in the
following sections.    


\subsection{Spin-Torsion Theory} \label{spintorsiontheory}

The concept of torsion is introduced as an extension to Einstein's
theory of relativity for various reasons \cite{garecki}. These
theories of gravitation included with 
torsion are built upon Cartan's earlier suggestion in 
\cite{cartan} that the torsion (the anti-symmetric part of the
connection) should be identified as a possible physical field. The
first reason, perhaps a historical one, is that it creates an
analogy of general relativity 
with the theory of dislocations for the study of materials in
continuum mechanics. Yet, there is no reason to believe that the
structures are inherently similar for both general relativity and
continuum mechanics. The next and perhaps more plausible reason is
that torsion theories are good candidates to formulate gravity as a
gauge theory, for e.g. \cite{cho, eguchi, hayashi,  ivanenko, ivanov,kibble, mansouri, utiyama}. Proceeding from the premise that torsion theories
are possible candidates of a gauge theory of gravity, one possible
extension of general relativity, known as {\bf spin-torsion}
theory can be found by the gauge-theoretic approach in \cite{hehl}. In
spin-torsion theory, the connection between torsion and quantum
spin was identified in \cite{kibble, sciama, weyl} when it was
realised later that the stress energy tensor for a massive fermion field was
not symmetric \cite{costa, weyssenhoff}.

In the presence of torsion, the standard cosmology would be
significantly altered. The first results were discussed in 
\cite{hehl2, kerlick1, wolf}. The singularity theorems in the presence of
torsion may be suppressed in a wide class of models \cite{hehl}. In
addition, Kerlick ~\cite{kerlick2} has shown 
that if a Dirac field provides the source of matter, then the energy
condition in the singularity theorems is weakened by the presence of
torsion, leading to an enhanced singularity formation rate. Hence it
is of considerable interest to study cosmological solutions and their
implications in such scenarios.

Recently, a gauge theory of gravity (GTG) has been developed by
Lasenby, Doran and Gull~\cite{doran1}. The gauge theory of gravity is
based on the reformulation of physics in Clifford algebra pioneered by Hestenes
and his collaborators in \cite{hestenes3,
hestenes4, hestenes1, hestenes2}. In the GTG approach, the gravitational effects are
described by a pair of gauge fields defined over a flat Minkowski
background spacetime. These gauge fields ensure the invariance of the
theory under arbitrary local displacements and rotations in the
background spacetime, and the physical predictions are all extracted
in a gauge invariant manner, i.e. the background spacetime does not
play any dynamic role in the physics. In \cite{doran1}, the torsion
sector was set to zero; generally, the GTG then reproduces
most of the standard results in general relativity, but differs in the
interpretation of global issues such as the role of topology and
horizons. 

Further work has been done in \cite{challinor, doran2} to explore the
torsion sector of GTG. One of the movitations was to look at
spinor-driven inflationary models and see the comparisons with
possible experiments and observations. Unlike in general relativity where torsion is
interpreted as a property of a non-Riemannian manifold, the torsion in
GTG is a physical field which is derived from the gravitational gauge
fields. The added feature of GTG is that it constrains physically the
form which the torsion can take. This constraint is a consequence of the
demanding minimally coupled equations for the matter fields when derived
from a minimally-coupled, gauge-invariant action. The spin-torsion
theory in GTG led to interesting developments in the problem of
finding cosmological solutions for a Dirac field coupled to gravity in
a self consistent model. A new exact solution to spin
torsion theory (or Einstein-Cartan-Dirac equations) was found in
\cite{challinor} that describes a homogeneous Universe with spin induced
anisotropy.  

In Chapter \ref{chapter2}, we extend the work in \cite{challinor}
to find the cosmological solutions for a Dirac field 
coupled self-consistently to gravity with a cosmological constant. We
also discuss the implications of these cosmological solutions.

\subsection{Extra Dimensions} \label{extradimensions}

The holy grail of modern physics is to 
unify the four fundamental forces namely, gravity, electromagnetism,
the strong and weak nuclear forces. Unfortunately, the two great pillars
of theoretical physics, namely general relativity which is the theory
of gravitation and quantum theory which describes the sub-atomic
regime of matter, are mutually incompatible. The search for the theory
of quantum gravity is still ongoing. 

The early Universe provides a testing ground for theories of
gravity. The standard cosmological model, based on general
relativity with an inflationary era, is very effective in
accounting for a broad range of observed features of the Universe.
However, the lack of a consistent theoretical framework for
inflation, together with the ongoing puzzles on the nature of dark
matter and dark energy, indicate that cosmology may be probing the
limits of validity of general relativity.

M-theory (or superstring theory) is considered to be a promising
potential path to quantum gravity. It assumes the basic premise that
matter is made up of strings instead of point particles at the Planck
scale (i.e. the scale where all the four forces are unified). However, 
in order for M-theory to be a theory of quantum gravity, there are two
pre-requisites, namely supersymmetry and extra 
dimensions. Both pre-requisites have not been experimentally
verified. The Large Hadron Collider (LHC) in CERN (which will run in 2006)
would test the possibility of finding supersymmetric particles at high
energies. The question is whether we can find experiments that could
demostrate the existence of extra dimensions~\footnote{For a more
technical review on the subject of extra dimensions in the high 
energy physics framework, see \cite{rubakov}.}. 

As such, it is an important candidate for cosmological
testing. In the absence of a sufficiently general M-theoretic
model of cosmology, we can use phenomenological models that share
some of the key features of M-theory, including branes. In brane
cosmology, the observable Universe is a 1+3-dimensional ``brane"
surface moving in a higher-dimensional ``bulk" spacetime.
Standard-model fields are confined to the brane, while gravity
propagates in the bulk. The simplest, and yet sufficiently
general, phenomenological braneworld models are those based on the
Randall-Sundrum~II scenario~\cite{randall}. These models have the
additional advantage that they provide a framework for
investigating aspects of holography and the AdS/CFT
correspondence.

\begin{figure}[!bth]
\begin{center}
\includegraphics[scale=1.0]{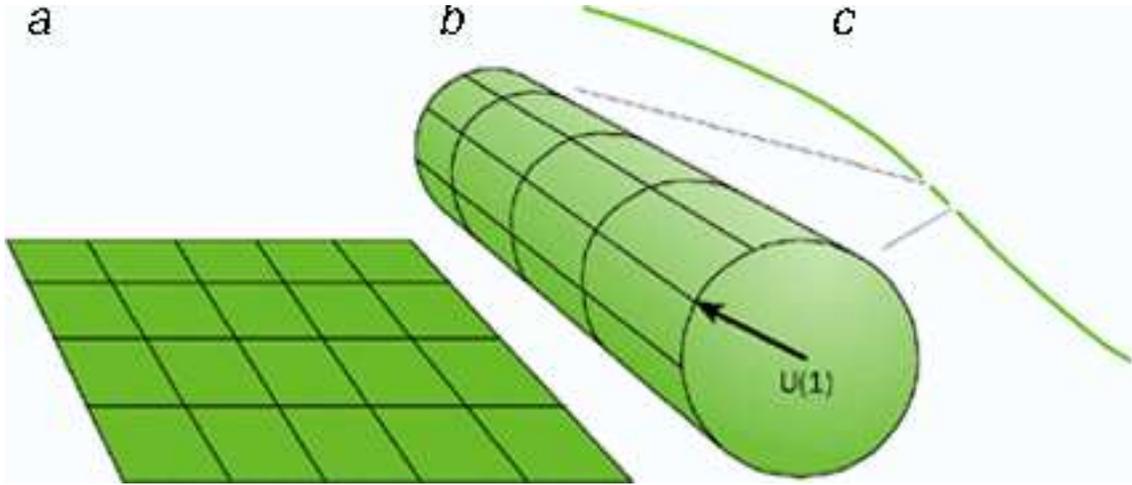}
\end{center}
\caption{A   diagram   depicting   the   Kaluza-Klein   concept   of
compactification of extra dimensions: (a) Compactifying a 3-D Universe
with two space dimensions and  one time dimension which simplifies the
picture of  Kaluza and Klein.  (b)  The Lorentz symmetry  of the large
dimension is  broken by the  compactification and what remains  is the
2-D  space plus the  U(1) symmetry  represented by  the arrow.  (c) On
large  scales we  see only  a 2-D  Universe (one  space plus  one time
dimension)     with     the     ``internal''    U(1)     symmetry     of
electromagnetism. The diagram is reproduced from the article ``The
search for extra dimensions'' by Abel and Russell in \cite{abel}.}
\label{kk}
\end{figure}

The idea  of using  extra dimensions to  unify fundamental  forces was
first  explored by  Kaluza~\cite{kaluza} and  Klein~\cite{klein}  in the
1920s.  Kaluza proposed  that one  can start  from a  five dimensional
theory of general relativty, and  assume that one of the dimensions is
being  curled up into  a circle,  leading  to a  theory that
unifies gravity and electromagnetism.  Klein took a step further after
rediscovering  Kaluza's  theory,  he  noted the  quantisation  of  the
electric charge  and hoped that Kaluza's theory  would underlie quantum
mechanics. He came close to the modern point of view in his discussion
of higher  harmonics and  the size of  the small circle,  and insisted
that the  5th dimension should  be treated seriously. He  assumed that
the coordinate  of the 5th dimension  is periodic. It  is difficult to
envisage a 5D  spacetime with such a topology. However,  one can use a
simple  analogy  provided  by  a  hosepipe: at  large  distances,  the
hosepipe looks like  a line but close  inspection reveals at  every point on
the line, there is a circle (see Fig. \ref{kk}). The Kaluza Klein picture
requires the dimensions to be compact and essentially homogeneous. In
this picture, it is the compactness which ensures that the spacetime
is effectively 4-dimensional at distances exceeding the
compactification scale, i.e. the size of the extra dimensions. As a
result, the size of the extra-dimensions must be very small, and the
size was of the order of the Planck scale. With the Planck length 
$l_P \sim 10^{-35}$m and the corresponding Planck energy scale $M_P
\sim 10^{19}$GeV, it remains almost impossible to probe the extra
dimensions with present technology. 

\begin{figure}[!bth]
\begin{center}
\includegraphics[scale=1.0]{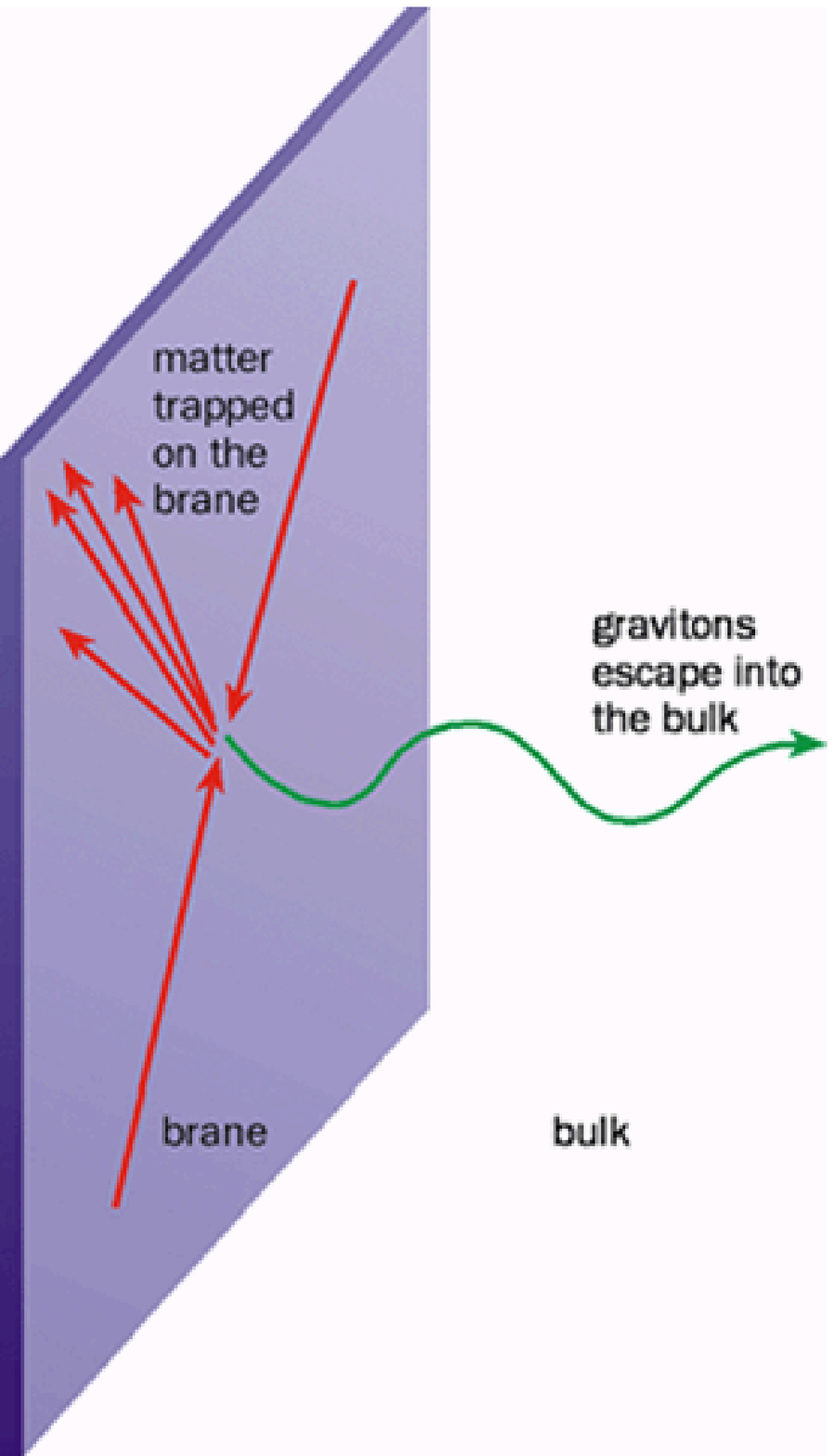}
\end{center}
\caption{The matter fields (the electromagnetic, strong and weak
nuclear forces) in the standard model are confined on the
3-brane (our observable Universe). Only gravity is allowed to
propagate into the extra dimensions (i.e. the bulk). A collision
between a proton and an antiproton can produce, for example, 
a single jet of matter particles plus graviton emission into the bulk.
Such collisions might be seen in high-energy physics
experiments. The diagram is reproduced from the article ``The
search for extra dimensions'' by Abel and Russell in \cite{abel}.
} 
\label{braneworld}
\end{figure}

In recent years, much attention has been focused on {\bf brane
world scenairos}. They are based on the idea that the standard model of
particle physics (which unifies the electromagnetic, strong and weak
nuclear forces) is confined to a 3-brane (an object or membrane which
is a solution in string theory that corresponds to our observable
Universe) and only gravity lives in the higher dimensional bulk. In
general, the spatial dimensions transverse to the branes must be very
small in order to avoid large deviations from Newton's law of
gravitation (see \ref{braneworld}). One interesting suggestion was put
forward in \cite{hamed} that two of the extra dimensions are large (mm
scale). Since the four dimensional effective gravitational constant is
related to the fundamental higher dimensional gravitational constant
through the size of the extra dimensions, the fundamental gravitational
constant can be reduced to the TeV scale, provided that the large
extra dimensions have a size of $\sim 1$ mm. The connection to superstring
theory was made in \cite{antoniadis}. This new proposal allows us to
deal with the mass hierarchy problem in elementary particle physics
without invoking supersymmetry. 

By projecting the brane world models to an effective 4D theory, we can
study how the extra dimensions may affect the standard observational
cosmology. Such an approach is the first step to understanding  these new
models and seeing  whether one can find any new features in them that
may show up in future experiments in observational cosmology. In
Chapter \ref{chapter3}, we will set up the formalism of the
(1+3)-covariant approach for studying cosmological perturbations
in these braneworld models. Subsequently, in Chapter
\ref{chapter4}, we will explore the effects of the extra
dimensions on the cosmic microwave background for both scalar and
tensor perturbations. 

\subsection{Changing Global Symmetry} \label{globalsymmetry}

In standard general relativity, the Friedmann-Robertson-Walker (FRW)
metric is a well-defined metric describing an expanding, homogeneous,
isotropic Universe. The symmetry of a FRW spacetime is both translational and
rotational about every point and there is no preferred centre (Copernican). 

However, local inhomogeneities grow  to form galaxies and clusters
of  galaxies and  hence generate  the large  scale structure  which we
observe  today.  In  an  Einstein-de Sitter  (flat, dust dominated) 
Universe,  these
clustered structures would grow continuously, leading to the merger of
galaxy clusters to form superclusters. Subsequently, this leads to the
formation  of  an   ever-changing  hierarchical  structure.   Such  an
astrophysical mechanism results in a changing equation of state from a
pressure-free  expansion  to  a  transition state  where  gravitational
clustering induces  a cosmologically significant  pressure which feeds
back  into the  metric.  At this  stage,  the symmetry  of the  metric
alters,  and   depends  on   the  exact  form   of  the   equation  of
state. Extrapolating this picture into the future, we can envisage the
following possible development.  By taking the clustering model to
the extreme, one could imagine a time in the far future when the whole
Universe is  basically one large cluster. The internal  constituents of
this  cluster exchange  energy. The  whole Universe  virializes  on the
largest  possible  scale  and we  get  an asymptotic state represented
approximately by an isothermal Universe. For such a  state, the metric
is well-defined, as the   final  symmetry  is   rotational  about
one  point   only  and translational    nowhere   and    there   is
a    definite   centre (Anti-Copernican).

The effectiveness of gravitational clustering is related to the global
structure  of the standard  Einstein-Friedmann Universes.  There exist
three general cases mentioned earlier for the future evolution of the
Universe, and they depend on the total energy density.  In a closed
geometry the Universe 
will  eventually recollapse  into  a singularity  and all  large-scale
structure will be destroyed in the big crunch. In an open geometry the
rate  of   universal  expansion  is  too  great   to  allow  complete
clustering.  In a  flat  geometry (Einstein-de  Sitter) the  expansion
timescale  is comparable  to the  clustering  timescale. Gravitational
clustering  grows  continuously.   Therefore,  although  gravitational
clustering of  galaxies occurs in  all Einstein-Friedmann cosmological
models,   only  in  the   Einstein-de  Sitter   model  does   it  grow
continuously. Only the Einstein-de Sitter model allows the formation
of the isothermal asymptotic state. However, the existence  of a
cosmological constant would change this evolution.

Saslaw  and his collaborators\cite{saslaw2, saslaw4, saslaw} have
found a solution to Einstein's field equations, which corresponds to a
class of isothermal inhomogeneous Universes in which the nonzero pressure balances
gravity and discussed whether the infinite and
unbounded isothermal sphere could provide a possible state of our
Universe. They suggested various arguments which demonstrate the
possibility that such spherical and static models might represent the
ultimate state of an Einstein-de Sitter Universe (with $\Lambda=0$).  
If that is the case, the Einstein-de Sitter Universe will undergo a
dynamical symmetry breaking into the isothermal Universe, i.e. from
Copernican to anti-Copernican, which may correspond to a possible phase
transition in the far future. This is (at least to our knowledge) the
first time the question has been asked about what happens when a local
condensation with one symmetry grows to eventually fill a Universe
with another symmetry. 

In Chapter \ref{chapter5}, we examine the changing global symmetry
of the Einstein-de Sitter Universe and  see  how  it could  lead  to
a possible phase transition in the future.   

      \clearpage{\pagestyle{empty}\cleardoublepage}
      \part{Spin-Torsion Theories}	
\chapter{Spin-Torsion Cosmology} \label{chapter2}

\begin{quote}
{\it ``There could be  no fairer destiny for any  ... theory than that
it should point the way to a  more comprehensive theory which it lives
on as a limiting case.  ''}
\par{\bf - Albert Einstein}
\end{quote}

{\it In this chapter, we extend the gauge theory of gravity (GTG)
approach of finding cosmological solutions in spin-torsion theories
to include a cosmological constant. We find a new anti-particle solution in
addition to the two solutions found in \cite{challinor}. We also discuss the
cosmological implications of these solutions.}

\section{Introduction}

The gauge theory of gravity (GTG) provides an alternative formulation to the
standard spin-torsion or Einstein-Cartan-Kibble-Sciama theories. In
the GTG approach developed in \cite{doran2, lasenby1, doran1}, one can
identify the torsion with the spin of the matter field that follows
naturally from the gauge-theoretic approach to gravity (see
\cite{hehl2}). Fundamentally in GTG, the gravitational effects
are described by a pair of gauge fields which are defined over a flat Minkowski background
spacetime. The gauge fields ensure the invariance of the theory under
arbitrary local displacements and rotations in the background
spacetime. All the physical predictions are extracted in a
gauge-invariant manner, thus ensuring that the background spacetime 
plays no dynamical role in the physics. In \cite{doran2}, the equations
describing the Dirac field coupled self-consistently to gravity were given, and
the analagoues of the Einstein-Cartan-Dirac (ECD) equations were derived
from a minimially-coupled, gauged-invariant action. It is known that
the Einstein-Dirac equations, which describe a Dirac field coupled to
gravity through the (symmetrised) stress-energy tensor only, cannot be
derived from a minimally coupled action. Hence this serves as a
compelling reason to regard the ECD equations as being more fundamental.

One of the main applications of GTG is in finding cosmological
solutions for a Dirac field coupled to gravity in a self-consistent
model. This problem was considered by 
several authors \cite{ale, chimento, dimakis1, dimakis2, henneaux,
isham, ochs, pullin, radford,seitz} in the past. One of the main motivations
was to see whether the Dirac field could offer an alternative type of
inflationary model over	standard scalar field models. However, the
solutions found by these authors 
either solve the Einstein-Dirac equations \cite{ale, chimento,
henneaux, ochs, radford} which do not include the effect of torsion
induced by the spin of the Dirac field, or they do include the
spin-induced torsion (i.e. they solve the 
Einstein-Dirac-Cartan equations) but only for massless fields 
\cite{henneaux, isham, seitz} or for ghost solutions
\cite{dimakis1, dimakis2, pullin}. (A ghost solution 
has a vanishing stress energy tensor for the Dirac field.) The first
massive, non-ghost solution was recently found for a Dirac field
coupled self-consistently to gravity in \cite{challinor}. 

The origin of a cosmological constant, $\Lambda$, is a mystery in
gravitation and modern particle physics.  It was introduced originally
by Einstein to allow static homogeneous cosmological solutions
to the field equations in the presence of matter. The
discovery of the expanding Universe by Hubble, made it
unnecessary, but it did not change its status as a legitimate
addition to the gravitational field equations, or as a parameter
constrained by observation. In particle physics, the cosmological
constant turns out to be a measure of the energy density of the vacuum
i.e. the state of the lowest energy. This identification becomes
crucial in dealing with two present unsolved problems: firstly, why the
vacuum energy is so small and secondly, to 
understand why it is comparable to the present day density of the
Universe. It turns out that there are compelling reasons to believe
why our Universe should have a non-zero $\Lambda$ in both theory and experiment.\footnote{A good
review of the cosmological constant is given by 
Carroll~\cite{carroll}.} Recent results from the cosmic microwave
background, supernova data and large scale structure seem to favour a
cosmological constant with $\Omega_{\Lambda}=0.7$.      

In this chapter, we will extend the work in \cite{challinor} to find a
massive non-ghost solution for a Dirac field coupled self-consistently
to gravity with the inclusion of a cosmological constant. We will
briefly summarize the geometric algebra formalism and the gauge theory
of gravity in \ref{sec:gagtg}. Subsequently, we will introduce
the minimally coupled Einstein-Cartan-Dirac (ECD) action in
\ref{sec:diracfield}, and the GTG analogue of the field equations. We
will derive the cosmological solutions in \ref{sec:cosmosol}
and discuss their implications in \ref{sec:cosmoimp}. 

\section{Geometric Algebra and Gauge Theory of Gravity} \label{sec:gagtg}

To find the cosmological solutions for a Dirac field
coupled to gravity in a self consistent model with a cosmological
constant, we shall employ the tools of Geometric Algebra (GA), 
which was originally developed from the spacetime
algebra (STA) approach by Hestenes (see \cite{hestenes3, hestenes4,
hestenes1, hestenes2}). We briefly summarize the geometric algebra
formalism in \ref{sec:ga} and the gauge theory of gravity in
\ref{sec:gtg}. Note that only in this chapter, we will use the other
metric signature convention $(+,~-,~-,~-)$.

\subsection{Geometric Algebra} \label{sec:ga}

We start with the definition of a geometric (or Clifford) product in
\cite{clifford}  that combines both symmetric inner product and
antisymmetric outer product whuch is defined as follows:
\begin{equation}
ab = a \dt b + a \w b
\end{equation}
where both $a$ and $b$ are vectors. 

In general, an arbitrary real superposition of the basis elements $\eqref{e:sta2}$
is called a {\it multivector}. A multivector can be decomposed into
sumns of elements of different grades, for example, scalars are grade
zero objects, vectors grade one, bivectors grade two and etc. These
elements inherit the associative Clifford product of the $\{\g_{\mu}
\}$ generators. For a grade-$r$ multivector $A_r$ and a grade-$s$
multivector $B_s$, we define the inner and out products via 
\begin{align}
\label{e:ga1} A_r \dt B_s &\equiv \l<A B \r>_{|r-s|} \;,\\
\label{e:ga2} A_r \w B_s &\equiv  \l<A B \r>_{r+s}   \;,
\end{align} 
where $\l< M \r>_r$ denotes the grade-$r$ part of M. The subscript 0
will be left implicit when taking the scalar part of a multivector. We
shall make use of the commutator product
\begin{equation} \label{e:ga3}
A \times B \equiv \frac{1}{2} (AB - BA) \;.
\end{equation}
The operation of reversion, denoted by a tilde, is defined by 
\begin{equation} \label{e:ga4}
(AB)^{\sim} \equiv \tilde{B} \tilde{A} \;,
\end{equation}
and the rule that vectors are unchanged under reversion. We adopt
the convention that in the absence of brackets, inner, outer and
commutator products take precedence over Clifford products. 

The geometric algebra of spacetime is familiar to
physicists in the form of the algebra generated from the Dirac-$\g$
matrices. The spacetime algebra (STA) is generated by four vectors,
$\{ \g_{\mu} \}$, $\mu=0...3$, equipped with an associative (Clifford)
product, denoted by juxtaposition. The symmetrised and antisymmetrised
products define the inner and outer products between vectors, and are
denoted by a dot and a wedge respectively:
\begin{equation}
\begin{split}
\label{e:sta1}
\g_{\mu} \dt \g_{\nu} &\equiv \frac{1}{2} (\g_\mu \g_\nu + \g_\nu
\g_\mu) = \eta_{\mu \nu} =~\text{diag}(+~-~-~-) \;, \\
\g_{\mu} \w \g_{\nu} &\equiv \frac{1}{2} (\g_\mu \g_\nu - \g_\nu
\g_\mu) \;.
\end{split}
\end{equation}
The outer product of two vectors defines a bivector --- a directed plane
segement including the two vectors. 

The full basis for a 16-dimensional STA is provided by
\begin{equation}
\label{e:sta2}
\begin{array}{ccccc}
1 & \{ \g_\mu \} & \{ \sigma_k, \ps \sigma_k \} & \{ \ps \g_\mu \} & \ps \\
1~\text{scalar} & 4~\text{vectors} & 6~\text{bivectors} &
4~\text{trivectors} & 1~\text{pseudoscalar} 
\end{array}
\end{equation}
where $\sigma_k \equiv \g_k \g_0$, $k=1 \dots 3$, and $\ps \equiv \g_0 \g_1
\g_2 \g_3 = \sigma_1 \sigma_2 \sigma_3$. The pseudoscalar $\ps$ squares
to $-1$ and anticommutes with all odd-grade elements. The $\{ \sigma_k
\}$ generate the geometric algebra of Euclidean 3-space, and are
isomorphic to Pauli-matrices.

We denote vectors in lower case Latin, $a$, or Greek for a set of
basis vectors. A (coordinate) frame of vectors $\{ e_{\mu} \}$ is
generated from a set of coordinates $\{ x^{\mu} (x) \}$ via $e_{\mu}
\equiv \p_{\mu} x$, where $\p_{\mu} \equiv \pp{}{x^{\mu}}$. The
reciprocal frame, denoted by $\{e^{\mu} \}$, satisfies $e_{\mu} \dt
e^{\nu} = \delta^{\nu}_{\mu}$. The vector derivative $\n (\equiv
\p_x)$ is then defined as 
\begin{equation}\label{e:ga5}
\n \equiv e^{\mu} \p_{\mu} \; .
\end{equation}
More generally, the vector derivative with respect to $a$ is denoted
$\p_{a}$. Further details concerning geometric algebra and the STA may
be found in \cite{hestenes4, hestenes2} and a summary of useful
identities is given in appendix \ref{appendix2}. 

\subsection{Gauge Theory of Gravity} \label{sec:gtg}

The language of geometric algebra is used by Lasenby, Doran and Gull
\cite{doran1} to develop an alternative theory of gravity based on gauge
principles alone. The theory is known as {\bf Gauge Theory of Gravity}
(GTG). The key ideas in GTG are summarized as follows:
\begin{enumerate}
\item The gravitational effects are described by a pair of gauge
fields, defined over a flat Minkowski spacetime namely, the position
gauge field, $\bar{h} (a)$ and the rotation gauge field,
$\Omega(a)$. 
\begin{enumerate}
\item The first of these, the position gauge field, $\bar{h}(a)$
 describes a position dependent linear
function mapping the vector $a$ to vectors. The overbar serves to
distinguish the linear function, $\underline{h}(a)$ from its adjoint,
$\bar{h}(a)$, where
\begin{equation} \label{e:ga6}
\underline{h} (a) \equiv \p_b \bar{h} (b) \dt a \;.
\end{equation}
The gauge-theoretic purpose of $\bar{h}(a)$ is to ensure covariance
of the equations under arbitrary local displacements of the matter
fields in the background spacetime. \\
\item The second gauge field, the rotation gauge field, $\Omega(a)$,
is a position dependent linear 
function mapping the vector $a$ to bivectors. The
purpose of this gauge field is to ensure covariance under local
(Lorentz) rotations of fields at a point in the background
spacetime \cite{doran1}. A gauge field $\omega(a)$ is introduced to map vectors to
bivectors and is defined as
\begin{equation} \label{e:ga7}
\omega(a) = \Omega \underline{h}(a) \;.
\end{equation}
\end{enumerate} 
\item Physical predictions are extracted in a gauge invariant manner
which ensures that the background spacetime plays no dynamical role in
the physics.
\item The two gauge fields are minimally coupled to the Dirac field to
enforce invariance under local displacements and both spacetime and
phase rotations~\footnote{This requires a vector gauge field for
electromagnetism.}. One can formulate the Dirac action in a way that it
ensures that the internal phase rotations and spacetime rotations
assume equivalent forms. 
\item In the process, one can construct a Lagrangian density for
gravitational fields. The first surprise of adopting such an approach
is that the demand of the minimally-coupled Dirac equation restricts
us to a single action integral. There is no freedom except for the
inclusion of a cosmological constant. 
\item This leads to a set of field equations which are completely
independent of how we choose to label the position of fields by a
vector $x$.
\end{enumerate} 
The resulting theory based on the above key ideas is a first order
theory which is consistent with quantum mechanics at the first
quantized level. One can develop additional and alternative insight to
the general theory of relativity from the GTG. The ideas are applied
in the context of black holes, collapsing matter and cosmology in the
second part of \cite{doran1}. 

The covariant derivative $\mathcal{D}$ is assembled from the (flat
space) vector derivative $\n$ and the gravitational gauge fields. The
action of $\mathcal{D}$ on a general multivector $M$ is given by 
\begin{equation}
\begin{split} \label{e:ga7a}
\mathcal{D} M &\equiv \p_a a \dt \mathcal{D} M \\
              &\equiv \p_a [a \dt \bar{h} (\n) M + \omega (a) \times
              M] \;,
\end{split} 
\end{equation}
where we have introduced the rotational gauge field $\omega (a)$
defined in $\eqref{e:ga7}$.

The covariant derivative contains a grade-raising and lowering component, so that
we may write 
\begin{equation} \label{e:ga8}
\mathcal{D} M = \mathcal{D} \dt M + \mathcal{D} \w M   \; ,
\end{equation}
where 
\begin{equation} \label{e:ga9}
\mathcal{D} \dt M \equiv \p_a \dt (a \dt \mathcal{D} M) \;, \quad \quad
\mathcal{D} \w M \equiv \p_a \w (a \dt \mathcal{D} M) \;.
\end{equation} 
The single-sided transformation law of spinors under rotations
requires us to introduce a seperate spinor covariant derivative $D
\psi \equiv \p_a a \dt D \psi$, where
\begin{equation} \label{e:ga10}
a \dt D \psi \equiv a \dt \bar{h} (\n) \psi + \frac{1}{2} \omega(a)
\psi \;.
\end{equation}

The field strength corresponding to the $\Omega (a)$ gauge field is
defined by 
\begin{equation} \label{e:gariemann0}
R( a \w b) \equiv a \dt \n \Omega (b) - b \dt \n \Omega (a) +
\Omega(a) \times \Omega(b) \;. 
\end{equation}
 From this, we define the covariant Riemann tensor
\begin{equation}
\label{e:gariemann}
\mathcal{R} (a \w b) \equiv R \underline{h} (a \w b) \;.
\end{equation}
The Ricci tensor, Ricci scalar and Einstein tensor are then given by
\begin{align}
\label{e:garicciten}
& \mathcal{R}(a) = \p_b \dt \mathcal{R} (b \w a) \;, \quad (\text{Ricci
tensor}) \\
\label{e:gariccisca}
& \mathcal{R} = \p_a \dt \mathcal{R} (a) \;, \quad (\text{Ricci scalar})
 \\
\label{e:gaeinstein}
& \mathcal{G}(a) = \mathcal{R}(a) - \frac{1}{2} a \mathcal{R} \;. \quad
(\text{Einstein tensor}) 
\end{align}

\section{The Self-Consistent Dirac Field \\
and the Field equations}
\label{sec:diracfield} 
The equations of motion for the Dirac field and the gauge fields are
derived from a form of the minimally coupled action from \cite{doran2,
doran1}:  
\begin{equation} \label{e:SECDaction}
S_{\mathcal{E} \mathcal{C} \mathcal{D}} = \int |d^4 x|~ \text{det}(\underline{h})^{-1}~
\l[\frac{\mathcal{R}}{2} - \kappa^2 \langle D \psi i \g_3 \tilde{\psi} -
m \psi \tilde{\psi} \rangle - \Lambda \r] \;.
\end{equation}
Varying Eq. $\eqref{e:SECDaction}$ with respect to $\Omega(a)$,
$\bar{h}(a)$ and $\psi$ respectively (refer to the procedure in \cite{doran1}), 
the following equations describe a Dirac field of mass $m$ coupled
self-consistently to gravity~\cite{doran2} with the inclusion of the
cosmological constant in this paper.    
\begin{align}
\label{e:stwedge}
\mathcal{D} \w \bar{h} (a) &= \kappa^2 \mathcal{S} \dt \bar{h}(a) +
\frac{\kappa^2}{2} [\p_b \cdot \mathcal{S}(b)] \w \bar{h}(a) \;, &
\quad \quad (\text{'wedge'}) \\ 
\label{e:ste}
\mathcal{G}(a) - \Lambda a &= \kappa^2 \mathcal{T} (a) \;, \quad  &\quad
(\text{Einstein})\\  
\label{e:stdirac}
D \psi i \sigma_3 &= m \psi \g_0 \;, & \quad \quad (\text{Dirac}) 
\end{align} 
where we set $\kappa^2=8 \pi G$, $\Lambda$ is the cosmological
constant. The term $\mathcal{S}(a)$ is the covariant spin tensor, and is
defined by
\begin{equation}
\mathcal{S}(a)= S \bar{h}^{-1} (a) \;,
\end{equation}
and $\mathcal{S}$ is the spin trivector and is defined by,  
\begin{equation} \label{e:spintri}
\mathcal{S} = \frac{1}{2} \psi i \g_3 \tilde{\psi} \;,
\end{equation}
and the matter stress-energy tensor is defined to be
\begin{equation}
\mathcal{T}(a) = \l< a \dt D \psi i \g_3 \tilde{\psi} \r>_1 \;.
\end{equation}
The above equations are the GTG analogue of the {\bf
Einstein-Cartan-Dirac} equations. The second term on the RHS in
Eq. $\eqref{e:stwedge}$ is set to zero because the
minimally-coupled equations for the matter fields are obtained from a
minimally-coupled Lagrangian only if the contraction of the spin
tensor vanishes, i.e. $\p_a \dt \mathcal{S}(a)=0$. One could include
cases for non-vanishing spin tensor since the spin tensor vanishes
for scalar fields and Yang-Mills gauge fields. For spin-$1/2$ fields,
we can use the form $\mathcal{S} (a) = \mathcal{S} \dt a$, where
$\mathcal{S}$ is the spin trivector. Further results for a
non-vanishing tensor are given in \cite{doran2}.  

The wedge equation in the spin-torsion theory is algebraic in
$\omega(a)$, and may be solved for a vanishing spin tensor to
give 
\begin{equation} \label{e:stw}
\omega (a) = - \bar{h} [\n \w \bar{h}^{-1} (a)] + \frac{1}{2} a \dt \{
\partial_b \w \bar{h} [\n \w \bar{h}^{-1} (b)] \}+ \frac{1}{2} \kappa^2 a
\dt \mathcal{S} \;.
\end{equation}

\section{Massive Non-Ghost Solutions to ECKS theory} \label{sec:cosmosol}

In this section, we derive a self-consistent solution for a massive Dirac
field $\psi$, which is both homogeneous and isotropic at the level of
classical fields, with the inclusion of a cosmological constant. We
note that the classical fields do not feel the anisotropy of the
$\omega$-function which arises due to the spin of the Dirac field, as
the classical fields couple to gravity via the $\bar{h}$-function
only. We first summarize the procedure in \cite{challinor} which we
adopt and thn extend the solutions to the case with a cosmological
constant. 

We begin by introducing a set of polar coordinates, 
\begin{equation}
\begin{split} 
t &\equiv x \dt \g_0 \;,  \qquad \quad \cos \theta \equiv \frac{x \dt
\g^3}{r} \;, \\
r &\equiv \sqrt{(x \w \g_0)^2} \;, \quad  \tan \phi \equiv \frac{x
\dt \g^2}{x \dt \g^1} \;.
\end{split}
\end{equation}
We also make use of the following vectors which are members of the polar
coordinate frame:
\begin{equation}
e_t \equiv \g_0 \;, \quad \quad \quad e_r \equiv \frac{x \w \g_0 \g_0}{r} \;.
\end{equation}

We choose the gauge based on the assumed symmetry for the
$\bar{h}$-function for a vector $b$, and giving~\cite{doran1}: 
\begin{equation}
\bar{h}(a) = b \dt e_t e_t + a(t) \l(1 + \frac{kr^2}{4} \r) b \w e_t
e_t  \;,
\end{equation}
where $a(t)$ is the scale factor of the Universe, and the curvature
constant $k =-1,0,+1$ defined for open, flat and closed Universes
respectively.  Note that $\bar{h}(a)$ is globally defined for all
$k$.  

The isotropic line element is generated by the $\bar{h}$-function
\begin{equation}
ds^2 = dt^2 -  [a(t)]^{-2} \l(1 + \frac{kr^2}{4} \r)^{-2} [dr^2 + r^2
(d \theta^2 + \sin^2 \theta d \phi^2)] \;.
\end{equation}
In this gauge, the fundamental observers have covariant velocity $e_t$
where $t$ is the cosmic time, and their surfaces of homogeneity are
$t=$constant surfaces.

From Eq. $\eqref{e:stw}$, we obtain $\omega(b)$ in the following form :
\begin{equation}
\omega (b) = H(t) b \w e_t - \frac{1}{2} kr a(t) e_r (b \w e_t
e_t) + \frac{1}{2} \kappa^2 b \dt \mathcal{S} \;,
\end{equation}
where $H(t)$ is the Hubble parameter in cosmic time, i.e. $H(t) \equiv
\dot{a} (t) / a(t)$ and $\dot{a}$ is the derivative of the scale
factor with respect to cosmic time, $t$.

We can cast the Dirac equation from Eq. $\eqref{e:stdirac}$ in the following form
\begin{equation} \label{e:stdirac2} 
\l[ e_t \p_t + a(t) \l(1 + \frac{kr^2}{4} \r) e_t e_t \w \n +
\frac{3}{2} H e_t + \frac{1}{2} k r a(t) e_r + \frac{3}{4} \kappa^2
\mathcal{S} \r] \psi i \sigma_3 = m \psi \g_0 \;.
\end{equation} 
Following Isham and Nelson \cite{isham}, if the gauge
invariant observables (for example, the the projection of the
Dirac current onto the velocity of the fundamental observers
constructed from $\psi$) are to be homogeneous, we must have
$\psi=\psi(t)$. Then the only term with any dependence on the spatial
coordinates would be $(k/2) a(t) e_r \psi i \sigma_3$ in
Eq. $\eqref{e:stdirac2}$. We thus arrive at the conclusion that the 
Universe must be spatially flat (i.e. $k=0$), if we require the
observables associated with the Dirac field to be
homogeneous~\footnote{One can arrive at the same 
conclusion if one attempts to solve the Dirac equation
non-self-consistently on a homogeneous gravitational background. The
reason is due to the fact that the only change to the equation
$\eqref{e:stdirac2}$ would be that $\mathcal{S}$ is now the torsion trivector of
the background rather than the spin of the Dirac field.}~\cite{isham}. The motivation
for the above gauge choice is attributed to the fact that $\bar{h}$
must be globally defined, for all choices to $k$. 

From now onwards, we specialize to the flat Universe
($k=0$). It is then convenient to work in a new gauge reached by the
displacement, 
\begin{equation}
x' = f(x) = x \dt e_t e_t + a(t) x \w e_t e_t \;,
\end{equation}
which turns $\bar{h}$-function to the simple form 
\begin{equation}
\bar{h}(b) = b + r H(t) b \dt e_r e_t \;.
\end{equation}
This generates the line-element 
\begin{equation}
ds^2 = (1 - r^2 H^2) dt^2 + 2Hr dt dr - [dr^2 + r^2 (d \theta^2 +
\sin^2 \theta d \phi^2)] \;.
\end{equation}
The above gauge choice is known as the {\bf Newtonian gauge}, and has
an analogue in GR which employs the description of a set of geodesic
clocks in a radial freefall, comoving with the fluid. 

The $\omega(a)$ transforms to give 
\begin{equation} \label{e:omegaa}
\omega(b) = H(t) b \w e_t + \frac{\kappa^2}{2} (b \dt \mathcal{S}) \;,
\end{equation}
which leaves the homogeneous Dirac field, $\psi$ unchanged. The
surfaces of homogeneity still have $t=$constant. The 
covariant velocity of the fundamental observers remains $e_t$, but
their radial coordinates are proportional to the scale factor
$a(t)$.

We now consider the spin-torsion Universe with a non-zero cosmological
constant $\Lambda$. The Riemann tensor evaluates to be 
\begin{equation} \label{e:Rie1}
\mathcal{R}(B) = - \dot{H} B \dt e_t e_t - H^2 B + \frac{1}{4}
\kappa^4 B \dt \mathcal{S} 
\mathcal{S} - \frac{1}{2} \kappa^2 (B \dt \mathcal{D}) \dt \mathcal{S} \;,
\end{equation}
for any arbitrary bivector $B$, and the Einstein tensor $\mathcal{G}(b)$ is
found to be 
\begin{equation}
\mathcal{G}(b) = 2 \dot{H} b \w e_t e_t + 3 H^2 b - \frac{1}{4} \kappa^2 b
\dt (\mathcal{D} \dt \mathcal{S}) + \frac{1}{2} \kappa^4 (b \dt
\mathcal{S}) \mathcal{S} - \frac{3}{4} \kappa^4 \mathcal{S}^2 b  \;,
\end{equation}
The stress energy tensor of the self-consistent Dirac field
evaluates to
\begin{equation} \label{e:EM1}
T(a) = \l< b \dt e_t \dot{\psi} i \g_3 \tilde{\psi} + H b \w e_t \mathcal{S} +
\frac{\kappa^2}{2} (b \dt \mathcal{S}) \mathcal{S} \r>_1 \;,
\end{equation}
where we assumed that $\psi$ is a function of time alone. The Dirac
equation reduces to
\begin{equation} \label{e:dirac2}
\l(e_t \p_t + \frac{3}{2} H e_t + \frac{3}{4} \kappa^2 \mathcal{S} \r)
\psi i \sigma_3 = m \psi \g_0  \;,
\end{equation}
from which we deduce
\begin{equation}
\begin{split}
\label{e:dirac3}
\dot{\mathcal{S}} &= -3 H \mathcal{S} - m \l<\psi \tilde{\psi} \r>_4
\g_0 \;, \\
\mathcal{D} \dt \mathcal{S} &= -2 H e_t \dt \mathcal{S} \;.
\end{split}
\end{equation}
Substituting Eq. $\eqref{e:dirac3}$ into Eq. $\eqref{e:ste}$, we
obtain the following equation :  
\begin{equation}
2 \dot{H} b \w e_t e_t + 3 H^2 b + \frac{3}{4} \kappa^4 \mathcal{S}^2
e_t b e_t - m \kappa^2 b \dt e_t e_t \l< \psi \tilde{\psi} \r> -
\Lambda b = 0 \;,
\end{equation}
from which we extract the following set of scalar equations
\begin{align}
\label{e:torsion1}
2 \dot{H} + 3 H^2 - \frac{3}{4} \kappa^4 \mathcal{S}^2 - \Lambda &= 0 \;, \\
\label{e:torsion2}
3 H^2 + \frac{3}{4} \kappa^4 \mathcal{S}^2 - m \kappa^2  \l< \psi \tilde{\psi} \r> -
\Lambda &= 0 \;.
\end{align}

If we exclude torsion from the model, from Eq. $\eqref{e:torsion1}$, the
equation becomes 
\begin{equation}
3 H^2 + 2 \dot{H} - \Lambda = 0 \;. 
\end{equation}
Since $\mathcal{S}^2 \leq 0$, the effect of torsion would
make $\dot{H}$ more negative for any given value of the Hubble parameter
$H$. Consequently, torsion enhances singularity formation for
a Hubble parameter at a particular instant. It follows that the Universe will be
younger if we look at the singularity theorems in the presence of
torsion~\cite{kerlick2}. Torsion effects become significant when the
Compton wavelength of the field is larger than the Hubble radius. At
earlier times, where $\rho$ is singular, the torsion effects are
dominant. However, the inclusion of a positive cosmological constant
would provide a repulsive force which would dominate the late time 
behaviour of the Universe. The competing effects between the torsion
and the cosmological constant lead to interesting consequences in the
models which we are about to examine.

Suppose one sets $m=0$, then Eqs. $\eqref{e:torsion1}$ and
$\eqref{e:torsion2}$ would be solved immediately to give 
\begin{equation}
H(t) = \sqrt{\frac{\Lambda}{3}} \coth \frac{\sqrt{3 \Lambda}}{2} t \;.
\end{equation}
The cosmological constant alters the large $t$ behaviour
dramatically, as $H$ would tend to a finite constant, i.e. $H
\rightarrow \sqrt{\frac{\Lambda}{3}}$. 

We consider the case where $\psi \tilde{\psi} \neq 0$, and parametrise
$\psi$ by the following relation \cite{hestenes4}: 
\begin{equation} \label{e:psi1}
\psi = \sqrt{\rho} \exp \l(\frac{\ps \beta}{2} \r) R \;,
\end{equation}
where $\rho > 0$ and $\beta$ are scalar functions of $t$. The rotor $R$ is
even-grade (and time dependent) and satisfies $R
\tilde{R}=1$. Substituting Eq. $\eqref{e:psi1}$ into Eq. 
$\eqref{e:dirac2}$, and equating grades on either side, one finds the
following equations
\begin{align}
\label{e:dirac4}
\dot{\rho} &= 4 m \sin \beta \ps \g_0 \w \mathcal{S} - 3 \rho H \;, \\
\label{e:dirac5}
\rho \dot{\beta} &= 4 (m \cos \beta + 3 \pi \rho) \ps \g_0 \w
\mathcal{S} \;,  \\
\label{e:dirac6}
\rho \dot{R} \tilde{R} &= -2 (me^{- \ps \beta} + 3 \pi \rho) \g_0 \dt
\mathcal{S} \;.
\end{align}
where $\dot{R} \tilde{R}$ is a bivector. We simplify the calculation by taking
$\sin \beta = 0$. Since $\beta$ is then constant, a non-zero and
constant $\rho$ is forbidden by Eq. $\eqref{e:dirac4}$, since
Eq. $\eqref{e:dirac5}$ gives $\g_0 \w \mathcal{S}=0$. It follows that we
must solve
\begin{align}
\label{e:torsion3}
\dot{\rho} &= - 3 \rho H  \;, \\
\label{e:torsion4}
\dot{R} &= - (m \cos \beta + 3 \pi \rho) \g_0 R \ps \g_3 \;,
\end{align}
subject to the constraint 
\begin{equation}
\label{e:torsion5}
3 H^2 - 12 \pi^2 \rho^2 - 8 \pi m \rho \cos \beta -\Lambda = 0
\;.
\end{equation}
In these equations, $\cos \beta = \pm 1$. The solutions with $\cos
\beta=1$ are regarded as ``particle'' (positive energy) solutions and
those with $\cos \beta = -1$ as ``anti-particle'' (negative energy)
solutions in the absence of gravity \cite{doran4}. 

To solve the Eqs. $\eqref{e:torsion3}$ and $\eqref{e:torsion5}$, we
make use of the substitution $y=\frac{1}{6 \pi \rho}$ to find
\begin{equation} \label{e:diffeq1}
\begin{split}
\dot{y}^2 &= 1 \pm 4my + 3 \Lambda y^2 \\
          &= 3 \Lambda \l(y \pm \frac{2m}{3 \Lambda} + \frac{\sqrt{4m^2 - 3
          \Lambda}}{3 \Lambda} \r) \l(y \pm \frac{2m}{3 \Lambda} -
          \frac{\sqrt{4m^2 - 3 \Lambda}}{3 \Lambda} \r)  \;.
\end{split}
\end{equation}
One could obtain, by solving the above equation, the density parameter
$\rho$ and the Hubble parameter, $H(t)$, for all $t$ in both particle
and anti-particle sectors.

\subsection{Cosmological implications} \label{sec:cosmoimp}

In this section, we analyze how the inclusion of a cosmological
constant would modify the solutions found in \cite{challinor} for a
spin-torsion Universe (see table \ref{tab1} for the classification of
the solutions). Before we proceed to the solutions, we examine the
discriminant $\sqrt{4m^2-3\Lambda}$ and the limits which these
solutions can take. There are three possible cases:
(i) $4m^2 \gg 3 \Lambda$, (ii) $4m^2 = 3 \Lambda$ and (iii) $4 m^2 \ll
3 \Lambda$. In natural units, the cosmological constant, $\Lambda \sim
10^{-122}$. We compare this value to the known experimental values
from the masses of the fermions in natural units. The mass of the
electron, $m_e \sim 10^{-23}$ and the mass of the neutrino is
estimated approximately, $m_{\nu} \sim 10^{-7} m_e$ (see the
experimental bounds for the neutrino masses in \cite{super}). It is
found that the values of $m^2$ for the fermions are significantly much larger
than the cosmological constant, hence, the preferred model would be
case (i). For case (ii), $m \sim 10^{-38} m_e$. However no such
fermion is known to exist in nature, and hence case (ii) is not a
plausible model. Since case (ii) is not a plausible model, case (iii)
is effectively ruled out as well. In this section, we consider the
solutions of interest for both case (i) and (ii).   

\begin{figure}[!htb]
  \begin{center}
     \psfrag{A}{$(\dot{y})^2$}
     \includegraphics{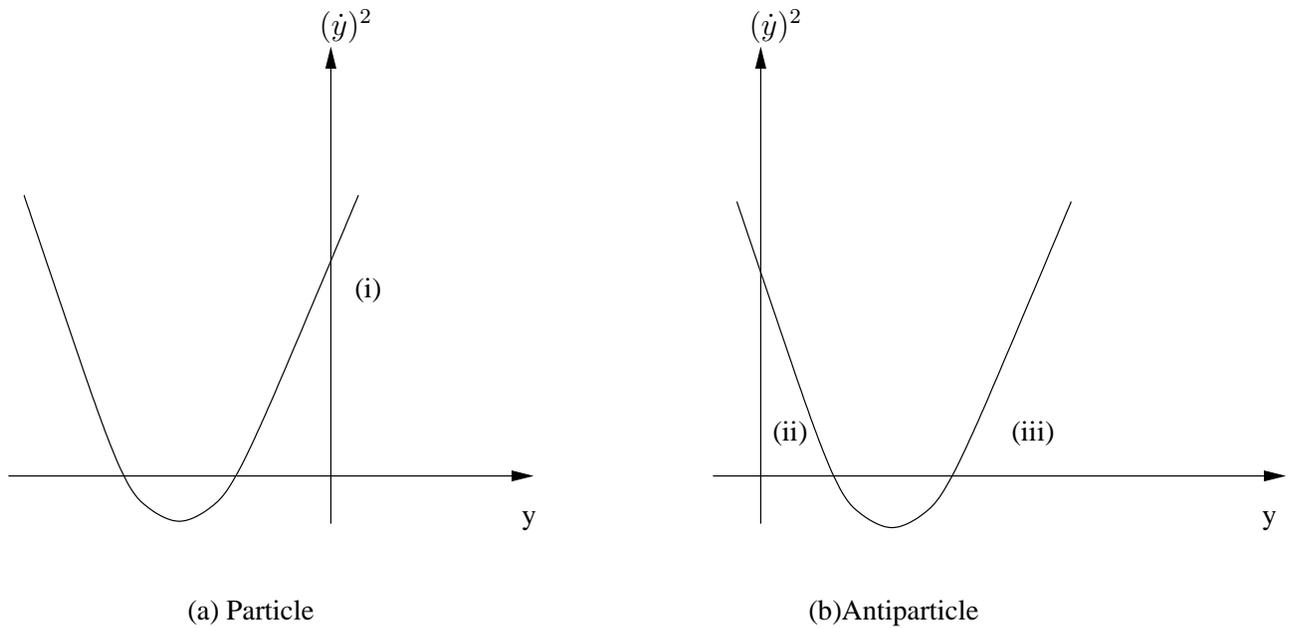}
  \end{center}
  \caption{A sketch of the equation $\eqref{e:diffeq1}$ with $\dot{y}^2$ vs $y$
     for both particle and anti-particle respectively, i.e. $\pm
     m$. This is for the case, $4m^2 \gg 3 \Lambda$. The three regions
     of interest : (i) an expanding Universe with 
     particle from an initial singularity (for $t>0$) or a Universe
     contracting to a singularity in some finite time (for $t<0$), 
     (ii) an anti-particle Universe which expands to a finite value and
     contracts to a singularity (for $t>0$), and (iii) 
     an expanding anti-particle Universe (for $t>0$) and a contracting
     Universe which bounces at $t=0$ and expands indefinitely (for
     $t<0$). 
  }
  \label{fig2.1}
\end{figure}

To solve Eq. $\eqref{e:diffeq1}$, we demand the condition $y \nleq 0$
since $\rho>0$. We note that these solutions would work for all values
of $t$. We start with case (i) with $4m^2 \gg 3 \Lambda$, and assume
a positive cosmological constant i.e. a repulsive force. From
Fig. \ref{fig2.1}, since $\rho > 0$, the requirement imposes an
asymmetry between the particle and anti-particle solutions. For the
particle case, we consider only one branch of the 
solution with $y > 0$ [region (i)]. On the other hand, we observe that
the anti-particle case has two branches of solutions. These two
branches are labelled as regions (ii) and (iii). For $\Lambda \to 0$,
the solutions found in \cite{challinor} would correspond to regions
(i) and regions (ii), with the appropriate limit taken. The presence
of the cosmological constant has introduced an additional solution.    

For regions (i) and (ii), the Hubble parameter and density parameter
for both particle and anti-particle respectively are:    
\begin{align}
\label{e:dp0} \rho (t) &= \frac{\alpha^2}{6 \pi \sinh \alpha t (\alpha \cosh \alpha t \pm
m \sinh \alpha t)} \;, \\
\label{e:Hp0} H(t) &= \frac{\alpha (\alpha \cosh 2 \alpha t \pm m
\sinh 2 \alpha t)}{3 \sinh \alpha t ( \alpha \cosh \alpha t \pm m
\sinh \alpha t )}  \;,
\end{align}
where $\alpha = \frac{\sqrt{3 \Lambda}}{2}$. The density is singular
at $t=0$ and at $t = \alpha^{-1} \coth^{-1} [\pm (m/\alpha)]$ for both
particle and anti-particle cases respectively. 

Starting from the particle case for $t >0$, the solution in region (i)
corresponds to an expanding Universe from an initial singularity at
$t=0$, and $H(t)$ tends to the value for a flat Universe
with a cosmological constant at later times. During this epoch, the
scale factor of the Universe asymptotically approaches $\exp(t
\sqrt{\Lambda/3})$. If we move from $t \to -t$, we found $\rho(t) > 0$,
but $H(t) <0$. For $t < 0$, this solution describes a Universe
contracting to a singularity in some finite time. 

Next, we examine the anti-particle solution in region (ii). For $t>0$, the
Universe would expand from an initial singularity, turn around at some
finite time and collapse to a singularity. A particle horizon is also
present in the model which continues to exist right up to the
singularity at the endpoint of the collapse. For Eqs. $\eqref{e:dp0}$
and $\eqref{e:Hp0}$, taking the limit $\Lambda \to 0$, the solutions
would be those found in \cite{challinor}. For $t \to -t$, the
solution describes the same cosmology as before. 

From Fig. \ref{fig2.1}, the quadratic nature of
Eq. $\eqref{e:diffeq1}$ introduces a new anti-particle solution [region
(iii)]. The Hubble parameter and density parameter for region (iii)
are found to be: 
\begin{align}
\label{e:dap1} 
\rho &= \frac{\alpha^2}{3 \pi} \frac{1}{m + M \cosh 2 \alpha t} \;,
 \\
\label{e:Hap1}
H(t) &= \frac{2 \alpha}{3} \l[ \frac{\sinh 2 \alpha t}{M +  \cosh
2\alpha t} \r]  \;,
\end{align} 
where $M=(m/\sqrt{m^2-\alpha^2})$. This solution satisfies
Eq. $\eqref{e:diffeq1}$ for the anti-particle case and is geodesically
complete. For $t>0$, the solution describes an expanding Universe
starting from some finite $y$. With the inclusion of a positive 
cosmological constant, the Universe generated by the anti-particle
sector appears to be an eternal Universe with no singularities. The
$t<0$ is an interesting case, since $\rho(-t) >0$. This solution
describes a contracting Universe from some finite $-t$, bounces at
the instant at $y=0$ and expands indefinitely like the case for
$t>0$. The solution has the property that it smoothly matches a
contracting phase ($t<0$) onto an expanding phase ($t>0$) which is
reminiscent of the pre-Big-Bang scenario~\cite{gasperini}.

In principle, we can incorporate a negative cosmological constant by
switching $\Lambda\to -\Lambda$ in the above solutions. For both
particle and anti-particle sectors, the solutions are similar to a
closed FRW model, which corresponds to an expanding Universe starting
from an initial singularity which turns around at a finite time, and
collapses back into a singularity. 

\begin{figure}[!htb]
  \begin{center}
     \psfrag{A}{$(\dot{y})^2$}
     \includegraphics{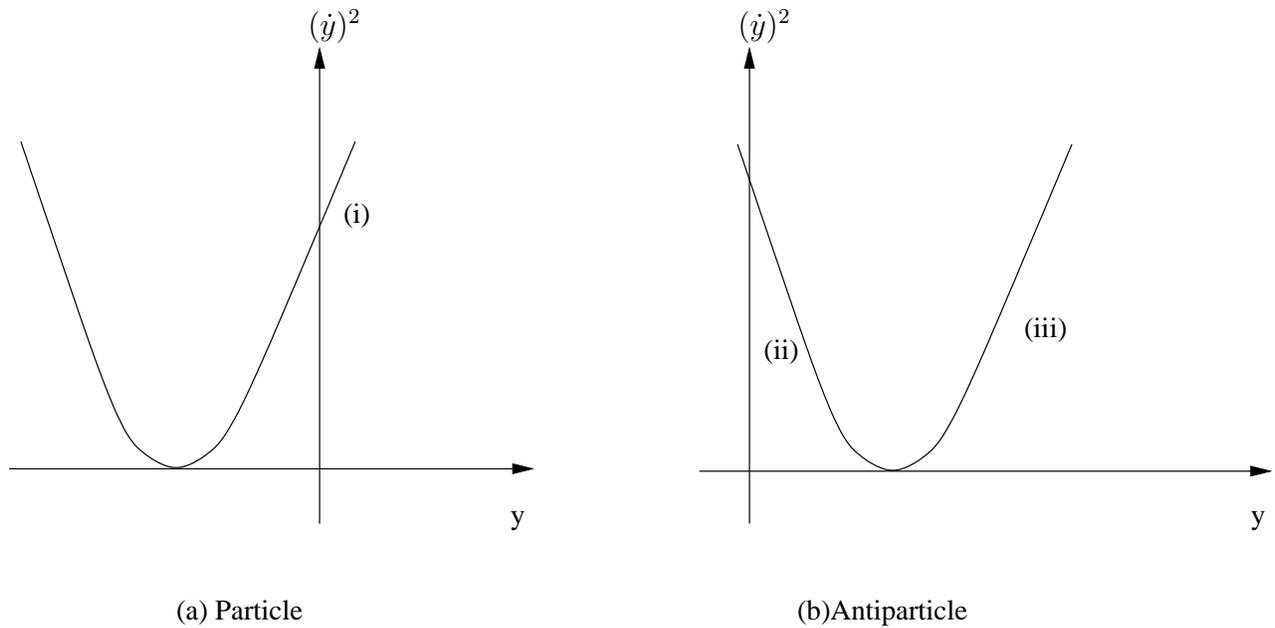}
  \end{center}
  \caption{A sketch of the equation $\eqref{e:diffeq1}$ with $\dot{y}^2$ vs $y$
     for both particle and anti-particle respectively, i.e. $\pm
     m$. This is for the extremal case: $4m^2=3 \Lambda$. Like
     \ref{fig2.1}, we have three regions of interest : 
     (i) an expanding Universe with
     particle from an initial singularity (for $t>0$) or a Universe
     contracting to a singularity in some finite time (for $t<0$), 
     (ii) an anti-particle Universe which expands to a finite value and
     contracts to a singularity (for $t>0$), and (iii) 
     an expanding anti-particle Universe (for $t>0$) and a static
     Universe which smoothly matches from $-t$ to $t$ (for $t<0$). 
  }
  \label{fig2.2}
\end{figure}

Another solution of interest is the extremal case where $4m^2 = 3
\Lambda$. From Fig. \ref{fig2.2}, we note that there are three regions
of interest, one solution for the particle case and two solutions for
the anti-particle case. The solution is degenerate at $\rho=m/(3\pi)$.

By setting the initial conditions when $t_0 = 0$, and the
initial density, $\rho_0 \rightarrow  \infty$, we obtain the solutions
of $\eqref{e:diffeq1}$ for $4m^2=3\Lambda$. Starting from the particle
case, for region (i), we obtain the density and Hubble parameter to be: 
\begin{align}
\label{e:dp4}
\rho &= \frac{m}{3 \pi (e^{2mt} -1)} \quad \text{(particle)} \;, \\
\label{e:Hp3} H(t) &= \frac{2m}{3(1 - e^{-2mt})} \quad
\text{(particle)} \;.
\end{align} 
From the above solutions, as $t \rightarrow \infty$, the particle
Universe, $H(t)$ would tend to a finite constant $\sqrt{(\Lambda/3)}$,
which is typical of a Friedmann dust model with a positive
cosmological constant. 

\begin{figure}[!htb]
  \begin{center}
     \includegraphics[scale=0.65]{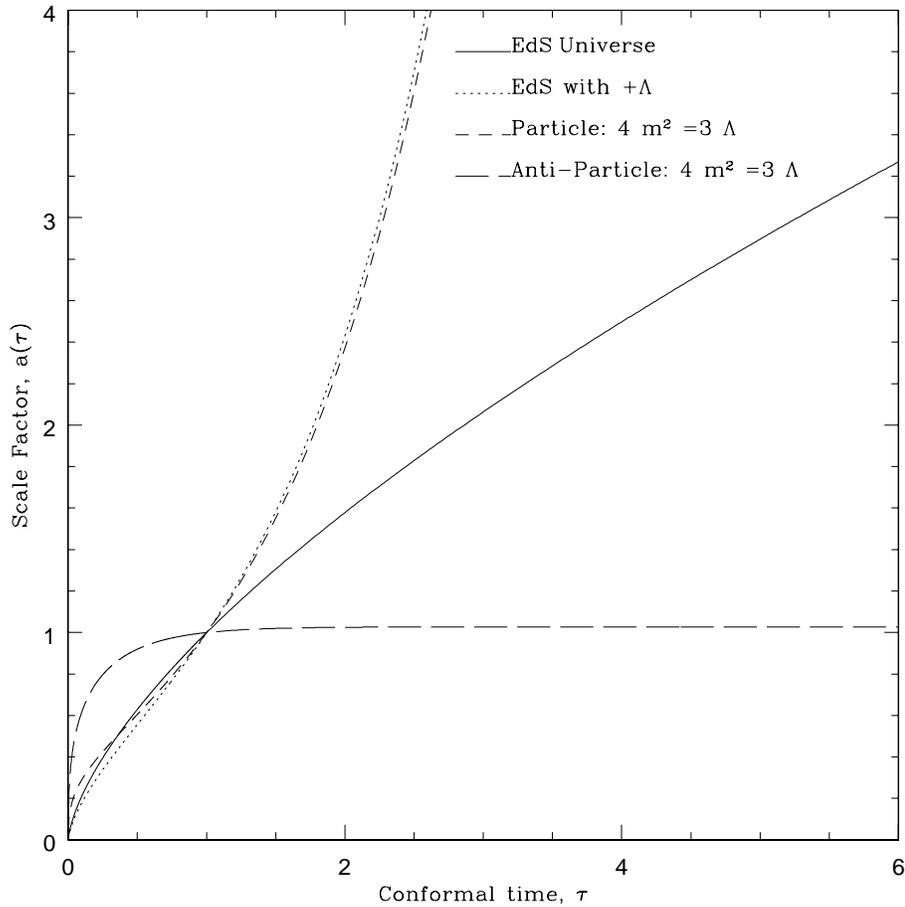}
  \end{center}
  \caption{Graph of the scale factor $a(\tau)$ vs conformal time (in
     units of $H_0 \tau$) for
     the particle and anti-particle solutions for the extremal
     Universes at $t>0$ i.e. $4m^2=3 \Lambda$ for $\Omega_{\Lambda}=0.7$. The
     scale factor for flat Universe (with and without
     the  cosmological constant) are plotted for comparison. 
     Note that the scale factor for an
     extremal particle solution is similar to a flat universe with positive
     $\Lambda$, whereas the scale factor for an extremal anti-particle
     asymptotes towards a constant.} 
  \label{chap2-3}
\end{figure}

For region (ii), the solutions are:
\begin{align}
\label{e:dap4}
\rho &= \frac{m}{3 \pi (1 - e^{-2mt})} \quad
\text{(anti-particle)} \;, \\
\label{e:Hap3} H(t) &= \frac{2m}{3 (e^{2mt} - 1)} \quad
\text{(anti-particle)} \;.
\end{align}
For the anti-particle Universe, $H(t)$ would tend to zero with the
same limit, which suggests the scale factor of the anti-particle
Universe tends asymptotically to a constant (see
Fig. \ref{chap2-3}). In this scenario, the Universe is a
Minkowski space with torsion at late times. The sole parameter of the
extremal Universe is dependent on the mass of the fermion, $m$. Both
models represent an expanding Universe with an initial singular
density which decays as it evolves in time (except that the
anti-particle decays to $m/3\pi$). In both cases, as we switch $t \to
-t$, we get collapsing particle and anti-particle solutions which
contract to a singularity. 

In addition, there is an additional anti-particle solution [region
(iii)] for  $\rho > m/(3 \pi)$ in the extremal case. 
The density parameter and the Hubble parameter are found to be:
\begin{align}
\label{e:dap5}
\rho &= \frac{m}{3 \pi (e^{2mt}+1)}  \\
\label{e:Hap5}
H(t) &= \frac{2m}{3 (1 + e^{-2mt})}
\end{align}
This solution is similar to the region (iii) for the case $4m^2 \gg 3
\Lambda$. For $t>0$, it is an expanding anti-particle Universe. The
$t<0$ solution appears to be a Universe which smoothly matches a
contracting phase onto an expanding phase ($t>0$) at a finite density,
with a bounce at $\rho=m/(6 \pi)$. However, for the $t<0$ solution,
since the anti-particle solution starts at $t= -\infty$, indicating that $H(t) \to
0$. This means that the Universe would not be expanding, and this
scenario is similar to the Einstein static Universe, but in this case,
the spin-torsion has balanced against the repulsive force of the
cosmological constant.  

The stress-energy tensor $\mathcal{T}(a)$ is found to be 
\begin{equation}
\begin{split}
\mathcal{T} (e_t) &= \rho (2 \pi \rho \pm m) e_t \;, \\
\mathcal{T} (\g_3) &= 0 \;, \\
\mathcal{T} (\g_1) &= +\frac{1}{2} \rho H \g_2 - \pi \rho^2 \g_1 \;, \\
\mathcal{T} (\g_2) &= -\frac{1}{2} \rho H \g_1 - \pi \rho^2 \g_2  \;.
\end{split}
\end{equation}
As expected, $e_t$ is a timelike eigenvector of
$\mathcal{T}(a)$. The only other real eigenvector is $\g_3$ which is
dual to the spin trivector $\mathcal{S}$. (Taking the dual is performed by
multiplying the pseudoscalar $\ps$.) The stress-energy tensor singles
out the two directions in space to be algebraically special,
and reflects the anisotropy of the solution at the level of the
gravitational gauge fields. The energy density as measured by the
fundamental observers is evaluated to be
\begin{equation}
\mathcal{T}(e_t) \dt e_t = \rho (2 \pi \rho \pm m) \;.
\end{equation}

From the energy density, we evaluate the density parameter which is
defined by the following relation
\begin{equation}
\Omega = \frac{8 \pi \mathcal{T}(e_t) \dt e_t }{3 H^2} \;,
\end{equation}
which in regions (i) and (ii) evaluates to
\begin{equation}
\Omega = \frac{4}{3} \frac{(\alpha^2 + 3m^2 \sinh^2 \alpha t \pm 3 m
\alpha \sinh \alpha t
\cosh \alpha t)}{(\alpha \cosh 2 \alpha t \pm m \sinh 2 \alpha t)^2} \;,
\end{equation}
and in region (iii) 
\begin{equation}
\Omega = \frac{2}{3} \l[\frac{3 m^2 + 2 \alpha^2 + 3m b \cosh
2 \alpha t}{b^2 \sinh 2 \alpha t} \r] \;,
\end{equation}
where $b=\sqrt{m^2 - \alpha^2}$. Setting $\Lambda \rightarrow 0$, we
obtain the result in \cite{challinor}. Hence at early times, $\Omega
\neq 1$ since $4m^2 \gg 3 \Lambda$.

\subsection{Massive Non-Ghost Solution}

Finally, we arrive at the final form of the non-ghost solution. 
To solve the rotor equation $\eqref{e:torsion4}$, we would note the
following result from \cite{challinor}:
\begin{equation} \label{e:torsion7}
\g_0 \w \mathcal{S} = 0 \quad \Rightarrow \quad \g_3 \dt (\tilde{R}
\g_0 R) = 0 \;.
\end{equation}
The rotor $R$ can be decomposed into a product of rotors $R \equiv
\Phi L$, where $\Phi$ and $L$ are uniquely determined by the
requirements $\Phi \g_0 = \g_0 \Phi$ and $L \g_0 = \g_0
\tilde{L}$. The condition in Eq. $\eqref{e:torsion7}$ restricts $L$
to the form 
\begin{equation} \label{e:torsion8}
L =  e^{\frac{1}{2} \xi (\sin \eta \sigma_1 + \cos \eta \sigma_2)} \;.
\end{equation}
where $\xi$ and $\eta$ are scalars. However from the rotor Eq.
$\eqref{e:torsion4}$, we find that the vector $\tilde{R} \g_0 R$ is
time independent. Since $L$ is determined uniquely by this vector and
$\g_0$, therefore $L$ must be constant. Now we substiture $R= \Phi L$
into Eq. $\eqref{e:torsion4}$, and note that $L$ commutes with $\ps
\g_3$, to find that 
\begin{equation} \label{e:torsion9} 
\dot{\Phi} = - \dot{\chi} \Phi \ps \sigma_3 \;,
\end{equation}
where the scalar $\chi(t)$ is defined by
\begin{equation} \label{e:torsion10}
\dot{\chi} \equiv m \cos \beta + 3 \pi \rho \;.
\end{equation}
Hence Eq. $\eqref{e:torsion9}$ has the solution 
\begin{equation} \label{e:torsion11}
\Phi (t) = \Phi_0 e^{-\ps \sigma_3 \chi (t)} \;,
\end{equation}
where $\Phi_0$ is a constant rotor which commutes with $\g_0$. The
rotor may be eliminated by combining the global position-gauge
transformation $x' = \Phi_0 x \tilde{\Phi}_0$, with the global
rotation-gauge transformation defined by the rotor
$\tilde{\Phi}_0$. The solution to Eq.
$\eqref{e:torsion10}$ is 
\begin{equation} \label{e:chi0}
\chi (t) = \pm mt + 3 \pi \int \rho(t) dt \;,
\end{equation}
where we could substitute the various forms of $\rho(t)$ found in Eqs.
$\eqref{e:dp0}$, $\eqref{e:dap1}$, $\eqref{e:dp4}$ or
$\eqref{e:dap4}$. The integration constant is chosen so that $\chi$ is
defined for $\rho > 0$. 

For regions (i) and (ii), we found that $\chi$ is evaluated to be 
\begin{equation} \label{e:chi1}
\chi(t) = \pm mt + \frac{1}{2} \ln \l[\frac{\alpha \l(1 + \tanh^2
\frac{\alpha t}{2} \r) \pm 2m \tanh\frac{\alpha t}{2}}{\tanh
\frac{\alpha t}{2}} \r] \;,
\end{equation}
Similarly, for region (iii), which only applies to the anti-particle
case,
\begin{equation} \label{e:chi2}
\chi(t) = \pm mt + \frac{1}{2} \ln \l(\frac{M+\cosh 2 \alpha
t}{\alpha} \r) \;.
\end{equation}

For the extremal Universes, for both particle and anti-particle cases
we find $\chi$ to be
\begin{align}
\label{e:chi3}
\chi(t) &= \pm mt + \frac{1}{2} \ln (1 - e^{-2mt}) \quad
\text{[particle, region (i)]} \;,  \\
\label{e:chi4}
\chi(t) &= \pm mt + \frac{1}{2} \ln (1 - e^{2mt})  \quad
\text{[anti-particle, region (ii)]} \;, \\
\chi(t) &= \pm mt - \frac{1}{2} \ln (1 + e^{-2mt}) \quad
\text{[anti-particle, region (iii)]} \;. 
\end{align}

Hence we obtain the general self-consistent massive non-ghost solution
for a Dirac field with the inclusion of a cosmological constant,
$\psi$ with $\sin \beta =0$, with $\beta=0$ or $\pi$ to be  
\begin{equation} \label{e:nonghostsol}
\psi (t) = \sqrt{\rho(t)} \exp \l(\frac{\ps \beta}{2} - \ps \sigma_3
\chi (t) \r) \exp \l(\frac{\xi}{2} (\sin \eta \sigma_1 + \cos \eta
\sigma_2) \r) \;.
\end{equation}

\section{Discussion}

Finally, we examine whether the solutions for the spin-torsion
Universes are inflationary or not. From $\eqref{e:inflation}$ in
Chapter \ref{prologue}, the condition for inflation is 
\begin{equation} \label{e:inflation2}
\ddot{a} > 0 \quad \Rightarrow \dot{H} + H^2 > 0 \;.
\end{equation}
Using $\eqref{e:torsion5}$, we obtain
\begin{equation} \label{e:inflation3}
\dot{H} + H^2 = \frac{\Lambda}{3} + g(\rho, \beta) \;,
\end{equation}
where 
\begin{equation}
g(\rho, \beta) = - 8 \pi^2 \rho^2 \mp \frac{4\pi m \rho}{3} \;,
\end{equation}
for the cases of particle and anti-particle respectively. For the
particle case, $g(\rho, \beta) < 0$, and hence there is no inflation
at earlier times since $g(\rho, \beta) \gg \Lambda$. However, at late
times, there will be de-Sitter inflation induced by the cosmological
constant. On the other hand, it is observed for the anti-particle case
that $g(\rho, \beta) >0$ if $\rho < (m/6\pi)$. For $\rho < (m/6 \pi)$,
the spin-torsion Universe would have an inflationary phase driven by
the anti-particle. From Eq. $\eqref{e:chal2}$ in Table
\ref{tab1}, the lower bound for the density of a spin-torsion
Universe is $\rho=(m/3 \pi)$. Therefore, the spin-torsion solutions do not
incorporate an inflationary phase at earlier times. The only
inflationary phase is provided by the cosmological
constant at late times.

\newpage
\section*{A Table of the cosmological solutions in Spin Torsion Theories}
\begin{center}
\begin{longtable}{|p{4.8cm}|p{10cm}|} 
\hline
{\bf Solution Type} & {\bf Hubble Parameter, $H(t)$} \\ \hline \hline
{\bf Flat}:
 & 
(i) ($K=0, \Lambda=0$)   
\begin{equation} \label{e:eds1}
H = \frac{2}{3t} \;.
\end{equation} 
(ii) with $\Lambda > 0$  
\begin{equation} \label{e:eds2}
H = \frac{2 \alpha}{3} \coth \alpha t \;.
\end{equation}
 
(iii) with $\Lambda <0$  
\begin{equation} \label{e:eds3}
H = \frac{2 \alpha}{3} \cot \alpha t \;,
\end{equation} 
where 
\begin{equation}
\alpha=\frac{3\Lambda}{2} \;.
\end{equation}
\\ \hline \hline
{\bf Spin Torsion Universe}: & \\ \hline 
(a)Particle   
& General solution [region (i)]:
\begin{equation} 
H(t) = \frac{\alpha (\alpha \cosh 2 \alpha t + m \sinh 2\alpha t)}{3 \sinh
\alpha t ( \alpha \cosh \alpha t + m \sinh \alpha t )} \;.
\end{equation}  
(i) Setting $4m^2 \gg 3 \Lambda$, we obtain Eq. $\eqref{e:eds2}$ for
$\Lambda >0$ and Eq. $\eqref{e:eds3}$ for $\Lambda<0$. If we set
$\Lambda \to 0$, we get the solution in \cite{challinor}:
\begin{equation} \label{e:chal1}
H(t) = \frac{1+2mt}{3t(1+mt)} \;.
\end{equation}
(ii) For $4m^2=3 \Lambda$, we get
\begin{equation}
\frac{2m}{3 [1- \exp (-2mt)]} \;.
\end{equation}  
\\ \hline \hline
(b) Anti-particle 
&  General solution [region (ii)]
\begin{equation} 
H(t) = \frac{\alpha (\alpha \cosh 2 \alpha t - m \sinh 2\alpha t)}{3 \sinh
\alpha t ( \alpha \cosh \alpha t - m \sinh \alpha t )} \;.
\end{equation}  
(i) Setting $4m^2 \gg 3 \Lambda$, we obtain $\eqref{e:eds2}$ for
$\Lambda > 0$ and the inverse of $\eqref{e:eds3}$ 
for $\Lambda <0$. If we set $\Lambda \to 0$, we get the other solution
in \cite{challinor}: 
\begin{equation} \label{e:chal2}
H(t) = \frac{1-2mt}{3t(1-mt)} \;.
\end{equation}
(ii) For $4m^2=3 \Lambda$, we get
\begin{equation}
\frac{2m}{3 [\exp (2mt)-1]} \;.
\end{equation}  
General solution II (Region (iii))

(i) Setting $4 m^2 \gg 3 \Lambda$, we obtain
\begin{equation}
H(t) = \frac{2 \alpha}{3} \l[\frac{\sinh 2 \alpha
t}{M + \cosh 2 \alpha t} \r] \;,
\end{equation}
where
\begin{equation}
M = \frac{m}{\sqrt{m^2 - \alpha^2}} \;.
\end{equation}
(ii) For the extremal case, we have
\begin{equation}
H(t) = \frac{2m}{3 [\exp(2mt) + 1]}
\end{equation} 
There is no analogue for this case as $\Lambda \to 0$. 

\\ \hline
\caption{A summary of the solutions for the flat
Universes and the spin-torsion Universes}
\label{tab1}
\end{longtable}
\end{center}

      \clearpage{\pagestyle{empty}\cleardoublepage}
      \part{Braneworld Cosmology}	
\chapter{Braneworld Cosmology I: Formalism} \label{chapter3}
\begin{quote}
{\it ``I do not wish to hide the fact that I can only look with
 repugnance .. upon the puffed up pretentiousness of all these volumes
 filled with wisdom, such as are fashionable nowadays. For I am fully
 satisfied that .. the accepted methods must endlessly increase these
 follies and blunders, and that even the complete annihilation of all
 these fanciful achievements could not possibly be as harmful as this
 fictitious science with its accursed fertility.''}~{\bf - Immanuel Kant}
\end{quote}

{\it In this chapter, we discuss the dynamics of linearized scalar and
tensor perturbations in an almost Friedmann-Robertson-Walker braneworld  cosmology of 
Randall-Sundrum type II using the 1+3 covariant approach based on
\cite{leong1} and \cite{leong4}. We derive
a complete set of frame-independent equations for the total matter variables,
and a partial set of equations for the non-local variables which arise
from the projection of the Weyl tensor in the bulk. The latter equations are
incomplete since there is no propagation equation for the non-local
anisotropic stress. As a first step, we derive the
covariant form of the line of sight solution for the CMB temperature
anisotropies in braneworld cosmologies. We discuss possible
mechanisms by which braneworld effects may remain in the low-energy
universe in Chapter \ref{chapter4}.} 

\section{Introduction to Braneworld Cosmology}

It is understood that Einstein's theory of general relativity is an
effective theory in the low-energy limit of a more general
theory. Recent developments in theoretical physics, particularly in
string theory or M-theory, have led to the idea that gravity is a
higher-dimensional theory which would become effectively
four-dimensional at lower energies.

Braneworlds, which were inspired by string and M-theory, provide simple,
yet plausible, models of how the extra dimensions might affect the
four-dimensional world we inhabit. There is the exciting possibility that
these extra dimensions might reveal themselves through specific
cosmological signatures that survive the transition to the low-energy
universe. It has been suggested that in the context of braneworld models 
the fields that govern the basic interactions in the standard model 
of particle physics are confined to a 3-brane, while the 
gravitational field can propagate in $3+d$ dimensions (the {\it
bulk}). It is not necessarily true that the extra dimensions are
required to be small or even compact. It was shown recently by Randall
and Sundrum~\cite{randall} for the case $d=1$, that gravity could be
localised to a single 3-brane even when the fifth dimension was
infinite. As a result, the Newtonian potential is recovered on large
scales, but with a leading-order correction on small scales:
\begin{equation}
V(r) = -\frac{GM}{r} \left( 1 + \frac{2l^2}{3r^2} \right)\;,
\end{equation}
where the 5-dimensional cosmological constant $\tilde{\Lambda} \propto
-l^{-2}$. As a result, general relativity is recovered in 4 dimensions
in the static weak-field limit, with a first-order correction which is
believed to be constrained by sub-millimeter experiments at the TeV
level \cite{maartens3, randall}. 

The aim of this chapter is to set up the evolution and constraint equations
for perturbations in a cold dark matter (CDM) brane cosmology,
presenting them in such a way that they can be readily compared with the
standard four-dimensional results, and to provide approximate solutions
in the high and low-energy universe under certain restrictions on how the
bulk reacts on the brane. Our equations are clearly incomplete
since they lack a propagation equation for the non-local
anisotropic stress that arises from projecting the bulk Weyl tensor onto
the brane, and our solutions are only valid under the neglect of this stress.
However, our presentation is such that we can easily include effective
four-dimensional propagation equations for the non-local stress should
such equations arise from a study of the full bulk perturbations.
The lack of a four-dimensional propagation equation for the non-local
stress means that it is currently not possible to obtain general results
for the anisotropy of the CMB in braneworld models. Such a calculation
would require solving the full five-dimensional perturbation equations which
is non-trivial since the equations can only be reduced to two-dimensional
partial differential equations on Fourier transforming the 3-space dependence.

The theory of cosmological perturbations using the metric-based
approach, or the Bardeen approach has been extensively studied
in \cite{binetruy, bridgman, deffayet, deruelle, dorca,
kanno, kanno2, kodama, kodama2, koyama2, koyama, langlois3, mukohyama,
mukohyama2, neronov, sago, sasaki, soda, vandebruck2, vandebruck}
(with $Z_2$ symmetry) and in \cite{riazuelo} (without $Z_2$
symmetry). On the other hand, it has been 
also explored by various authors within the 1+3 covariant
methods in \cite{bruni2, gordon, leong1, leong4, maartens, maartens2}. We will further
summarise some of the qualitative implications for observational
cosmology of the braneworld scenarios in the literature, and 
explore the possible implications for the CMB anisotropies in braneworld cosmology in
the next chapter.    

We give a brief summary of the (1+3)-covariant approach in
\ref{sec:covariantapproach}. In \ref{sec:bw1}, we 
look at the field equations of the braneworld cosmology. Following
that, we will look at the linearized perturbation equations in 
\ref{sec:equations}, where the local and nonlocal conservation
equations are given in \ref{sec:equations1}, and the propagation,
divergence and constraint equations are given in
\ref{sec:equations2}. Following that, we discuss the cosmological
perturbations on the brane in \ref{sec:perturbations}, distinguishing between
scalar perturbations (\ref{sec:scalarpert}), vector perturbations
(\ref{sec:vectorpert}) and tensor perturbations
(\ref{sec:tensorpert}). To conclude this chapter, we will derive a 
covariant expression for the temperature anisotropy in
\ref{sec:aniso}. 

\section{(1+3)-Covariant Approach in Cosmological \\ Perturbation
Theory} \label{sec:covariantapproach}
The (1+3)-covariant approach, pioneered by Ehlers~\cite{ehlers},
Ellis~\cite{ellis5} and Hawking~\cite{hawking} provided an alternative
gauge-invariant treatment of cosmological pertubations 
\cite{ellis89, ellis2} other than the standard metric-based approach
formulated by Bardeen\cite{bardeen}\footnote{The 
Bardeen approach is inherently linear. It assumes a nonlocal
decomposition of the perturbations into scalar, vector and tensor
modes at the outset, with each of them treated
seperately. Even though the Bardeen variables are gauge-invariant, they are not
physically transparent, i.e. they do not characterise perturbations in
a way that is amenable to simple physical interpretation.}. In the (1+3)
treatment, based on the covariant formulation of hydrodynamics and
gravito-dynamics of Ehlers and Ellis (for an extensive review of the subject, see
\cite{bertschinger, dunsby, ehlers, ellis5, ellis6, ellis7}), the cosmological 
perturbations are described by gauge-invariant variables which have
simple physical interpretations in terms of inhomogeneous measures of anisotropy
of the Universe. 

The covariant approach has been applied in many areas of cosmology,
for example, gravitational waves (see e.g. \cite{dunsby4}), inflation
\cite{bruni1, dunsby3}, magnetic fields \cite{tsagas1,
tsagas2, tsagas3} and radiative transport theory \cite{ellis3,
ellis4}. This develop a strong case for its application in 
the study of cosmic microwave background physics. The line of sight
calculation of CMB anisotropies under the instantaneous recombination
based on the covariant approach was obtained by Dunsby~\cite{dunsby2}
and Challinor and Lasenby~\cite{challinor1} based on earlier work in
\cite{dunsby5, ellis89, ellis3, ellis4, ellis2,
stoeger}. Subsequently, the covariant approach was extended to studying CMB
anisotropies in the CDM model in \cite{challinor2}, tensor anisotropies
induced by gravitational waves in \cite{challinor4}, polarization in
\cite{challinor5, challinor6}, the mode and multipole representations
and their application to temperature anisotropies in \cite{gebbie1,
gebbie2} and obtaining model-independent limits 
on the inhomogeneity and anisotropy of the CMB anisotropy on large
scales in \cite{maartens7}. 

The covariant approach has the advantage that it does not employ the
non-local decomposition into scalar, vector and tensor modes. If
required, that can be adopted at later stages of the calculation in aid
to solve the equations. Even if one denies that working with gauge-invariant
quantities is a significant merit, a key advantage of the
covariant approach is that one is able to work exclusively with
physically relevant quantities, satisfying equations which make
manifest their physical consequences. Another advantage of the
covariant approach is that the full extension to
non-linear methods \cite{maartens4} is straightforward.  

We start by choosing a 4-velocity $u^a$. This must be physically
defined in such a way that if the universe is exactly FRW, the velocity
reduces to that of the fundamental observers to ensure gauge-invariance of
the approach. From the 4-velocity $u^a$,
we construct a projection tensor $h_{ab}$ into the space perpendicular
to $u^a$ (the instantaneous rest space of observers whose
4-velocity is $u^a$):
\begin{equation} \label{e:projection1}
h_{ab} \equiv g_{ab} + u_a u_b\;,
\end{equation}
where $g_{ab}$ is the metric of spacetime. The operation of projecting
a tensor fully with $h_{ab}$, symmetrizing, and removing the trace on every
index (to return the projected-symmetric-trace-free or PSTF part) is
denoted by angle brackets, i.e.\ $T_{\langle ab\ldots c\rangle}$.

The symmetric tensor $h_{ab}$ is used to define a projected (spatial)
covariant derviative $D^a$ which when acting on a tensor
$T^{b\ldots c}{}_{d\ldots e}$ returns a tensor that is orthogonal to $u^a$ on
every index, 
\begin{equation} \label{e:projection2}
D^a T^{b\ldots c}{}_{d\ldots e}
\equiv h^{a}{}_{p} h^{b}{}_{r}
\ldots  h^{c}{}_{s} h^{t}{}_{d} \ldots  h^{u}{}_{e} \n^p T^{r..s}{}_{t..u}\;,
\end{equation}
where $\n^a$ denotes the usual covariant derivative.

The covariant derivative of the 4-velocity can be decomposed
as follows:
\begin{equation} \label{e:covariant1}
\n_a u_b = \omega_{ab} + \sigma_{ab} + \frac{1}{3} \Theta h_{ab} - u_a
A_b\;,
\end{equation}
where $\omega_{ab}$ is the vorticity tensor which satisfies $u^a
\omega_{ab}=0$, $\sigma_{ab}=\sigma_{\langle ab \rangle}$
is the shear which is PSTF, $\Theta \equiv \n^{a} u_a = 3H$
measures the volume expansion rate (where $H$ is the local Hubble
parameter), and $A_a \equiv u^b \n_b u_a$ is the acceleration. In
addition, we define the vorticity vector, $\omega_a = -(1/2)
\text{curl} u_a$. 

Gauge-invariant quantities can be constructed from scalar variables
by taking their projected gradients. Such quantities vanish in the
FRW limit by construction. The comoving fractional projected gradient
of the density field $\rho^{(i)}$ of a species $i$ (for example, photons) is
one important example of this construction:
\begin{equation} \label{e:sg1}
\Delta_a^{(i)} \equiv \frac{a}{\rho^{(i)}} D_a \rho^{(i)} \;,
\end{equation}
where $a$ is a locally defined scale factor satisfying
\begin{equation} \label{e:hubble}
\dot{a} \equiv u^b \n_b a = Ha\;,
\end{equation}
which is included to remove the effect of the expansion on the projected
gradients. Another important vector variable is the 
comoving projected gradient of the expansion,
\begin{equation} \label{e:sg2}
{\cal Z}_a \equiv a D_a \Theta\;,
\end{equation}
which provides a measure of the inhomogeneity of the expansion.

The matter stress-energy tensor $T_{ab}$ can be decomposed irreducibly
with respect to $u^a$ as follows:
\begin{equation} \label{e:emequation1}
T_{ab} \equiv \rho u_a u_b + 2 u_{(a}q_{b)} + P h_{ab} + \pi_{ab}\;,
\end{equation}
where $\rho \equiv  T_{ab} u^a u^b$ is the energy density measured
by an observer moving with 4-velocity $u^a$,
$q_a \equiv - h^{b}{}_{a} T_{bc} u^c$ is the energy flux or momentum density
(orthogonal to $u^a$), $P \equiv h_{ab} T^{ab}/3$ is the isotropic
pressure, and the PSTF tensor $\pi_{ab} \equiv T_{\langle a b\rangle}$ is
the anisotropic stress.

The remaining gauge-invariant variables are formed
from the Weyl tensor $C_{abcd}$ which vanishes in an exact FRW universe
because these models are conformally flat. The ten degrees of freedom in the
4-dimensional Weyl tensor can be encoded in two PSTF tensors: the electric
and magnetic parts defined respectively as
\begin{align}
E_{ab} &= C_{abcd} u^b u^d\;,  \label{e:eweyl} \\
H_{ab} &= \frac{1}{2} C_{acst} u^{c} \eta^{st}{}_{bd} u^d\;,
\label{e:bweyl}
\end{align}
where $\eta_{abcd}$ is the 4-dimensional covariant permutation
tensor.

In the covariant approach, we defer from making the frame choice until
all the relevant equations, valid for $u^a$ are derived. There are many
different frame choices and we state the two main
examples\footnote{Other frame choices are  discussed in a review
paper by Gebbie, Dunsby and Ellis \cite{gebbie2}.} which we
employ in this thesis: (i)the pressure-free CDM frame, where the rest
frame of the CDM defines a geodesic frame i.e. the peculiar velocity
of the CDM is zero, and it defines a zero-acceleration frame, $A_a=0$;
and (ii) the energy frame, which is defined by vanishing total energy
flux $q_a=0$.  

\section{Field Equations of Braneworld Cosmology} \label{sec:bw1}

In a recent paper, Maartens~\cite{maartens} introduced a formalism for
describing the non-linear, intrinsic dynamics of the brane in Randall-Sundrum
type II braneworld models in the form of bulk corrections to the 1+3 covariant
propagation and constraint equations of general relativity.
This approach is well suited to identifying the geometric and physical
properties which determine homogeneity and anisotropy on the brane, and serves
as a basis for developing a gauge-invariant description of
cosmological perturbations in these models.

An important distinction between braneworlds and general relativity
is that the set of 1+3 dynamical equations does not close on the brane.
This is because there is no propagation equation for the non-local effective
anistropic stress that arises from projecting the bulk Weyl tensor onto the
brane. The physical implication is that the initial value problem cannot be
solved by brane-bound observers. The non-local Weyl variables
enter crucially into the dynamics (for example,
the Raychaudhuri equation) of the intrinsic geometry of the brane.
Consequently, the existence of these non-local effects leads to the violation
of several important results in theoretical cosmology, such as the connection
between isotropy of the CMB and the Robertson-Walker geometry.

The field equations induced on the brane are derived by Shiromizu et
al~\cite{shiromizu} using the Gauss-Codazzi equations, together with
the Israel junction conditions and $Z_2$ symmetry. We will discuss this
approach in detail in the later part of this section. In their elegant
geometrical approach which they employed to analyze the dynamics of
the RSII model, the bulk is a 1+4-dimensional spacetime with a
non-compact extra spatial  dimension. What prevents gravity from
`leaking' into the extra dimension at low energies is the negative
bulk cosmological constant $\Lambda_5=-6/\ell^2$, where $\ell$ is a
curvature scale of the bulk. 

In the weak-field static limit, null results in tests
for deviations from Newton's law impose the limit $\ell \lesssim
1$~mm. The negative $\Lambda_5$ is offset by the positive brane
tension $\lambda$, which defines the energy scale dividing low
from high energies. The limit $\ell<1~$mm implies 
$\lambda>(100~{\rm GeV})^4$, and the energy scales i.e. the effective
cosmological constant on the brane are related by the following
relations: 
\begin{align}
\label{e:constant1} \lambda &= 6 \frac{\kappa^2}{\tilde{\kappa}^4}\;, \\
\label{e:cc} \Lambda &=\frac{1}{2} (\Lambda_5+\kappa^2\lambda)   \;,
\end{align}
where $\tilde{\kappa}^2= 8 \pi/M_{5}^2$, with $M_5$ being the 
fundamental 5-dimensional Planck mass, $\Lambda_5$ is the cosmological
constant in the bulk and $\lambda$ is the tension of the brane. The
constant $\kappa^2= 8\pi G=8\pi/M_4^2$, and $M_4\sim 10^{19}~$GeV 
is the effective Planck scale on the brane. A further intriguing
feature of the braneworld scenario is that, because of the large extra
dimensions, the fundamental energy scale of gravity can be 
dramatically lower than the effective Planck scale on the brane --
as low as $\sim$~TeV in some scenarios \cite{maartens3}. In
generalised RSII models, the lower bound arising from current tests
for deviations from Newton's law is
\begin{equation}
M_5>10^5~\text{TeV}, \quad \lambda > (100 \text{GeV})^4
\end{equation}
and is related to $M_4$ via $M_5^3=M_4^2/\ell\,$. 

A less stringent constraint given by nucleosynthesis \cite{bratt,
maartens3} implies that $\lambda
\gtrsim(1$Mev)$^4$ which gives 
\begin{equation} \label{e:nucleo}
M_5 \gtrsim \l(\frac{1~\text{MeV}}{M_4} \r)^{\frac{2}{3}}, \quad M_5
\gtrsim 10~\text{TeV} \;.
\end{equation}
At energies well above the brane tension $\lambda$, gravity
becomes 5-dimensional and significant corrections to general
relativity occur. There are also corrections that can operate at
low energies, mediated by bulk graviton or Kaluza-Klein (KK)
modes.

From the above discussion, we proceed to summarise the approach in
\cite{shiromizu} to derive the effective Einstein equations induced on
the brane. The induced metric on the brane on 
all the hypersurfaces orthogonal to the unit normal $n^{A}$ (A
spacelike vector), is defined to be
\begin{equation} \label{e:5dmetric}
g_{AB} = \tilde{g}_{AB} - n_A n_B
\end{equation}
where we use tildes to denote the 5-dimensional generalization of
standard general relativity quantities. We adopt the choice of
coordinates to be $x^A = (x^a,~y)$, where $x^{a}=(t,~x^i)$ are
spacetime coordinates on the brane and $n_A = \delta^{y}{}_A$. 

Starting from the 5-dimensional Einstein equations, 
\begin{equation} \label{e:5deinstein}
\tilde{G}_{AB} = \tilde{\kappa}^2 [- \Lambda_5 ~\tilde{g}_{AB} + \delta
(y) (-\lambda g_{AB} + T_{AB}) ] \;,
\end{equation}
where the fields are confined to the brane that make up the brane
energy momentum tensor, with $T_{AB} n^{B} =0$. 

Applying the Gauss-Codazzi relations to $\eqref{e:5deinstein}$, 
\begin{align} 
\label{e:gauss}
& R_{ABCD} = \tilde{R}_{EFGH} g_{A}{}^{E} g_{B}{}^{F} g_{C}{}^{G}
g_{D}{}^{H} + 2 K_{A[C} K_{D]B} 
\;,  \\
\label{e:codazzi}
& \tilde{R}_{BC} g_A{}^B n^C = \n_B K^{B}{}_{A} - \n_A K_{B}{}^{B} \;,
\end{align} 
we obtain the following equation,
\begin{equation} \label{e:5deinstein2}
G_{AB} = -\frac{\tilde{\kappa}^2}{2} \Lambda_5 g_{AB} + K_{C}{}^{C}
K_{AB} - K_A{}^{C} K_{CB} + \frac{1}{2} [K^{CD} K_{CD} -
(K_{C}{}^{C})^2] g_{AB} - {\cal E}_{AB} \;, 
\end{equation}
where 
\begin{equation} \label{e:5dweyl}
{\cal E}_{AB} = \tilde{C}_{ACBD} n^C n^D \;,
\end{equation}
is the projection of the bulk Weyl tensor orthogonal to $n^A$, with
${\cal E}_{[AB]}= 0 = {\cal E}_{A}{}^{A}$. 

We evaluate $\eqref{e:5deinstein2}$ n the brane as $y \to \pm 0$ to
get the standard Einstein equation on the brane. To do that, $K_{AB}$
has to be determined at the brane. The Israel junction conditions across the
brane imply that $g_{AB}$ is continuous, while $K_{AB}$ undergoes a
jump due to the energy-momentum on the brane:
\begin{equation}\label{e:junction}
K_{AB}^{+} - K_{AB}^{-} = - \tilde{\kappa}^2 \l[ T_{AB} + \frac{1}{3} g_{AB}
(\lambda - T_{C}{}^{C}) \r] \;.
\end{equation}

We impose $Z_2$ symmetry\footnote{The choice of $Z_2$ symmetry is
motivated by a result in string theory by Horava and Witten in
\cite{horava}. They showed that the 10-dimensional $E_8 \times E_8$
heterotic string is related to an 11 dimensional theory on the
orbifold $R^{10} \times S^1/Z_2$. In that model, the standard model is
confined on the 4-dimensional spacetime while the gravitons propagate
in the full spacetime.} which implies that
\begin{equation} \label{e:z2}
K_{AB}^{-} = - K_{AB}^{+} \;,
\end{equation}
and obtain
\begin{equation}
K_{AB} =  - \frac{\tilde{\kappa}^2}{2} \l[ T_{AB} + \frac{1}{3} g_{AB}
(\lambda - T_{C}{}^{C}) \r] \;.
\end{equation}
where we dropped the $(+)$ and evaluate the quantities on the brane by
taking the limit $y \to +0$. 

Finally, we arrive at the result in \cite{shiromizu}. The standard
Einstein equation which is  
modified with new terms carrying the bulk effects on the brane:
\begin{equation} \label{e:einstein1}
G_{ab} = - \Lambda g_{ab} + \kappa^2 T_{ab} + \tilde{\kappa}^4 {\cal S}_{ab}
- {\cal E}_{ab}\;,
\end{equation}
where $\kappa^2=8 \pi/M_{4}^2$. The bulk corrections to the
Einstein equations on the brane are made up of 
two parts: (i) the matter fields which contribute local quadratic
energy-momentum corrections via the symmetric tensor ${\cal S}_{ab}$;
and (ii) the non-local effects from the free gravitational field in the
bulk transmitted by the (symmetric) projection ${\cal E}_{ab}$ of the bulk
Weyl tensor. The matter corrections are given by
\begin{equation} \label{e:emtensor2}
{\cal S}_{ab} = \frac{1}{12} T_{c}{}^{c} T_{ab} - \frac{1}{4} T_{ac}
T^{c}{}_b + \frac{1}{24} g_{ab} [3 T_{cd} T^{cd} - (T_{c}{}^{c})^2]\;.
\end{equation}
We note that the local part of the bulk gravitational field is the
five dimensional Einstein tensor $\tilde{G}_{AB}$, which is determined
by the bulk field equations. Consequently, ${\cal E}_{ab}$
transmits non-local gravitational degrees of freedom from the
bulk to the brane that includes both tidal (or Coulomb),
gravito-magnetic, and transverse traceless (gravitational wave)
effects.

The bulk corrections can all be consolidated into an effective total
energy density, pressure, anisotropic stress and energy flux. The
modified Einstein equations take the standard Einstein form with a
re-defined energy-momentum tensor:
\begin{equation} \label{e:einstein2}
G_{ab} = - \Lambda g_{ab} + \kappa^2 T^{\text{tot}}_{ab}\;,
\end{equation}
where
\begin{equation} \label{e:emtensor3}
T^{\text{tot}}_{ab} = T_{ab} + \frac{\tilde{\kappa}^4}{\kappa^2}
{\cal S}_{ab} - \frac{1}{\kappa^2} {\cal E}_{ab}\;.
\end{equation}
Decomposing ${\cal E}_{ab}$ irreducibly with respect to $u^a$ by analogy
with Eq.~(\ref{e:emequation1}), in \cite{gordon,maartens,maartens2}
\begin{equation}
{\cal E}_{ab} = - \left(\frac{\tilde{\kappa}}{\kappa}\right)^4
\left( {\cal U} u_a u_b + 2 u_{(a}{\cal Q}_{b)} + \frac{\cal U}{3} h_{ab} +
{\cal P}_{ab} \right), 
\end{equation}
(the prefactor is included to make e.g. ${\cal U}$ have dimensions of energy
density), it follows that the total density, pressure, energy flux and
anisotropic pressure are given as follows:
\begin{align}
\label{e:rhototal1}
\rho^{\text{tot}} &= \rho + \frac{\tilde{\kappa}^4}{\kappa^6}
\left[\frac{\kappa^4}{24} (2 \rho^2 - 3 
\pi^{ab} \pi_{ab}) + {\cal U} \right]\;, \\
\label{e:ptotal1}
P^{\text{tot}} &= P  + \frac{\tilde{\kappa}^4}{\kappa^6}
\left[\frac{\kappa^4}{24} \left(2 \rho^2 + 4 P \rho + \pi^{ab} \pi_{ab} - 4
q_{a} q^{a} \right) + \frac{1}{3} {\cal U} \right]\;, \\
\label{e:fluxtotal1}
q^{\text{tot}}_{a} &= q_a + \frac{\tilde{\kappa}^4}{\kappa^6}
\left[\frac{\kappa^4}{24} (4 \rho q_a - 6\pi_{ab} q^{b}) + {\cal Q}_a \right]\;,
\\
\label{e:pressuretotal1}
\pi^{\text{tot}}_{ab} &= \pi_{ab} +  \frac{\tilde{\kappa}^4}{\kappa^6}
\left[\frac{\kappa^4}{12} \left[ -(\rho + 3P) \pi_{ab} -3 \pi_{c
\langle a}\pi_{b\rangle}{}^{c} + 3q_{\langle a}q_{b\rangle} \right] +
{\cal P}_{ab} \right]\;.
\end{align}
Making use of eq. $\eqref{e:constant1}$, the above equations can be recast into
the following form which is more convenient for our purposes:
\begin{align}
\label{e:rhototal2}
\rho^{\text{tot}} &= \rho + \frac{1}{4 \lambda} (2 \rho^2 - 3 \pi_{ab}
\pi^{ab}) + \frac{6}{\kappa^4 \lambda} {\cal U} \;, \\
\label{e:ptotal2}
P^{\text{tot}} &= P + \frac{1}{4 \lambda} (2 \rho^2 + 4 \rho P + \pi_{ab}
\pi^{ab} - 4 q_a q^a) + \frac{2}{\kappa^4 \lambda} {\cal U} \;, \\
\label{e:fluxtotal2}
q^{\text{tot}}_{a} &= q_a + \frac{1}{4 \lambda} (4 \rho q_a - \pi_{ab}
q^b) + \frac{6}{\kappa^4 \lambda} {\cal Q}_a \;, \\ 
\label{e:pressuretotal2}
\pi^{\text{tot}}_{ab} &= \pi_{ab} + \frac{1}{2 \lambda}
[-(\rho+ 3 P)\pi_{ab} + \pi_{c \langle a} \pi_{b \rangle}{}^{c} +
q_{\langle a} q_{b \rangle}] + \frac{6}{\kappa^4 \lambda} {\cal
P}_{ab}  \;.
\end{align}
It is immediately obvious from the above equations that we can regain
the 4-dimensional general relativity results, when $\lambda^{-1} \to
0$. 

For the braneworld case, it is useful to introduce an additional
dimensionless gradient which describes inhomogeneity in the non-local energy
density ${\cal U}$:
\begin{equation}
\label{e:nonlocal0}
\Upsilon_a \equiv \frac{a}{\rho} D_a {\cal U}\;.
\end{equation}

The Gauss-Codazzi scalar equation for the 3-curvature defined by ${\cal
R}$ is given by
\begin{equation} \label{e:curvature1}
{\cal R} = \kappa^2 \rho \l(2 + \frac{\rho}{\lambda} \r) +
\frac{12}{\kappa^2 \lambda} {\cal U} - \frac{2}{3} \Theta^2
+ 2 \Lambda\;,
\end{equation}
where
\begin{equation} \label{e:curvature2}
{\cal R} \equiv ~^{(3)}R = h^{ab}~^{(3)}R_{ab}
\end{equation}
with ${}^{(3)} R_{ab}$ the intrinsic curvature of the surfaces orthogonal
to $u^a$\footnote{If the
vorticity is non-vanishing flow-orthogonal hypersurfaces will not exist, and
${\cal R}$ cannot be interpreted as the spatial curvature scalar.}.
In braneworld models, the Gauss-Codazzi constraint reduces to the modified
Friedmann equation
\begin{equation}\label{e:friedmann1}
H^2 + \frac{K}{a^2} = \frac{1}{3}\kappa^2 \rho \l(1+ \frac{\rho}{2
\lambda} \r) + \frac{1}{3} \Lambda + \frac{2}{\kappa^2 \lambda} {\cal U},
\end{equation}
where the 3-curvature scalar is ${\cal R}=6K/a^2$.
In non-flat models ($K \neq 0$) ${\cal R}$ is not gauge-invariant since it
does not vanish in the FRW limit. However, the comoving projected gradient
\begin{equation} \label{e:curvature3}
\eta_b \equiv \frac{a}{2} D_b {\cal R}\;
\end{equation} 
is a gauge-invariant measure of inhomogeneity in the intrinsic three
curvature of the hypersurfaces orthogonal to $u^a$.

\section{Linearised perturbation equations for the
total matter variables}  \label{sec:equations}

In this section we derive the linearized perturbation
equations\footnote{ 
The exact, non-linear local and nonlocal conservation equations,
propagation and constraint equations are found in \cite{maartens,
maartens2} and these equations can be also used for both cosmological
and astrophysical modelling, including strong 
gravity effects.} 
for the study of cosmological perturbations (scalar, vector
and tensor) on the brane. These linearized equations are essential for
the study of CMB anisotropies and large scale structure, which provide
an indirect probe of early universe cosmology.
\subsection{Local and non-local conservation equations} \label{sec:equations1}

Based on the form of the bulk energy-momentum tensor and $Z_2$
symmetry, the brane energy-momentum tensor is still covariantly conserved:
\begin{equation} \label{e:em1}
\n^b T_{ab} = 0\;.
\end{equation}
The contracted Bianchi identities on the brane ensure conservation of the total
energy-momentum tensor, which combined with conservation of the
matter tensor gives
\begin{equation} \label{e:bianchi1}
\n^{a} {\cal E}_{ab} = \tilde{\kappa}^4 \n^{a} {\cal S}_{ab} = \frac{6
\kappa^2}{\lambda} \n^{a} {\cal S}_{ab} \;.
\end{equation}
The longitudinal part of ${\cal E}_{ab}$ is sourced by quadratic
energy-momentum terms including spatial gradients and time derivatives.
As a result any evolution and inhomogeneity in the matter fields
would generate non-local Coulomb-like gravitational effects
in the bulk which back react on the brane. The conservation
equation~(\ref{e:em1}) implies evolution equations for the energy and
momentum densities, and these are unchanged from their general relativistic
form. To linear order in an almost-FRW brane cosmology we have
\begin{equation} \label{e:em2}
\dot{\rho} + \Theta (\rho + P) + D^a q_a = 0 \;,
\end{equation}
and
\begin{equation} \label{e:em3}
\dot{q}_a + \frac{4}{3} \Theta q_{a}  + (\rho +P )A_a + D_a P + D^b
\pi_{ab} = 0\;.
\end{equation}
The linearised propagation equations for ${\cal U}$ and ${\cal Q}$ follow from
Eq.~(\ref{e:bianchi1}) in \cite{maartens}:
\begin{equation} \label{e:nonlocal1}
\dot{{\cal U}} + \frac{4}{3} \Theta {\cal U} + D^a {\cal Q}_a =0 \;,
\end{equation}
and
\begin{equation} \label{e:nonlocal2}
\dot{{\cal Q}_{a}} +\frac{4}{3} \Theta {\cal Q}_{a} +
\frac{1}{3} D_a {\cal U} + D^{b} {\cal P}_{ab} + \frac{4}{3} {\cal U}A_a
= \frac{\kappa^4}{12} (\rho + P) \left(-2 D_a \rho + 3 D^b
\pi_{ab} + 2 \Theta q_a \right)\;.
\end{equation}

Taking the projected derivative of Eq.~(\ref{e:em2}) we obtain the
propagation equation for $\Delta_a$ at linear order:
\begin{equation}
\rho \dot{\Delta}_a + (\rho +P)({\cal Z}_a + a \Theta A_a) + a
D_a D^b q_b + a \Theta D_a P - \Theta P \Delta_a = 0.
\end{equation}
From equation $\eqref{e:nonlocal1}$, we obtain the evolution equation
of the spatial gradient of the non-local energy density: 
\begin{equation}
\dot{\Upsilon}_a = \left(\frac{P}{\rho} - \frac{1}{3} \right) \Theta
\Upsilon_a - \frac{4}{3} \frac{{\cal U}}{\rho}({\cal Z}_a + a\Theta A_a)
- \frac{a}{\rho} D_a D^b {\cal Q}_b.
\end{equation}

From the propagation equations for ${\cal U}$ and ${\cal Q}$ it can be seen
that the energy of the
projected Weyl fluid is conserved while the momentum is not conserved;
rather it is driven by the matter source terms on the right of
Eq.~(\ref{e:bianchi1}). Note that no propagation equation for ${\cal P}_{ab}$
is implied so the set of equations will not close.

\subsection{Propagation and constraint equations} \label{sec:equations2}

In this section we give the general linearised gravito-electric,
gravito-magnetic, shear and vorticity propagation and constraint
equations on the brane, which follow from the Bianchi
identities, and the equations for the kinematic variables
$\sigma_{ab}$, and $\Theta$ and its gradient ${\cal Z}_a$ which follow
from the Ricci identity. 
\begin{enumerate}
\item Gravito-electric propagation:
\begin{equation} \label{e:propagation1}
\begin{split}
&\dot{E}_{ab} + \Theta E_{ab} + \frac{1}{2} \kappa^2 (\rho + P)
\sigma_{ab} + \frac{1}{2} \kappa^2 D_{\langle a}q_{b \rangle} +
\frac{1}{6} \kappa^2 \Theta \pi_{ab} + \frac{1}{2} \kappa^2
\dot{\pi}_{ab} - \text{curl} H_{ab} \\
&= \frac{1}{12 \kappa^2 \lambda} [\kappa^4 \{-6\rho
(\rho + P) \sigma_{ab} + 3 (\dot{\rho} + 3\dot{P}) \pi_{ab}
+ 3 (\rho + 3 P) \dot{\pi}_{ab}   \\
&- 6\rho D_{\langle a} q_{b\rangle}
+ \Theta [\rho + 3 P] \pi_{ab}\} - 48 {\cal U} \sigma_{ab} - 36
\dot{{\cal P}}_{ab} - 36 D_{\langle a}{\cal
Q}_{b\rangle} - 12 \Theta {\cal P}_{ab}]\; ;
\end{split}
\end{equation}
\item Gravito-electric divergence:
\begin{equation} \label{e:constraint2}
\begin{split}
&D^{b} E_{ab} + \frac{1}{2} \kappa^2 D^{b} \pi_{ab} - \frac{1}{3}
\kappa^2 D_{a} \rho + \frac{1}{3} \kappa^2 \Theta q_a \\
&= \frac{1}{8 \kappa^2 \lambda} \bigg[\kappa^4 \left(- \frac{8}{3}
\rho \Theta q_a + 2 (\rho + 3P) D^{b}\pi_{ab} + \frac{8}{3} \rho D_a
\rho \right)  + 16 D_a {\cal U}  - 16 \Theta {\cal Q}_a - 24 D^b {\cal
P}_{ab} \bigg]\; ; 
\end{split}
\end{equation}
\item Gravito-magnetic propagation:
\begin{equation} \label{e:Hpropagation}
\dot{H}_{ab}  + \Theta H_{ab} + \text{curl} E_{ab} -
\frac{\kappa^2}{2} \text{curl} \pi_{ab} = \frac{3}{\kappa^2 \lambda}
\text{curl}{\cal P}_{ab} - \frac{\kappa^2}{4
 \lambda} \text{curl} \l[(\rho+3 P) \pi_{ab} \r]   \; ;
\end{equation}
\item Gravito-magnetic divergence:
\begin{equation} \label{e:Hdivergence}
D^{b} H_{ab} - \kappa^2 (\rho+P) \omega_a + \frac{\kappa^2}{2}
\text{curl} q_a  = \frac{\kappa^2
\rho}{\lambda} \omega_a + \frac{1}{\kappa^2 \lambda} (8 {\cal U}
\omega_a - 3 \text{curl} {\cal Q}_a ) \; ;
\end{equation}
\item Gravito-magnetic constraint
\begin{equation} \label{e:Hconstraint}
D_{\langle a} \omega_{b \rangle} + \text{curl} \sigma_{ab} - H_{ab} = 0 \; ;
\end{equation}
\item Shear propagation:
\begin{equation} \label{e:propagation2}
\dot{\sigma}_{ab} + \frac{2}{3} \Theta \sigma_{ab} +
E_{ab} - \frac{1}{2} \kappa^2 \pi_{ab} - D_{\langle a} A_{b\rangle} 
= \frac{1}{4 \kappa^2 \lambda} \{ \kappa^4 [ - (\rho + 3P)
\pi_{ab} ] + 12 {\cal P}_{ab}\}\; ;
\end{equation}
\item Shear constraint:
\begin{equation} \label{e:constraint1}
D^{b} \sigma_{ab} - \text{curl}\omega_a-  \frac{2}{3} D_{a} \Theta + \kappa^2 q_a =
-\frac{1}{\kappa^2 \lambda} (\kappa^4 \rho q_a + 6 {\cal Q}_{a})\; ;
\end{equation}
\item Vorticity Propagation: 
\begin{equation} \label{e:vorticity1}
\dot{\omega}_{\langle a \rangle} + \frac{2}{3} \Theta \omega_a +
\frac{1}{2} \text{curl} A_a = 0 \; ; 
\end{equation}
\item Vorticity Constraint: 
\begin{equation}\label{e:vorticity2}
D^a \omega_a - A^a \omega_a = 0 \; ;
\end{equation}
\item Modified Raychaudhuri equation:
\begin{equation} \label{e:raychaudhuri}
\dot{\Theta}= -\frac{1}{3}\Theta^2 - \frac{1}{2}\kappa^2(\rho+3P)
+ \Lambda - \frac{1}{2 \kappa^2 \lambda} [\kappa^4 \rho
(2\rho+3P) + 12 \mathcal{U}] + D^a A_a\; ; 
\end{equation}
\item Propagation equation for the comoving expansion gradient
${\cal Z}_a$ which follows from Eq.~$\eqref{e:raychaudhuri}$:
\begin{equation} \label{e:propagation3}
\dot{{\cal Z}}_a + \frac{2}{3} \Theta \mathcal{Z}_{a} - a \dot{\Theta} A_a
+ \frac{\kappa^2}{2} aD_a(\rho+3P) - aD_a D^b A_b
= -\frac{1}{2 \kappa^2 \lambda} \{
\kappa^4 aD_a[\rho(2\rho+3P)] + 12 a D_a {\cal U} \} \;.
\end{equation}
\end{enumerate}

The above propagation and constraint equations reduce to
 general relativity when $\lambda^{-1} \to 0$. These linearized
equations together with Eqs.~$\eqref{e:em1}$--$\eqref{e:nonlocal2}$,
govern the dynamics of the matter and gravitational fields on the
brane, and incorporate both the quadratic energy-momentum
(local) and the projected effects from the bulk (nonlocal). The local
terms are proportional to $\rho/\lambda$, and they are dominant at
high energies. The nonlocal terms would introduce imperfect fluid
effects onto the brane, even if the matter has a perfect fluid form. 

It was shown in \cite{maartens, maartens2} that the bulk effects
give rise to new driving and source terms from
Eqs.~$\eqref{e:propagation1}$--$\eqref{e:propagation3}$. However, the
vorticity propagation and constraint equations, together with the
gravito-magnetic constraint have no direct bulk effects. The local and
nonlocal energy densities act as driving terms in the expansion
propagation. The spatial gradients of the local and nonlocal energy
densities provide sources for the gravito-electric
field. The nonlocal energy flux  provides a source for the shear
and gravito-magnetic field, and finally, the nonlocal anistropic
stress acts a driving term in the shear propagation and the
gravito-electric and gravito-magnetic fields. 

The spatial gradient of the 3-curvature scalar is an auxiliary
variable. It can be related to the other gauge-invariant variables
using Eqs.~$\eqref{e:curvature1}$ and $\eqref{e:curvature3}$:
\begin{equation} \label{e:curconstraint}
\eta_a = \kappa^2 \rho \Delta_a \l(1 + \frac{\rho}{\lambda} \r) 
+ \frac{6}{\kappa^2 \lambda} \rho \Upsilon_a - \frac{2}{3}
\Theta {\cal Z}_a\;.
\end{equation}

Taking the time derivative of Eq.~$\eqref{e:curconstraint}$,
commuting the spatial and temporal
derivatives, and then making use of Eqs.~$\eqref{e:raychaudhuri}$ and
$\eqref{e:propagation3}$, we obtain the evolution of the spatial
gradient of the 3-curvature scalar:
\begin{equation} \label{e:evolution}
\dot{\eta}_a + \frac{2}{3} \Theta \eta_a + \frac{1}{3} {\cal R}
({\cal Z}_a + a \Theta A_a) + \frac{2}{3}\Theta a D_a D^b A_b
= - \kappa^2 \left(1 + \frac{\rho}{\lambda} \right) a D_a D^b q_b
- \frac{6}{\kappa^2 \lambda} a D_a D^b {\cal Q}_b.
\end{equation}
In general relativity, propagating $\eta_a$ is a useful device to avoid
numerical instability problems when solving for isocurvature modes in a
zero acceleration frame (such as the rest-frame of the CDM)~\cite{lewis}.

\section{Cosmological Perturbations on the Brane} \label{sec:perturbations}

In this section, we study the covariant, gauge-invariant 
splitting of linear cosmological perturbations into scalar,
vector and tensor modes. In particular, we concentrate on the scalar and
tensor equations on the brane, which are used for the analysis in
the following chapter. The vector perturbations are placed here for
completeness. 

The limiting case of the background FRW brane is characterized
by homogeneity and isotropy, i.e.:
\begin{equation}
D_a f = V_a = W_{ab} = 0  \;,
\end{equation}
where the quantities $f=\rho, P, \Theta, {\cal U}$,
$V_a=A_a,~\omega_a,~{\cal Q}_a$, and $W_{ab}=\sigma_{ab}, E_{ab}, 
H_{ab}, {\cal P}_{ab}$. 

The covariant and gauge-invariant splitting into scalar, vector and
tensor modes is given by:
\begin{equation}
V_a = D_a V + \hat{V}_a \;,
\end{equation}
and 
\begin{equation}
W_{ab} = D_{\langle a} D_{b \rangle} W + D_{\langle a} \hat{W}_{b
\rangle} + \hat{W}_{ab} \;,
\end{equation}
where a hat denotes a transverse (divergence free) quantity (and
$W_{ab}$ is assumed trace-free). We note that $V_a$, $W_{ab}$ and the
other derived quantities, e.g. $f$ and $\hat{V}_a$ are first order.

\subsection{Scalar Perturbations} \label{sec:scalarpert}

The scalar perturbations are covariantly characterized by 
\begin{equation}
\hat{V}_a = \hat{W}_a = \hat{W}_{ab}=0 \;,
\end{equation}
and consequently, we obtain
\begin{equation}
\text{curl} V_a = 0 = \text{curl} W_{ab}, \quad D^b W_{ab} =
\frac{2}{3} D^2 (D_a W)\;,
\end{equation}

The vorticity constraint equation and gravitomagnetic constraint
equation show that
\begin{equation}
\quad \omega_a = 0 = H_{ab} \;.
\end{equation}

For scalar perturbations, the magnetic part of the Weyl tensor
$H_{ab}$ and the vorticity tensor $\omega_{ab}$ vanish identically.
The electric part of the Weyl tensor $E_{ab}$ and the
shear $\sigma_{ab}$ need not vanish. The non-vanishing variables
satisfy the propagation and constraint equations on the brane.

\subsubsection{Scalar Harmonics}

The tensor-valued, partial differential equations presented in the earlier
sections can be reduced to scalar-valued, ordinary differential equations
by expanding in an appropriate complete set of eigentensors. For scalar
perturbations all gauge-invariant tensors can be constructed from
derivatives of scalar functions . Thus it is natural to expand in STF tensors
derived from the scalar eigenfunctions $Q^{(k)}$ of the projected Laplacian:
\begin{equation} \label{e:helmholtz}
D^2 Q^{(k)} = - \frac{k^2}{a^2} Q^{(k)}\;,
\end{equation}
satisfying $\dot{Q}^{(k)} = O(1)$\footnote{The notation $O(n)$ is short
for $O(\epsilon^n)$ where $\epsilon$ is some dimensionless quantity
characterising the departure from FRW symmetry. A list of identities
for the scalar, vector and tensor harmonics are given in Appendix \ref{appendix3}.}.
We adopt the following harmonic expansions of the gauge-invariant variables:
\begin{equation} \label{e:harmonics1}
\begin{split}
\Delta^{(i)}_a = \sum_k k \Delta_k^{(i)} Q^{(k)}_a\;,
\quad&\quad {\cal Z}_a = \sum_k \frac{k^2}{a} {\cal Z}_k
Q^{(k)}_a\;, \\
q^{(i)}_a = \rho^{(i)} \sum_k q_k^{(i)} Q_a^{(k)}\;,
\quad&\quad \pi^{(i)}_{ab} =  \rho^{(i)} \sum_k \pi_k^{(i)}
Q_{ab}^{(k)}\;, \\
E_{ab} = \sum_k \frac{k^2}{a^2} \Phi_k Q_{ab}^{(k)}\;,
\quad&\quad \sigma_{ab} =   \sum_k \frac{k}{a} \sigma_k Q_{ab}^{(k)}\;,
\\ v^{(i)}_{a} =  \sum_k v^{(i)}_k Q_a^{(k)}\;,  \quad&\quad A_a =  \sum_k
\frac{k}{a} A_k Q_a^{(k)}\;.
\end{split}
\end{equation}
Here $v^{(i)}_a$ is the 3-velocity of species $i$ relative to $u^a$; for the
CDM model considered here we shall make use of $v^{(i)}_a$ for baryons $b$
and CDM $c$. For photons $\gamma$ and neutrinos $\nu$ we continue to work with
the momentum densities which are related to the peculiar velocity of the
energy frame for that species by e.g.\ $q^{(\gamma)}_a = (4/3)\rho^{(\gamma)}
v^{(\gamma)}_a$ in linear theory.
The scalar expansion coefficients, such as $\Delta_k^{(i)}$ are
first-order gauge-invariant variables satisfying  e.g.\ $D^a
\Delta_k^{(i)} = O(2)$. \\

We expand the non-local perturbation variables in scalar harmonics in the
following manner:
\begin{equation} \label{e:harmonics2}
\begin{split}
\Upsilon_a &= \sum_k k \Upsilon_k Q_a^{(k)}\;, \\
{\cal Q}_a &= \sum_k \rho {\cal Q}_k Q_a^{(k)}\;, \\
{\cal P}_{ab} &= \sum_k \rho {\cal P}_k Q_{ab}^{(k)}\;.
\end{split}
\end{equation}
In addition, we can expand the projected gradient of the 3-curvature term:
\begin{equation}
\eta_a = \sum_k 2 \left(\frac{k^3}{a^2} \right) \left(1 - \frac{3K}{k^2}
\right) \eta_k Q_{a}^{(k)} \;. 
\end{equation}
The form of this expansion is chosen so that if we adopt the energy
frame (where $q_a=0$) the variable $\eta_k$ coincides with the curvature
perturbation usually employed in gauge-invariant calculations.

\subsubsection{Scalar equations on the brane}
It is now straightforward to expand the 1+3 covariant propagation and
constraint equations in scalar harmonics (see Appendix
\ref{appendix3}). We shall consider the
CDM model where the particle species are baryons (including
electrons), which we model as an ideal fluid with pressure $p^{(b)}$ and
peculiar velocity $v^{(b)}_a$, cold dark matter, which has vanishing
pressure and peculiar velocity $v^{(c)}_a$, and photons and (massless)
neutrinos which require a kinetic theory description.
We neglect photon polarization, although this can easily be included in the 1+3 covariant
framework~\cite{challinor3}. Also, we assume that the entropy perturbations
are negligible for the baryons, so that $D_a P^{(b)} = c_s^2 D_a \rho^{(b)}$
where $c_s^2$ is the adiabatic sound speed. 
A complete set of 1+3 perturbation
equations for the general relativistic model can be found
in~\cite{challinor2}. We extend these equations to braneworld models here.

In the following, perturbations in the total matter variables are related to
those in the individual components by
\begin{equation}
\rho \Delta_k = \sum_i \rho^{(i)} \Delta^{(i)}_k, \quad
\rho q_k = \sum_i \rho^{(i)} q^{(i)}_k, \quad
\rho \pi_k = \sum_k \rho^{(i)} \pi^{(i)}_k, 
\end{equation}
where $q^{(b)}_k=(1+P^{(b)}/\rho^{(b)})v^{(b)}_k$, $q^{(c)}_k = v^{(c)}_k$, and
$\pi^{(i)}_k$ vanishes for baryons and CDM. Similarly, the total density
and pressure are obtained by summing over components, e.g.\
$P = \sum_i P^{(i)}$. It is also convenient to write 
\begin{equation} \label{e:prho}
P=(\gamma-1)\rho \; ,
\end{equation}
but $\gamma$ should not be assumed constant (in space or time).

We begin with the equation for the gravito-electric field:
\begin{equation} \label{e:propagation1a}
\begin{split}
&\left(\frac{k}{a}\right)^2 \left(\dot{\Phi}_k + \frac{1}{3} \Theta \Phi_k \right) +
\frac{1}{2} \frac{k}{a} \kappa^2 \rho (\g \sigma_k - q_k) +
\frac{1}{6} \kappa^2 \rho \Theta (1-3\gamma) \pi_k + \frac{1}{2} \kappa^2
\rho \dot{\pi}_k  \\
&= \frac{1}{12 \kappa^2 \lambda} \bigg\{ -\kappa^4
\bigg[6 \left(\frac{k}{a} \right) \rho^2 (\g \sigma_k -q_k)- 3(\dot{\rho} + 3\dot{P})
\rho \pi_k - 3 (3 \g -2) \rho (\rho \dot{\pi}_k + \dot{\rho} \pi_k) \\ 
&\quad - (3\g -2) \rho^2 \Theta \pi_k \bigg] - 12 \left(\frac{k}{a} \right)(4 {\cal U}
\sigma_k - 3 \rho \mathcal{Q}_k) - 36
(\dot{\rho} {\cal P}_k + \rho \dot{{\cal P}}_k) - 12 \rho \Theta
{\cal P}_k \bigg\}\;.
\end{split}
\end{equation}
We have written this equation in such a form that every term is
manifestly frame-independent.
The shear propagation equation is
\begin{equation} \label{e:propagation2a}
\frac{k}{a} \left(\dot{\sigma}_k + \frac{1}{3} \Theta \sigma_k \right) +
\left(\frac{k}{a}\right)^2 (\Phi_k + A_k) - \frac{\kappa^2}{2} \rho \pi_k =
\frac{1}{4 \kappa^2 \lambda} [-(3 \g - 2) \kappa^4
\rho^2 \pi_k + 12 \rho {\cal P}_k]\;.
\end{equation}
The shear constraint is given by
\begin{equation} \label{e:constraint1a}
\kappa^2 \rho q_k - \frac{2}{3} \left(\frac{k}{a} \right)^2 \left[{\cal Z}_k - \left(1
- \frac{3K}{k^2} \right) \sigma_k \right] = - \frac{1}{\kappa^2 \lambda}
(\kappa^4 \rho^2 q_k + 6\rho  {\cal Q}_k).
\end{equation}
The gravito-electric divergence is
\begin{equation} \label{e:constraint2a}
\begin{split}
& 2 \left(\frac{k}{a} \right)^3 \left(1 - \frac{3K}{k^2} \right) \Phi_k -
\kappa^2 \rho \left(\frac{k}{a} \right) \left[\Delta_k - \left(1 -
\frac{3K}{k^2} \right) \pi_k \right] + \kappa^2 \Theta \rho q_k \\
&= \frac{3}{8 \kappa^2 \lambda} 
\bigg\{\kappa^4 \bigg[-\frac{8}{3} \rho^2 \Theta q_k + \frac{4}{3} (3\g -2)
\rho^2 \left(1 - \frac{3K}{k^2}\right) \frac{k}{a} \pi_k + \frac{8}{3}
\frac{k}{a} \rho^2 \Delta_k \bigg]  \\
&\quad  + 16 \frac{k}{a} \rho \Upsilon_k - 16 \Theta \rho {\cal Q}_k - 16 \rho 
\left(\frac{k}{a} \right) \left(1 -
\frac{3K}{k^2} \right) {\cal P}_k \bigg\} \;.
\end{split}
\end{equation}
The  propagation equation for the comoving expansion gradient
${\cal Z}_a$ is given by
\begin{equation} \label{e:propagation3b}
\begin{split}
&\dot{{\cal Z}}_k + \frac{1}{3} \Theta \mathcal{Z}_k - \frac{a}{k}
\dot{\Theta} A_k + \frac{k}{a} A_k + \frac{\kappa^2}{2} \frac{a}{k}
\left[2 (\rho^{(\g)}
\Delta^{(\g)}_k + \rho^{(\nu)} \Delta^{(\nu)}_k ) + (1 + 3 c_s^2)
\rho^{(b)}
\Delta^{(b)}_k + \rho^{(c)} \Delta^{(c)}_k  \right] \\
&= -\frac{1}{2 \kappa^2 \lambda} \frac{a}{k}
\bigg\{ \kappa^4[(2\rho + 3P)\rho\Delta_k + \rho(3\rho^{(\g)}\Delta^{(\g)}_k
+3\rho^{(\nu)}\Delta^{(\nu)}_k + (2+3c_s^2)\rho^{(b)}\Delta_k^{(b)} +
2 \rho^{(c)} \Delta^{(c)}_k)] \\
&\quad  + 12 \rho \Upsilon_k \bigg \} \;.
\end{split}
\end{equation}
The non-local evolution equations for $\Upsilon_k$ and ${\cal Q}_k$ are
\begin{equation} \label{e:nonlocala}
\dot{\Upsilon}_k = \frac{1}{3}(3 \g -4 ) \Theta \Upsilon_k - \frac{4}{3} \Theta
\frac{{\cal U}}{\rho} A_k - \frac{4}{3}
\frac{{\cal U}}{\rho} \frac{k}{a} {\cal Z}_k + \frac{k}{a}
{\cal Q}_k ,
\end{equation}
and
\begin{equation} 
\begin{split}
\label{e:nonlocal2a}
& \dot{{\cal Q}}_k - \frac{1}{3}(3 \g - 4) \Theta
{\cal Q}_k + \frac{1}{3} \frac{k}{a} \left[\Upsilon_k + 2 \left(1
- \frac{3 K}{k^2} \right) {\cal P}_k \right]  + \frac{4}{3}\frac{k}{a}
\frac{\mathcal{U}}{\rho} A_k \\
&=
\frac{\kappa^4}{6} \g \rho \left\{ \Theta q_k
+ \frac{k}{a} \left[ \left(1 - \frac{3 K}{k^2} \right) \pi_k -
\Delta_k \right] \right\}\;. 
\end{split}
\end{equation}
The spatial gradient of the 3-curvature scalar is
\begin{equation} \label{e:3curvature}
\left(\frac{k}{a} \right)^2 \left(1 - \frac{3 K}{k^2} \right) \eta_k 
= \frac{\kappa^2 \rho}{2} \Delta_k \l(1 + \frac{\rho}{\lambda} \r) 
+ \frac{3}{\kappa^2 \lambda} \rho \Upsilon_k - \frac{1}{3}\frac{k}{a}
\Theta {\cal Z}_k, 
\end{equation} 
and it evolves according to
\begin{equation}
\frac{k}{a} \left(1 - \frac{3 K}{k^2} \right) \left(\dot{\eta}_k - \frac{1}{3}
\Theta A_k \right) + \frac{K}{a^2} {\cal Z}_k - \frac{1}{2} \kappa^2 \rho
q_k = \frac{1}{2 \kappa^2 \lambda} (\kappa^4
\rho^2 q_k + 6 \rho {\cal Q}_k).
\end{equation}

The evolution equations for the scalar harmonic components of the
comoving, fractional density gradients for photons, neutrinos, baryons and
cold dark matter (CDM) are
\begin{align}
\label{e:photons0}
\dot{\Delta}_{k}^{(\g)} &= -\frac{k}{a} \left(\frac{4}{3} {\cal Z}_{k} -
q_{k}^{(\g)} \right)  - \frac{4}{3} \Theta A_k \quad \text{(photons)}\;, \\
\label{e:neutrinos0}
\dot{\Delta}_{k}^{(\nu)} &= -\frac{k}{a} \left(\frac{4}{3} {\cal Z}_{k} -
q_{k}^{(\nu)} \right) - \frac{4}{3} \Theta A_k \quad \text{(neutrinos)}\;, \\
\dot{\Delta}_{k}^{(b)} &= \left(1+ \frac{P^{(b)}}{\rho^{(b)}} \right)
\label{e:baryons0}
\left[-\frac{k}{a}({\cal Z}_k - v_k^{(b)}) - \Theta A_k  \right] +
\left(\frac{P^{(b)}}{\rho^{(b)}} -  c_s^2 \right) \Theta \Delta_{k}^{(b)}
\quad \text{(baryons)}\;, \\
\label{e:CDM0}
\dot{\Delta}_{k}^{(c)} &= -\frac{k}{a} ({\cal Z}_k - v^{(c)}_{k}) -
\Theta A_k \quad \text{(CDM)}\;.
\end{align}
The evolution equations for the momentum densities and peculiar velocities
are
\begin{align}
\dot{q}^{(\g)}_k & = -\frac{1}{3}\frac{k}{a}\left[\Delta^{(\g)}_k+4A_k
+2\left(1-\frac{3K}{k^2}\right)\pi^{(\g)}_k\right] + n_e \sigma_T \left(\frac{4}{3}
v^{(b)}_k - q^{(\g)}_k \right) \; ,\\
\dot{q}^{(\nu)}_k & = -\frac{1}{3}\frac{k}{a}\left[\Delta^{(\nu)}_k+4A_k
+2\left(1-\frac{3K}{k^2}\right)\pi^{(\nu)}_k\right] \; ,\\
(\rho^{(b)}+P^{(b)}) \dot{v}^{(b)}_k &= -
(\rho^{(b)}+P^{(b)})\left[\frac{1}{3}(1-3c_s^2)\Theta v^{(b)}_k + \frac{k}{a}
A_k \right] - \frac{k}{a}(1+c_s^2) \Delta^{(b)}_k \\
&\quad - n_e \sigma_T \frac{\rho^{(\gamma)}}{\rho^{(\nu)}} \left(
\frac{4}{3}v^{(b)}_k - q^{(\g)}_k \right) \; , \notag \\
\dot{v}^{(c)}_k & = - \frac{1}{3}\Theta v^{(c)}_k - \frac{k}{a}A_k \;,
\label{eq:cdmvel}
\end{align}
where the Thomson scattering terms involving the electron density
$n_e$ and Thomson cross section $\sigma_T$ arise from the interaction between
photons and the tightly-coupled baryon/electron fluid. The remaining equations
are the propagation equations for the anisotropic stresses of photons and
neutrinos, and the higher moments of their distribution functions. These
equations can be found in~\cite{challinor2}, and with polarization included
in~\cite{challinor5, challinor3, challinor6}, since they are unchanged
from general relativity. However, we shall not require these
additional equations at the level of approximation we make in our
subsequent calculations.  

\subsection{Vector Perturbations} \label{sec:vectorpert}
The vector perturbations are covariantly characterized by
\begin{equation}
V_a = \hat{V}_a, \quad W_{ab}= D_{\langle a} \hat{W}_{b \rangle},
\quad \text{curl} D_a f = - 2 \dot{f} \omega_a \;,
\end{equation}
and it follows that
\begin{equation}
D^a W_{ab} = \frac{1}{2} D^2 \hat{W}_b, \quad \text{curl} W_{ab} =
\frac{1}{2}  D_{\langle a} \text{curl} \hat{W}_{b \rangle} \;.
\end{equation}

\subsubsection{Vector equations on the brane}
There are no bulk effects in the linearized vorticity propagation
equation $\eqref{e:vorticity1}$, and hence the vorticity decays away with
the expansion like the general relativity, and this can be interpreted as
angular momentum conservation equation on the
brane. However the gravito-magnetic divergence equation becomes
\begin{equation}
D^b H_{ab} = \kappa^2 (\rho+P) \omega_a -\frac{\kappa^2}{2}
\text{curl} q_a  + \frac{1}{\kappa^2 \lambda}
\{ \kappa^2 \rho (\rho+P)+ 8 {\cal U} \} \omega_a - 3 \text{curl}
{\cal Q}_a \;.
\end{equation}
The bulk terms provide additional sources for the gravito-magnetic
field. The existence of the bulk terms made it possible to 
source vector perturbations even when the vorticity
vanishes, since $\text{curl} {\cal Q}_a$ may be non-zero.

\subsection{Tensor Perturbations} \label{sec:tensorpert}

The (1+3)-covariant description of gravitational waves in a
cosmological context has been considered in detail by
Challinor~\cite{challinor4} and Dunsby et al~\cite{dunsby4} for the
non-flat and flat cases respectively. In this section, we will extend to
the case of braneworld cosmology. 

Tensor perturbations are covariantly characterized by
\begin{equation}
D_{a} f=0, \quad \quad A_a = \omega_a = {\cal Q}_a=0, \quad D^a W_{ab}
=0, \quad W_{ab}= \hat{W}_{ab} \;.
\end{equation}

In this description, the linearized
gravitational waves are described by the tranverse degrees of freedom
in the electric and magnetic parts of the Weyl tensor. $D^{a} W_{ab}$
indicates that the shear and anisotropic stress (both local and
nonlocal) are transverse. Furthermore, the vorticity and all projected
vectors vanish at linear order. The
individual matter components all possess the same four-velocity
which defines the fundamental velocity $u^a$ in a pure tensor
mode. 

The constraint equation (Eq. $\eqref{e:Hconstraint}$) becomes 
\begin{equation} \label{e:Hconstraint2}
H_{ab} = \text{curl} \sigma_{ab} \;,
\end{equation}
which determines the magnetic part of the Weyl tensor from the
shear. 

For convenience, we define a rescaled nonlocal anisotropic stress
${\pi}_{ab}^{*}$ by the following relation:
\begin{equation}
\kappa^2 \pi_{ab}^{*} = \frac{6}{\kappa^2 \lambda} {\cal P}_{ab} \;.
\end{equation}

The propagation equations for the tensor perturbations from
Eqs~$\eqref{e:propagation1}$--$\eqref{e:propagation3}$ are reduced to
the following set:
\begin{align}
\label{e:dotrho}
\dot{\rho} &= -(\rho+P) \Theta \;, \\
\label{e:dottheta}
\dot{\Theta} &= -\frac{1}{3}\Theta^2 - \frac{1}{2}\kappa^2(\rho+3P)
+ \Lambda - \frac{1}{2 \kappa^2 \lambda} [\kappa^4 \rho
(2\rho+3P) + 12 \mathcal{U}] \;,\\
\label{e:dotsigma}
\dot{\sigma}_{ab} &= - \frac{2}{3} \Theta \sigma_{ab} - E_{ab} +
\frac{\kappa^2}{2} (\pi_{ab} + \pi^{*}_{ab}) - \frac{\kappa^2}{4 \lambda}
 (\rho + 3P) \pi_{ab}  \;, \\
\label{e:dotH}
\dot{H}_{ab} &= - \Theta H_{ab} - \text{curl} E_{ab} +
\frac{\kappa^2}{2} \text{curl} (\pi_{ab} + \pi_{ab}^{*}) - \frac{\kappa^2}{4
\lambda} \text{curl} \l[(\rho+3P) \pi_{ab} \r]  \;, \\
\label{e:dotE}
\dot{E}_{ab} &= - \Theta E_{ab} + \text{curl} H_{ab} -
\frac{\kappa^2}{2} \l[(\rho+ P) \sigma_{ab} + \dot{\pi}_{ab} +
\frac{1}{3} \Theta \pi_{ab} + \dot{\pi}_{ab}^{*} + \frac{1}{3} \Theta
\pi_{ab}^{*} \r] \\ 
&\quad + \frac{1}{12 \kappa^2 \lambda}
 \bigg[3 \kappa^4 \{ 
(\dot{\rho} + 3 \dot{P}) \pi_{ab} + (\rho + 3P) \dot{\pi}_{ab} \}+
\kappa^4 (\rho + 3P) \Theta \pi_{ab} \notag \\
&\quad - 6 \kappa^4 \rho (\rho + P)
\sigma_{ab} -48 {\cal U} \sigma_{ab}  \bigg] \;. \notag
\end{align}

The inhomogeneous wave equations for the shear and the magnetic part
of the Weyl tensor follow from differentiating Eqs. $\eqref{e:dotsigma}$
and $\eqref{e:dotH}$ along the flow lines of $u^a$, making use of
Eq.$\eqref{e:prho}$ (assuming $\dot{\g} \neq 0$) and the identities
$\eqref{e:c26}$--$\eqref{e:c29}$ from Appendix \ref{appendix3}, we obtain:
\begin{equation}
\begin{split} \label{e:ddotsigma}
&\ddot{\sigma}_{ab} + D^2 \sigma_{ab} + \frac{5}{3} \theta
\dot{\sigma}_{ab} + \l[\frac{1}{2} (4 - 3 \g) \kappa^2 \rho -
\frac{K}{a^2} \r] \sigma_{ab} \\
&= \kappa^2 \bigg[\l(\dot{\pi}_{ab} + \frac{2}{3} \Theta \pi_{ab} \r) +
\l(\dot{\pi}_{ab}^{*} + \frac{2}{3}
\Theta \pi_{ab}^{*} \r) \\
&\quad -  (3 \g-2) \frac{\rho}{6 \lambda} \bigg(3 \dot{\pi}_{ab} -
(3 \g -2) \Theta \pi_{ab} 
+ \frac{9 \dot{\g}}{3 \g -2} \pi_{ab} - 3 \rho
\sigma_{ab} \bigg) \bigg]  \;,
\end{split}
\end{equation}
and 
\begin{equation}
\begin{split}
&\ddot{H}_{ab} + D^2 H_{ab} + \frac{7}{3} \theta \dot{H}_{ab} + 2
\l[(2- \g) \kappa^2 \rho - \frac{3K}{a^2} \r] H_{ab} \\
&= \kappa^2 \l[\text{curl} \dot{\pi}_{ab} + \frac{2}{3} \Theta
\text{curl} \pi_{ab} \r] + \kappa^2
\bigg[\text{curl} \dot{\pi}_{ab}^{*} + \frac{2}{3} \Theta \text{curl}
\pi_{ab}^{*} \bigg] \\
&\quad -\frac{\kappa^2}{12} (3 \g -2) \rho \bigg[ \l(\frac{3
\dot{\g}}{3 \g -2} \r)\text{curl} \pi_{ab} - \g \Theta \text{curl}
\pi_{ab} + \text{curl} \dot{\pi}_{ab} + \frac{2}{3} \Theta \rho \text{curl}
\pi_{ab} \bigg] \;.
\end{split}
\end{equation}

\subsubsection{Tensor Harmonics}

Analogous to the scalar case, the electric and magnetic parts of the
Weyl tensor, the shear, and the local and non-local anistropic stress
can be expanded directly in the electric and magnetic parity tensor
harmonics defined in \cite{challinor4}:
\begin{align}
E_{ab} &= \sum_{k} \l(\frac{k}{a}\r)^2 \l(E_k Q_{ab}^{(k)} + \bar{E}_k
\bar{Q}_{ab}^{(k)} \r) \;, \\
H_{ab} &= \sum_{k} \l(\frac{k}{a}\r)^2 \l(H_k Q_{ab}^{(k)} + \bar{H}_k
\bar{Q}_{ab}^{(k)} \r) \;, \\
\sigma_{ab} &= \sum_{k} \l(\frac{k}{a}\r) \l(\sigma_k Q_{ab}^{(k)} +
\bar{\sigma}_k \bar{Q}_{ab}^{(k)} \r) \;,  \\
\pi_{ab}^{(i)} &= \sum_{k} \rho \l( \pi_{k}^{(i)}  Q_{ab}^{(k)} +
\bar{\pi}_k^{(i)} \bar{Q}_{ab}^{(k)} \r) \;, \\
{\cal P}_{ab} &=  \sum_{k} \rho \l( {\cal P}_{k}^{(i)}  Q_{ab}^{(k)} +
\bar{\cal P}_k^{(i)} \bar{Q}_{ab}^{(k)} \r) \;.
\end{align}
where $\sum_k$ denotes a sum over the harmonic modes. Note that the
electric and magnetic parity tensor harmonics are related by
Eqs~$\eqref{e:c30}$ and $\eqref{e:c31}$ in Appendix \ref{appendix3}.

\subsubsection{Tensor Equations on the Brane}

Using the gravitomagnetic constraint in Eq.~$\eqref{e:Hconstraint}$
and together with tensor harmonics above, we expand the first order equations
(eqs. $\eqref{e:dotsigma}$ and $\eqref{e:dotE}$), and obtain
\footnote{Note that from eq. $\eqref{e:Hconstraint2}$, we find that
$H_k$ is algebraic in $\bar{\sigma}_k$:
\begin{equation}
H_k = \l(1+ \frac{3K}{k^2} \r)^{\frac{1}{2}} \bar{\sigma}_k \;.
\end{equation}
These curl terms lead to a coupling of
different polarization states.}
,
\begin{equation} \label{e:dotsigma0}
\l(\frac{k}{a}\r) \l(\dot{\sigma}_k + \frac{1}{3} \Theta \sigma_k \r)
+ \l(\frac{k}{a}\r)^2 E_k - \frac{\kappa^2}{2} \rho \pi_k 
= \frac{1}{4 \kappa^2 \lambda} \rho [- \kappa^4 (3 \g
-2 ) \rho \pi_k ] + \frac{\kappa^2}{2} \rho \pi_k^{*} \;,
\end{equation}
and 
\begin{equation} \label{e:dotE0}
\begin{split}
& \l(\frac{k}{a} \r)^2 \bigg(\dot{E}_k + \frac{1}{3} \Theta E_k \bigg) -
 \l(\frac{k}{a}\r) \l[\l(\frac{k}{a} \r)^2 + \frac{3K}{a^2} -
 \frac{\kappa^2}{2} \g \rho  \r] \sigma_k +
\frac{\kappa^2}{2} \rho \dot{\pi}_k -\frac{\kappa^2}{6} (3 \g -1)
\rho \Theta \pi_k \\
&= \frac{1}{12 \kappa^2 \lambda} \bigg\{ -\kappa^4
\bigg[6 \left(\frac{k}{a} \right) \g \rho^2  \sigma_k -
3(\dot{\rho} + 3\dot{P}) \rho \pi_k - 3 (3 \g -2) \rho (\rho
\dot{\pi}_k + \dot{\rho} \pi_k) - (3\g -2) \rho^2 \Theta \pi_k \bigg]
\\ 
&\quad  - 48 \left(\frac{k}{a} \right) {\cal U} 
\sigma_k \bigg\} - \frac{\kappa^2}{2} \l(\dot{\rho} \pi_k^{*} + \rho
 \dot{\pi}_k^{*} + \frac{1}{3} \rho \Theta \pi_k^{*} \r) \;.
\end{split}
\end{equation}

From Eq.$\eqref{e:ddotsigma}$, we get
\begin{equation} \label{e:shearwave}
\begin{split} 
&\ddot{\sigma}_k + \Theta \dot{\sigma}_k + \l[\frac{k^2}{a^2} +
\frac{2K}{a^2} -\frac{\kappa^2 \rho}{3} (3 \g -2) \r] \sigma_k 
+ \kappa^2 \rho \l(\frac{a}{k} \r) \l[\frac{1}{3} (3 \g -2) \Theta
\pi_k - \dot{\pi}_k \r] \\
&=    \frac{a}{k} \kappa^2 \rho \l(\dot{\pi}_k^{*} + \l(\frac{2}{3} - \g
\r) \Theta \pi^{*}_k \r) + \frac{1}{3 \kappa^2 \lambda} 
(12 {\cal U} + (3 \g -1) \kappa^4
\rho^2) \sigma_k  \\
&\quad  + \frac{1}{6 \kappa^2 \lambda} \l(\frac{a}{k} \r)
\bigg[9(2 \g^2 \Theta -  \dot{\g}) \kappa^4 \rho^2 \pi_k  -3(3 \g -2) \kappa^4
\rho^2 \dot{\pi}_k - 2(2-9 \g) \kappa^4 \rho^2 
\Theta \pi_k  \bigg]  \;.
\end{split}
\end{equation}
This equation is equivalent to the form found by Maartens in \cite{maartens,
maartens2} for the tranverse traceless modes on the brane. From the
above equation, the nonlocal bulk effects would provide driving terms
which would act like the anisotropic stress in general relativity. For
the local anisotropic stress, the evolution is determined by the
Boltzmann equation. However, the evolution equation
of the non-local anisotropic stress is not known. We will make an
local approximation to the evolution equation of the non-local
anisotropic stress in the subsequent chapter to see its effects on the CMB tensor
power spectra.   

\section{A covariant expression for the temperature \\ anisotropy}
\label{sec:aniso}

In this final section, we discuss the line of sight solution to the
Boltzmann equation for the scalar contribution to the gauge-invariant
temperature anisotropy $\delta_T(e)$ of the CMB in braneworld models.
We employ the 1+3 covariant approach, and show that our result is equivalent
to that given recently by Langlois et al~\cite{langlois} using the
Bardeen formalism.

Over the epoch of interest the individual matter
constituents of the universe interact with each other under gravity only,
except for the photons and baryons (including the electrons), which
are also coupled through Thomson scattering. The variation of
the gauge-invariant temperature perturbation $\delta_T (e)$, where $e^a$ is
the (projected) photon propagation direction, along the
line of sight is given by the (linearized) covariant Boltzmann
equation (valid for scalar, vector, and tensor modes)~\cite{challinor1}:
\begin{equation} \label{e:temperature1}
\begin{split}
\delta_T (e)' + \sigma_T n_e \delta_T (e) &= -\sigma_{ab} e^a e^b - A_a
e^a - \frac{e^a D_a \rho^{(\gamma)}}{4\rho^{(\gamma)}} -
\frac{D^a q^{(\gamma)}_a}{4\rho^{(\gamma)}} \\
& + \sigma_T n_e \left(v^{(b)}_a e^a + \frac{3}{16} \rho^{(\g)}
\pi_{ab}^{(\g)} e^a e^b \right)\;,
\end{split}
\end{equation}
where the prime denotes the derivative with respect to a parameter $\lambda$
defined along the line of sight by $d \lambda = - u_a dx^a$.

Following the steps in \cite{challinor1}, we expand the right-hand side of
Eq.~(\ref{e:temperature1}) in scalar harmonics and integrate along the line
of sight from the early universe to the observation point $R$. Neglecting
effects due to the finite thickness of the last scattering surface, on
integrating by parts we find that the temperature anisotropy involves
the quantity
\begin{equation}
\left(\frac{a}{k} \sigma_k' \right)' + \frac{1}{3} \frac{k}{a} (\sigma_k -
{\cal Z}_k) + A_k' - H A_k = - 2 \dot{\Phi}_k + \left(\frac{a}{k} \right)^2 I\;
\label{e:temperature10}
\end{equation}
integrated along the line of sight (after multiplying with $Q^{(k)}$).
In simplifying Eq.~(\ref{e:temperature10}) we have made use of the derivative
of the shear propagation equation
$\eqref{e:propagation2a}$, substituted for $q_k$ and
${\cal Z}_k$ from Eqs. $\eqref{e:propagation1a}$ and
$\eqref{e:constraint1a}$, and finally used Eqs.
$\eqref{e:friedmann1}$ and $\eqref{e:raychaudhuri}$.
The quantity $I$ is the total sum of all the braneworld corrections:
\begin{equation}
I  = \left(\frac{a}{k} \right)^2 \left[\dot{I}_1 + \frac{1}{3} \Theta I_1 + I_2
+ \frac{1}{3}  \left(\frac{k}{a} \sigma_k \right) I_3 +
\frac{1}{2} \left(\frac{k}{a} \right) I_4 \right]\;,
\end{equation}
where
\begin{equation}
\begin{split}
I_1 &= \frac{1}{4 \kappa^2 \lambda} [-(3 \g - 2) \kappa^4
\rho^2 \pi_k + 12 \rho {\cal P}_k]\;,   \\
I_2 &= \frac{1}{12 \kappa^2 \lambda}  \bigg\{
-\kappa^4 \bigg[6 \left(\frac{k}{a} \right) \g \rho^2 \sigma_k -
3(\dot{\rho} + 3 \dot{P}) \rho
\pi_k - 3 (3 \g -2) \rho (\rho \dot{\pi}_k + \dot{\rho} \pi_k) - 6
\left(\frac{k}{a} \right) \rho^2
q_k  \\
&\quad - (3\g -2) \rho^2 \Theta \pi_k \bigg] - 48 \left(\frac{k}{a} \right) {\cal U}
\sigma_k - 36 (\dot{\rho} {\cal P}_k + \rho \dot{{\cal P}}_k) +
36 \left(\frac{k}{a} \right) \rho {\cal Q}_k - 12 \rho \Theta
{\cal P}_k \bigg\} \;, \\
I_3 &= \frac{1}{2 \kappa^2 \lambda} [(3 \g-1) \kappa^4 \rho^2 + 12 {\cal U}]\;, \\
I_4 &= \frac{\kappa^2 \rho^2}{3 \lambda} \sigma_k + \frac{4}{\kappa^2
\lambda} {\cal U} \sigma_k -  \frac{1}{4 \kappa^2 \lambda}
(4 \kappa^4 \rho^2 q_k + 24 \rho {\cal Q}_k)\;.
\end{split}
\end{equation}
A lengthy calculation making use of the propagation and constraint equations
shows that $I=0$. The final result for the temperature anisotropies is then
\begin{equation} \label{e:temperature2}
\begin{split}
[\delta_{T}(e) ]_R &= -\sum_k \left[ \left( \frac{1}{4} \Delta_k^{(\g)} +
\frac{a}{k} \dot{\sigma}_k + A_k \right) Q^{(k)} \right]_A
+ \sum_k [(v_k^{(b)} - \sigma_k) e^a Q_a^{(k)} ]_A \\
& + \frac{3}{16} \sum_k (\pi_k^{(\g)} e^a e^b Q_{ab}^{(k)})_A
 + 2 \sum_k \int^{\lambda_R}_{\lambda_A}  \dot{\Phi}_k Q^{(k)} d\lambda\;,
\end{split}
\end{equation}
where the event $A$ is the intersection of the null geodesic with the last
scattering surface.
  
In retrospect, one could re-derive the result for the temperature anisotropy
in braneworld models much more simply by retaining the effective stress-energy
variables $\rho^{\text{tot}}$,
$P^{\text{tot}}$, $q_{a}^{\text{tot}}$ and $\pi_{ab}^{\text{tot}}$ in the
propagation and constraint equations used in the manipulation of the
left-hand side of Eq.~(\ref{e:temperature10}), rather than isolating the
braneworld contributions.

If we adopt the longitudinal gauge, defined by $\sigma_{ab}=0$, we find that
the electric part of the Weyl tensor and the acceleration are related by
$\Phi_k = -A_k$ if the total anisotropic stress $\pi_{ab}^{\text{tot}}$
vanishes. It follows that in this zero shear frame we recover the result
found by Langlois et al~\cite{langlois}.

      \clearpage{\pagestyle{empty}\cleardoublepage}
\chapter{Braneworld Cosmology II: \\ Initial Conditions and CMB
Anisotropies} \label{chapter4}

\begin{quote}
{\it ``Order and simplification are the first steps toward the mastery
of a subject - the actual enemy is the unknown.''}
\par{\bf - Thomas Mann}
\end{quote}

{\it In this chapter, we discuss the possible implications of
braneworld cosmology for the CMB, based on work described in \cite{leong1}
and \cite{leong4}. For the scalar perturbations, we supplement the
equations for the total matter variables with equations for the
independent matter components in a cold dark matter cosmology, and provide
solutions in the high and low-energy radiation-dominated phase under
the assumption that the non-local anisotropic stress vanishes. These
solutions reveal the existence of new modes arising from the two
additional non-local degrees 
of freedom. Our solutions should prove useful in setting up initial
conditions for numerical codes aimed at exploring the effect of braneworld
corrections on the cosmic microwave background (CMB) power
spectrum. We show that the 3-curvature is only constant in the high
energy limit for the modes which correspond similarly to the growing
and decaying mode in general relativity. The CMB temperature
anisotropies are insensitive to the brane tension $\lambda$ if the
dark energy contribution is ignored. For the tensor perturbations, we set out
the framework of a program to compute the tensor anisotropies in the
CMB that are generated in braneworld models. In the simplest
approximation, we show the braneworld imprint as a correction to the
power spectra for standard temperature and polarization anisotropies
and similarly show that the tensor anisotropies are also insensitive to
high energy effects.}

\section{From Perturbation Theory to CMB Anisotropies in Braneworlds}

The cosmic microwave background (CMB) currently occupies a central role
in modern cosmology. It is the cleanest cosmological observable, providing
us with a unique record of conditions along our past light cone back
to the epoch of decoupling when the mean free path to Thomson scattering
rose suddenly due to hydrogen recombination. Present
(e.g. BOOMERANG \cite{boomerang1}, MAXIMA \cite{maxima1}, and VSA \cite{vsa}) and
future (e.g. MAP and PLANCK) data on the CMB anisotropies, combined
with large-scale 
structure data,  provide extensive information on the amplitude and
evolution of cosmological perturbations. Potentially this allows us to infer the
spectrum of initial perturbations in the early universe and to
determine the standard cosmological parameters to high accuracy. An obvious
question to ask is whether there are any signatures of extra dimensions
which could be imprinted on the cosmic microwave sky.

There has been an explosion of interest in the theory of
cosmological perturbations in braneworlds and their implications
for observational cosmology. We will briefly summarize~\footnote{We
apologize in advance if we have missed some of the literature in this
thesis.} most of the work which has explored these issues. 

The relation between braneworld cosmology, the AdS/CFT
correspondence and quantum cosmology has been discussed in
\cite{anchordoqui}. Binetruy et al~\cite{binetruy} have studied the
background cosmological dynamics of a Friedmann brane in a
Schwarzschild-Anti-de Sitter bulk. The modifications to inflation are also explored
by various authors in \cite{copeland, garriga, giudice, liddle2}, in  
\cite{huey} (for quintesscence) and in \cite{gorbunov, langlois3,
sahni} (for gravitational waves). The large scale perturbations 
generated from quantum fluctuations during  de Sitter inflation on
the brane have been computed and studied in \cite{bridgman, garriga, hawking2,
langlois2, maartens3}. High-energy inflation on the brane generates a
zero-mode (4D graviton mode) of tensor 
perturbations, and stretches it to super-Hubble scales. This
zero-mode has the same qualitative features as in general
relativity, remaining frozen at constant amplitude while beyond
the Hubble horizon, but the overall amplitude is
higher~\cite{langlois5}. The massive KK modes (5D graviton modes)
remain in the vacuum state during slow-roll inflation. The
evolution of the super-Hubble zero mode is the same as in general
relativity, so that high-energy braneworld effects in the early
universe serve only to re-scale the amplitude. However, when the
zero mode re-enters the Hubble horizon, massive KK modes can be
excited. Qualitative arguments~\cite{gorbunov, hawking2} indicate
that this is a very small effect, but it remains to be properly
quantified, so that the signature on the CMB may be calculated,
and constraints may be imposed on the braneworld parameters.

Other authors~\cite{brax, dorca, kanno, kanno2, kodama, kodama2, koyama2, koyama,
sago, sahni, soda, vandebruck3} have considered the cosmological
consequences of interactions between the 
brane and the bulk. These authors constructed approximations from the bulk
equations and explored the implications for observational cosmology on
the brane. Their methods are based mainly on the Bardeen approach. 

For the study of cosmological perturbations and its relation to the
cosmic microwave background (CMB) and large scale structure (LSS), the
covariant formalism has been well described in the previous chapter. We will
summarize some of the main results for the scalar and tensor
perturbations based on earlier work by various authors. 

For scalar perturbations, the bulk effects introduce a non-adiabatic mode
on large scales. The additional non-adiabatic has been found in
\cite{bruni2, gordon, langlois, leong1, maartens}. In additional,
another growing mode was found in \cite{leong1} in $\Lambda$CDM
cosmology. The density perturbations on large scales can be solved on
the brane without solving for the bulk perturbations \cite{gordon},
but the Sachs-Wolfe effect cannot be found on the brane because of the non-local
anisotropic stress (see also \cite{langlois}), which is
underdetermined. There are possible changes on the scalar modes for
both large and small scales.

For tensor perturbations, we recall earlier that the bulk effects
generate a massless mode during 
inflation and a continuum of KK modes \cite{garriga,
langlois2}. The massive modes stay in the vacuum state, and on large 
scales, there is a constant mode with enhanced amplitude in
\cite{langlois2}. In this case, there is no qualitative change on
large scales in the low energy regime (4D general relativity), but
significant change in small scales.  

We have learned from the previous chapter that there are two types of
corrections which will arise from braneworld cosmology. The first type
of correction occurs at energies well above the brane tension
$\lambda$, where gravity becomes 5-dimensional. This will introduce 
significant corrections to general relativity. On the other hand,
there are also corrections that can operate at low energies, mediated
by bulk graviton or Kaluza-Klein (KK) modes. Both types of correction
play an important role in both scalar and tensor perturbations.

In \ref{sec:scalar}, we work out the perturbative dynamics and initial
conditions for the scalar perturbations in both the CDM (\ref{sec:CDM})
and energy frames (\ref{sec:energyframe}). We first show that there
are two additional non-local isocurvature modes present in 
braneworld cosmology and subsequently examine whether they will
contribute significantly to the CMB anisotropy in \ref{sec:modes}. 

We develop the formalism to compute the tensor anisotropies in
the CMB in \ref{sec:tensor}, which incorporates the
early-universe high-energy 
braneworld effects, and we carefully delineate what is known on
the brane from what is required from bulk equations. Once the 5D
solutions are provided, our formalism, with its modified CMB code
(based on CAMB~\cite{lewis1,lewis}), is able to compute these
anisotropies. We illustrate this by using a simple approximation in
\ref{sec:tensorapprox} to the 5D effects (refer to the
analysis of the braneworld scalar Sachs-Wolfe effect
in \cite{barrow}). We plot the braneworld correction to 
the CMB power spectra for temperature and polarization anisotropies and
discuss their implications in \ref{sec:tensorspectra}.

\section{Perturbation dynamics and Initial Conditions of Scalar Modes}  \label{sec:scalar}

\subsection{The CDM Frame} \label{sec:CDM}

In this section we specialize our equations to FRW
backgrounds that are spatially flat\footnote{More generally,
curvature effects can be ignored for modes with wavelength much shorter
than the curvature scale, $k \gg \sqrt{|K|}$, provided the curvature does not
dominate the background dynamics.} and we
ignore the effects of the cosmological constant in the early
radiation-dominated universe. To solve the equations it is essential
to make a choice of frame $u^a$. We adopted a frame
comoving with the CDM similar to \cite{challinor2}.
Since the CDM is pressure free, this $u^a$ is geodesic
($A_a=0$) which simplifies the equations considerably. We shall adopt this
frame choice here also, though we note it may be preferable to use a frame more
closely tied to the dominant matter component over the epoch of interest. This
can be easily accomplished by adopting the energy frame ($q_a=0$). For
completeness, we give equations in the energy frame in the following section.

We neglect baryon pressure ($c_s^2 \rightarrow 0$ and $P^{(b)} \rightarrow 0$)
and work to lowest order in the tight-coupling approximation ($n_e \sigma_T
\rightarrow \infty$; see e.g.\  ~\cite{ma}). At this order the energy frame
of the photons coincides with the rest frame of the baryons, so that
$v^{(b)}_a = 3 q^{(\g)}_a /(4 \rho^{(\g)})$, and all moments of the
photon distribution are vanishingly small beyond the dipole.

With these approximations and frame choice we obtain the following equations
for the density perturbations of each component:
\begin{align}
\label{e:photons1}
\dot{\Delta}_{k}^{(\g)} &= -\frac{k}{a} \left(\frac{4}{3} {\cal Z}_{k} -
q_{k}^{(\g)} \right) \quad \text{(photons)}\;, \\
\label{e:neutrinos1}
\dot{\Delta}_{k}^{(\nu)} &= -\frac{k}{a} \left(\frac{4}{3} {\cal Z}_{k} -
q_{k}^{(\nu)} \right) \quad \text{(neutrinos)}\;, \\
\label{e:baryons1}
\dot{\Delta}_{k}^{(b)} &= -\frac{k}{a} ({\cal Z}_k - v_k^{(b)})
 \quad \text{(baryons)}\;, \\
\label{e:CDM1}
\dot{\Delta}_{k}^{(c)} &= -\frac{k}{a} {\cal Z}_k \quad \text{(CDM)}\;.
\end{align}
The equations for the peculiar velocities and momentum densities are
\begin{align}
(4\rho^{(\g)} + 3 \rho^{(b)}) \dot{q}^{(\g)}_k & =
-\frac{4}{3}\frac{k}{a} \rho^{(\g)}\Delta^{(\g)}_k - \rho^{(b)}\Theta
q^{(\g)}_k \; , \label{e:tightcouple1}\\
\dot{q}^{(\nu)}_k & = -\frac{1}{3}\frac{k}{a}\left(\Delta^{(\nu)}_k
+2\pi^{(\nu)}_k\right) \; ,
\end{align}
along with $v^{(c)}_k = 0$ and $v^{(b)}_k = 3 q^{(\g)}_k /4$. The latter
equation, together with Eqs.~(\ref{e:photons1}) and (\ref{e:baryons1}),
implies that $\dot{\Delta}^{(b)}_k = 3 \dot{\Delta}^{(\g)}_k /4$ so
that any entropy perturbation between the photons and baryons is conserved
while tight coupling holds.
The effects of baryon inertia appear in Eq.~(\ref{e:tightcouple1})
because of the tight coupling between the baryons and photons.

The constraint equations are found to be:
\begin{equation} \label{e:constraint1c}
\kappa^2 \rho q_k - \frac{2}{3} \left(\frac{k}{a} \right)^2 ({\cal Z}_k
  - \sigma_k ) - \frac{1}{6}
  \left(\frac{\tilde{\kappa}}{\kappa}\right)^4 (\kappa^4 \rho^2 q_k + 6 \rho
  {\cal Q}_k) = 0\;,
\end{equation}
and
\begin{equation} \label{e:constraint2c}
\begin{split}
& 2 \left(\frac{k}{a} \right)^3 \Phi_k -
\kappa^2 \rho \left(\frac{k}{a} \right) (\Delta_k - \pi_k ) + \kappa^2
\Theta \rho q_k \\
&= \frac{1}{16}\left(\frac{\tilde{\kappa}}{\kappa}\right)^4
\bigg[\kappa^4 \left(-\frac{8}{3} \Theta \rho^2 q_k + \frac{4}{3} \frac{k}{a}
\rho^2 \left[(3\g -2)
 \pi_k  +  2 \Delta_k \right] \right)  \\
& + 16  \left(\frac{k}{a} \right) \rho (\Upsilon_k -{\cal P}_k)  - 16
\Theta \rho {\cal Q}_k \bigg]\;.
\end{split}
\end{equation}
The propagation equation for the comoving expansion gradient in the
CDM frame is
\begin{equation} \label{e:propagation3c}
\begin{split}
&\dot{{\cal Z}}_k + \frac{1}{3} \Theta \mathcal{Z}_k + \frac{\kappa^2}{2}
\frac{a}{k} \left[2 (\rho^{(\g)}
\Delta^{(\g)}_k + \rho^{(\nu)} \Delta^{(\nu)}_k ) + \rho^{(b)}
\Delta^{(b)}_k + \rho^{(c)} \Delta^{(c)}_k  \right] \\
&= -\frac{1}{12}  \left(\frac{\tilde{\kappa}}{\kappa}\right)^4 \frac{a}{k}
\bigg\{\kappa^4 [(2\rho+3P)\rho\Delta_k + \rho(3\rho^{(\g)}\Delta^{(\g)}_k
+3\rho^{(\nu)}\Delta^{(\nu)}_k + (2+3c_s^2)\rho^{(b)}\Delta^{(b)}_k
+ 2 \rho^{(c)}\Delta^{(c)}_k)] \\
&\quad + 12 \rho \Upsilon_k \bigg\}\;.
\end{split}
\end{equation}
The variables $\Phi_k$ and $\sigma_k$ can be determined from the constraint
equations so their propagation equations are not independent of the above
set. The propagation equation for $\Phi_k$ is unchanged from
Eq.~$\eqref{e:propagation1a}$ since that equation was already written in
frame-invariant form. The propagation equation for the shear in the CDM frame
is
\begin{equation}
\frac{k}{a} \left(\dot{\sigma}_k + \frac{1}{3} \Theta \sigma_k \right) +
\left(\frac{k}{a}\right)^2 \Phi_k - \frac{\kappa^2}{2} \rho \pi_k =
\frac{1}{24}\left(\frac{\tilde{\kappa}}{\kappa}\right)^4 [-(3 \g - 2) \kappa^4
\rho^2 \pi_k + 12 \rho {\cal P}_k]\;.
\end{equation}

Finally we have the non-local evolution equations for ${\Upsilon}_k$ and
${\cal Q}_k$ which in the CDM frame become
\begin{equation} \label{e:nonlocal1c}
\dot{\Upsilon}_k = \frac{1}{3}(3 \g -4 )\Theta
\Upsilon_k - \frac{4}{3} \frac{{\cal U}}{\rho} \frac{k}{a}
{\cal Z}_k + \frac{k}{a} {\cal Q}_k \;,
\end{equation}
and
\begin{equation} \label{e:nonlocal2c}
\dot{{\cal Q}}_k -\frac{1}{3} (3 \g - 4) \Theta {\cal Q}_k + \frac{1}{3}
\frac{k}{a} (\Upsilon_k +2 \mathcal{P}_k)=
\frac{\kappa^4}{6} \g \rho \left[ \Theta q_k + \frac{k}{a}
(\pi_k - \Delta_k) \right]\;.
\end{equation}

\subsubsection{Solutions for the CDM Frame in the radiation-dominated era} \label{sec:radiation}

We now use the above equations to extract the
solutions of the scalar perturbation equations in the
radiation-dominated era, $\g=4/3$. To simplify matters, as well
as neglecting the contribution of the baryons and CDM to the background
dynamics, we shall only consider those modes for which $D_a \rho^{(b)}$
and $D_a \rho^{(c)}$ make a negligible contribution to the total
matter perturbation $D_a \rho$. This approximation allows us to write
the total matter perturbations in the form
\begin{equation}
(\rho^{(\g)} + \rho^{(\nu)})\Delta_k = \rho^{(\g)}\Delta^{(\g)}_k
+ \rho^{(\nu)} \Delta^{(\nu)}_k, \quad
(\rho^{(\g)} + \rho^{(\nu)}) q_k = \rho^{(\g)}q^{(\g)}_k
+ \rho^{(\nu)} q^{(\nu)}_k,
\label{eq:approx}
\end{equation}
and effectively removes the back-reaction of the baryon and CDM perturbations
on the perturbations of the spacetime geometry. We note that in making this
approximation we lose two modes corresponding to the baryon and CDM
isocurvature (density) modes of general relativity, in which the sub-dominant
matter components make significant contributions to the total fractional
density perturbation (which vanishes as $t\rightarrow 0$). However, for
our purposes the loss of generality is not that important, while the
simplifications resulting from decoupling the baryon and photon perturbations
are considerable. We also neglect moments of the neutrino distribution
function above the dipole (so there is no matter anisotropic stress). This
approximation is good for super-Hubble modes, but fails due to neutrino
free streaming on sub-Hubble scales.

We shall also assume that the non-local
energy density ${\cal U}$ vanishes in the background for all
energy regimes \cite{gordon}. Physically, vanishing ${\cal U}$
corresponds to the background bulk being conformally flat and strictly
anti-de Sitter. Note that ${\cal U}=0$ in the background need not imply that 
the fluctuations in the non-local
energy density are zero, i.e.\ $\Upsilon_a \neq 0$.

With the above conditions the
following set of equations are obtained: 
\begin{align}
\label{e:ae1}
\left(\frac{k}{a} \right)^2 (\dot{\Phi}_k + H \Phi_k ) + \frac{\kappa^2
\rho}{2} \left(\frac{k}{a} \right) \left( \frac{4}{3} \sigma_k - q_k \right) \left( 1 +
\frac{\rho}{\lambda} \right) &= \frac{3}{\kappa^2} \frac{\rho}{\lambda}
\left[ \left(\frac{k}{a} \right) {\cal Q}_k + 3 H {\cal P}_k -
\dot{{\cal P}}_k
\right]\;, \\
\label{e:ae2}
\left(\frac{k}{a} \right) (\dot{{\cal Z}}_k + H
{\cal Z}_k) + \kappa^2 \rho \left(1+ \frac{3 \rho}{\lambda} \right) \Delta_k &= -
\frac{6}{\kappa^2} \frac{\rho}{\lambda} \Upsilon_k \\
\label{e:ae3}
\left(\frac{k}{a} \right) (\dot{\sigma}_k + H \sigma_k) + \left(\frac{k}{a}
\right)^2 \Phi_k &= \frac{3}{\kappa^2} \frac{\rho}{\lambda}
{\cal P}_k\;,  \\
\label{e:ae4}
\dot{q}_k^{(\g)} + \frac{1}{3} \frac{k}{a} \Delta_k^{(\g)} &=0\;, \\
\label{e:ae5}
\dot{q}_k^{(\nu)} + \frac{1}{3} \frac{k}{a} \Delta_k^{(\nu)} &=0\;, \\
\label{e:ae6}
\dot{\Delta}_{k}^{(\g)} + \frac{k}{a} \left(\frac{4}{3} {\cal Z}_{k} -
q_{k}^{(\g)} \right) &=0\;, \\
\label{e:ae7}
\dot{\Delta}_{k}^{(\nu)} +\frac{k}{a} \left(\frac{4}{3} {\cal Z}_{k} -
q_{k}^{(\nu)} \right) &=0,
\end{align}
where recall $H=\Theta/3$. For the constraint equations we find
\begin{align}
\label{e:ae8}
3 \kappa^2 \left(1 + \frac{\rho}{\lambda} \right) \rho q_k - 2
\left(\frac{k}{a}\right)^2 ({\cal Z}_k - \sigma_k) &= -\frac{18}{\kappa^2}
\frac{\rho}{\lambda} {\cal Q}_k\;, \\
\label{e:ae9}
2 \left(\frac{k}{a}\right)^3 \Phi_k + \kappa^2 \rho \left(1 + \frac{\rho}{\lambda}
\right) \left[ 3 H q_k - \left(\frac{k}{a} \right) \Delta_k \right] &= \frac{6}{\kappa^2}
\frac{\rho}{\lambda} \left[\left( \frac{k}{a} \right) (\Upsilon_k -
{\cal P}_k) - 3H {\cal Q}_k \right]\;.
\end{align}
Finally the non-local evolution equations are found to be :
\begin{align}
\label{e:ae10}
\dot{\Upsilon}_k &=   \frac{k}{a} {\cal Q}_k\;, \\
\label{e:ae11}
9 \dot{{\cal Q}}_k + 3 \left( \frac{k}{a} \right) (\Upsilon_k + 2
{\cal P}_k) &= -2 \kappa^4 \rho \left(\frac{k}{a} \Delta_k - 3H
q_k \right)\;.
\end{align}
It is easy to show by propagating the constraint equations
that the above set of equations are consistent. 

By inspection, there is a solution of these equations with
\begin{align}
\Phi_k &= 0 , \\
{\cal Z}_k &= \left[3 \dot{H}
\left(\frac{a}{k}\right)^2 -1\right]\frac{A}{a} ,\\
\sigma_k &= - \frac{A}{a} , \\
q^{(\gamma)}_k &= - \frac{4}{3} \frac{A}{a} ,\\
q^{(\nu)}_k &= - \frac{4}{3} \frac{A}{a} ,\\
\Delta^{(\gamma)}_k &= -4 H \frac{A}{k} , \\
\Delta^{(\nu)}_k &= -4 H \frac{A}{k} , \\
\Upsilon_k &= 0 , \\
{\cal Q}_k &= 0 , \\
{\cal P}_k &=0,
\end{align} 
where $A$ is a constant.
This solution describes a radiation-dominated universe that is exactly FRW
except that the CDM has a peculiar velocity $\bar{v}_a^{(c)} = (A/a)
Q_a^{(k)}$ relative to the velocity of the FRW
fundamental observers. [This form for $\bar{v}_a^{(c)}$ clearly satisfies
Eq.~(\ref{eq:cdmvel}) with $A_a=0$.] Such a solution is possible since we have
neglected the gravitational effect of the CDM (and baryon) perturbations
in making the approximations in Eq.~(\ref{eq:approx}). The same solution arises
in general relativity~\cite{challinor2}. Including the back-reaction of the
CDM perturbations, we would find additional small peculiar velocities in the
dominant matter components which compensate the CDM flux. We shall not
consider this irregular CDM isocurvature velocity mode any further here.

Another pair of solutions are easily found by decoupling the photon/neutrino
entropy perturbations. Introducing the photon/neutrino entropy
perturbation (up to a constant) $\Delta_2$ and relative flux $q_2$:
\begin{equation}
\begin{split}
\Delta_2 &=  \Delta^{(\g)}_k - \Delta^{(\nu)}_k\;, \\
q_2 &= q^{(\g)}_k - q^{(\nu)}_k\;,
\end{split}
\end{equation}
the equations for $\Delta_2$ and $q_2$ decouple to give
\begin{align}
\dot{\Delta}_2 - \frac{k}{a} q_2 &= 0 \;,\\
\dot{q}_2 + \frac{1}{3} \frac{k}{a} \Delta_2 &=0 \;.
\end{align}
Switching to conformal time ($d \tau = dt /a$) we can solve for $\Delta_2$ and
$q_2$ to find
\begin{align} 
\label{e:neutrinoiso1}
q_2 (\tau) &= B \cos \left(\frac{k \tau}{3} \right) + C \sin \left(\frac{k
\tau}{3} \right)\;, \\
\label{e:neutrinoiso2} 
\Delta_2 (\tau) &= B \sin \left(\frac{k \tau}{3} \right)  - C \cos \left(\frac{k
\tau}{3} \right) \;.
\end{align}
The constants $B$ and $C$ label the neutrino velocity and density isocurvature
modes respectively~\cite{bucher,challinor2}, in which the neutrinos and photons
initially have mutually compensating peculiar velocities and density
perturbations. The perfect decoupling of these
isocurvature modes is a consequence of our neglecting anisotropic stresses
(and higher moments of the distribution functions) and baryon inertia.

Having decoupled the entropy perturbations, we write the remaining equations
in terms of the total variables $\Delta_k$ and $q_k$. The propagation equations
for the non-local variables $\Upsilon_k$ and ${\cal Q}_k$ are redundant since
these variables are determined by the constraint equations~(\ref{e:ae8}) and
(\ref{e:ae9}):
\begin{align} \label{eq:cons1}
\frac{6}{\kappa^2} \frac{\rho}{\lambda} \Upsilon_k &= 2
\left(\frac{k}{a}\right)^2 \Phi_k + 2H \left(\frac{k}{a} \right) ({\cal Z}_k -
\sigma_k) - \kappa^2 \rho \left( 1+ \frac{\rho}{\lambda} \right) \Delta_k 
+ \frac{6}{\kappa^2} \frac{\rho}{\lambda} {\cal P}_k, \\
\frac{3}{\kappa^2} \frac{\rho}{\lambda} {\cal Q}_k &= \frac{1}{3}
\left(\frac{k}{a} \right)^2 ({\cal Z}_k - \sigma_k) - \frac{\kappa^2 \rho}{2}
\left( 1+ \frac{\rho}{\lambda} \right) q_k. \label{eq:cons2} 
\end{align}
Substituting these expressions in the right-hand sides of Eqs.~(\ref{e:ae1})
and (\ref{e:ae2}) we find 
\begin{align}
\label{e:be1}
\left(\frac{k}{a}\right)^2 \left(\dot{\Phi}_k + H \Phi_k \right) + \frac{2\kappa^2
\rho}{3} \left(\frac{k}{a} \right) \left(1+\frac{\rho}{\lambda}\right)
\sigma_k - \frac{1}{3} \left(\frac{k}{a}\right)^3
({\cal Z}_k - \sigma_k) &= \frac{3}{\kappa^2} \frac{\rho}{\lambda} (3H
{\cal P}_k - \dot{\cal P}_k) \;,\\
\label{e:be2}
\left(\frac{k}{a} \right) \l(\dot{{\cal Z}}_k + H {\cal Z}_k \r) 
+ \kappa^2 \rho \left(\frac{2 \rho}{\lambda} \right) \Delta_k +
2\left(\frac{k}{a}\right)^2 \Phi_k + 2H \left(\frac{k}{a}\right)({\cal Z}_k -
\sigma_k)  &= - \frac{6 \rho}{\kappa^2 \lambda}
{\cal P}_k \;,\\
\label{e:be3}
\left(\frac{k}{a} \right) (\dot{\sigma}_k + H \sigma_k) + \left(\frac{k}{a}
\right)^2 \Phi_k &= \frac{3}{\kappa^2} \frac{\rho}{\lambda}
{\cal P}_k\;,  \\
\label{e:be4}
\dot{\Delta}_k + \frac{k}{a} \left(\frac{4}{3} {\cal Z}_k - q_k \right) &=0 \;,
\\
\label{e:be5}
\dot{q}_k + \frac{1}{3} \frac{k}{a} \Delta_k &=0 \;.
\end{align}
These equations describe the evolution of the intrinsic perturbations
to the brane. The usual general relativistic constraint equations are now
replaced by the constraints~(\ref{eq:cons1}) and (\ref{eq:cons2}) which
determine two of the non-local variables. The lack of a propagation equation
for ${\cal P}_k$ reflects the incompleteness of the 1+3 dimensional
description of braneworld dynamics.

In the following it will prove convenient to adopt the dimensionless
independent variable
\begin{equation} \label{e:trans1}
x= \frac{k}{Ha}\;,
\end{equation}
which is (to within a factor of $2\pi$) the ratio of the Hubble length
to the wavelength of the perturbations. Using the (modified) Friedmann
equations for the background in radiation domination, and with
$\mathcal{U}=0$, we find that
\begin{equation}
\frac{dx}{dt} = \frac{k}{a}\left(\frac{2+3\rho/\lambda}{2+\rho/\lambda}\right).
\end{equation}
The relative importance of the local (quadratic) braneworld corrections
to the Einstein equation depends on the dimensionless ratio $\rho/\lambda$.
In the low-energy limit, $\rho\ll \lambda$, the quadratic local corrections can
be neglected although the non-local corrections $\mathcal{E}_{ab}$ may
still be important. In the opposite (high-energy) limit the quadratic
corrections dominate over the terms that are linear in the energy-momentum
tensor. We now consider these two limits separately.

\subsubsection{Low-energy regime}
\label{sec:lowenergy}

In the low-energy regime we have $dx/dt \approx k/a$ and $x \approx k \tau$.
The total energy density $\rho$ is proportional to $x^{-4}$. Denoting
derivatives with respect to $x$ with a prime, using $\rho \ll \lambda$,
and assuming that we can neglect the term involving $(\rho/\lambda)\Delta_k$
in Eq.~(\ref{e:be2}) compared to the other terms, we find
\begin{align}
\label{e:le1}
3x^2 \Phi_k'+ 3x \Phi_k + (6+x^2) \sigma_k - x^2 {\cal Z}_k &=
\frac{27}{\kappa^4 \lambda} (3 {\cal P}_k - x {\cal P}_k') \\
\label{e:le2}
x^2 {\cal Z}_k' + 3 x {\cal Z}_k + 2x^2 \Phi_k - 2x \sigma_k &=
- \frac{18}{\kappa^4 \lambda} {\cal P}_k \\
\label{e:le3}
x^2 \sigma_k' + x\sigma_k + x^2 \Phi_k &=
\frac{9}{\kappa^4 \lambda} {\cal P}_k \\
\label{e:le4}
\Delta_k' + \frac{4}{3} {\cal Z}_k - q_k &=0 \\
\label{e:le5}
q_k' + \frac{1}{3} \Delta_k &=0 
\end{align}
Combining these equations we find an inhomogeneous, second-order equation
for $\Phi_{k}$:
\begin{equation}
3x \Phi_k'' + 12 \Phi_k' + x \Phi_k = F_k(x) \;,
\label{eq:secondPhi}
\end{equation}
where
\begin{equation}
F_k(x) \equiv -\frac{27}{\kappa^4 \lambda}
\left[ {\cal P}_k'' - \frac{{\cal P}_k'}
{x} + \left(\frac{2}{x^3} - \frac{3}{x^2} + \frac{1}{x} \right) {\cal P}_k \right].
\end{equation}
In general relativity the same second-order equation holds for $\Phi_k$ but
with $F_k(x)=0$.

The presence of terms involving the non-local anisotropic stress on the
right-hand side of Eq.~(\ref{eq:secondPhi}) ensure that $\Phi_k$ cannot be
evolved on the brane alone. The resolution of this problem will require
careful analysis of the bulk dynamics in five dimensions. In this thesis,
our aims are less ambitious; we shall solve Eq.~(\ref{eq:secondPhi}) with
${\cal P}_k=0$.
Although we certainly do not expect ${\cal P}_{ab}=0$\footnote{We
have not investigated the consistency of the condition ${\cal P}_{ab}=0$
with the five-dimensional bulk dynamics in the presence of a perturbed brane.},
the solutions of the homogeneous equation may still prove a useful starting
point for a more complete analysis. For example, they allow one to construct
Green's functions for Eq.~(\ref{eq:secondPhi}) which could be used to assess
the impact of specific ansatze for ${\cal P}_{ab}$~\cite{barrow}.

With ${\cal P}_k=0$ we can solve Eqs.~(\ref{e:le1})--(\ref{e:le5})
analytically to find
\begin{align}
\Phi_k &= \frac{c_1}{x^3} \left[3 \sin \left(\frac{x}{\sqrt{3}} \right) -
x\sqrt{3} \cos \left(\frac{x}{\sqrt{3}} \right) \right] 
+ \frac{c_2}{x^3} \left[3 \cos  \left(\frac{x}{\sqrt{3}} \right)+ x \sqrt{3}
\sin \left(\frac{x}{\sqrt{3}} \right) \right] , \\
\sigma_k &= \frac{3}{x^2} \left[c_2 \cos  \left(\frac{x}{\sqrt{3}} \right)
+ c_1 \sin  \left(\frac{x}{\sqrt{3}} \right) \right] + \frac{c_3}{x}, \\
{\cal Z}_k &= \frac{c_3 (6+x^2)}{x^3} + \frac{6 \sqrt{3}}{x^3} \left[c_1
\cos\left(\frac{x}{\sqrt{3}} \right)  - c_2 \sin \left(\frac{x}{\sqrt{3}}
\right)\right] +\frac{6}{x^2} \left[c_2 \cos \left(\frac{x}{\sqrt{3}} \right)
\quad + c_1 \sin \left(\frac{x}{\sqrt{3}} \right) \right], \\
\Delta_k &= c_4 \cos\left(\frac{x}{\sqrt{3}}\right) + c_5 \sin
\left(\frac{x}{\sqrt{3}}\right) + \frac{4 c_3}{x^2} + \frac{4}{x}
\left[c_2 \cos\left(\frac{x}{\sqrt{3}} \right) + c_1\sin \left(\frac{x}
{\sqrt{3}} \right) \right] \notag \\ 
& \quad  + \left(\frac{4 \sqrt{3}}{x^2}-\frac{2}{\sqrt{3}}\right)
\left[c_1 \cos \left(\frac{x}{\sqrt{3}} \right)
- c_2 \sin\left(\frac{x}{\sqrt{3}} \right)  \right] , \\
q_k &= \frac{c_5}{\sqrt{3}} \cos \left(\frac{x}{\sqrt{3}} \right) -
\frac{c_4}{\sqrt{3}} \sin\left(\frac{x}{\sqrt{3}} \right)    + \frac{4
c_3}{3x} + \frac{4x}{\sqrt{3}} \left[c_1 \cos \left(\frac{x}{\sqrt{3}}\right)
- c_2 \sin \left(\frac{x}{\sqrt{3}} \right) \right] \notag \\
& \quad + \frac{2}{3}\left[c_2 \cos \left(
\frac{x}{\sqrt{3}}\right) + c_1 \sin \left(\frac{x}{\sqrt{3}}\right) \right].
\end{align}
The mode labelled by $c_3$ is the CDM velocity isocurvature mode discussed
earlier. The modes labelled by $c_1$ and $c_2$ are the same as in general
relativity; they describe the adiabatic growing and decaying solutions
respectively. However, in the low-energy limit we also find two additional
isocurvature modes ($c_4$ and $c_5$) that are not present in general
relativity. These
arise from the two additional degrees of freedom $\Upsilon_k$ and ${\cal Q}_k$
present in the braneworld model (with $P_k=0$). The mode $c_4$ initially
has non-zero but compensating gradients in the total matter and
non-local densities, and $c_5$ initially has compensated energy fluxes.
Formally these isocurvature solutions violate the assumption that the
term involving $(\rho/\lambda)\Delta_k$ be negligible compared to the
other terms in Eq.~(\ref{e:be2}) since all other terms vanish. In practice,
there will be some gravitational back-reaction onto the other gauge-invariant
variables controlled by the dimensionless quantity $\rho/\lambda$, but the
general character of these isocurvature modes will be preserved for
$\rho/\lambda\ll 1$.

\subsubsection{High-energy regime}

We now turn to the high-energy regime, where the quadratic terms in
the stress-energy tensor dominate the (local) linear terms. In this limit
the scale factor $a \propto t^{1/4}$. The modification to the expansion rate
leads to an increase in the amplitude of scalar and tensor fluctuations 
produced during high-energy inflation~\cite{barrow}. With
${\cal U}=0$ in the background, and $\rho \gg \lambda$, the Hubble parameter
is approximately
\begin{equation}
H^2 \approx \frac{1}{36} \tilde{\kappa}^4 \rho^2 = \frac{\kappa^2
\rho^2}{6 \lambda},
\end{equation}
and $dx/dt \approx 3 k/a$. In terms of conformal time $\tau$, $x \approx
3 k \tau$. The total energy density, $\rho$ is proportional to $x^{(-4/3)}$.
  
\subsubsection{Power series solutions for the high-energy regime}
\label{power}

It is convenient to rescale the non-local variables by the dimensionless
quantity $\kappa^4 \rho$. Thus we define
\begin{align}
\label{e:trans2}
\bar{\Upsilon}_k &\equiv \frac{\Upsilon_k}{\kappa^4 \rho}\;, \\
\label{e:trans3}
\bar{{\cal Q}}_k &\equiv \frac{{\cal Q}_k}{\kappa^4 \rho}\;, \\
\label{e:trans4}
\bar{{\cal P}}_k &\equiv \frac{{\cal P}_k}{\kappa^4 \rho}\;.
\end{align}
The fractional total (effective) density perturbation and energy flux
can be written in terms of the barred variables
[e.g.\ $\bar{\Upsilon}_a \equiv \Upsilon_a/ (\kappa^4
\rho)$]\footnote{A general and useful identity for these variables is employed in
deriving the equations for the high energy regime:
\begin{equation*}
\frac{\dot{\bar{\Upsilon}}_k}{\kappa^4 \rho} = \dot{\bar{\Upsilon}}_k
- 3 \g H \bar{\Upsilon}_k  
\end{equation*}
} in the high-energy limit as
\begin{align}
\frac{a D_a \rho^{\text{tot}}}{\rho^{\text{tot}}} &\approx 2(\Delta_a + 6
\bar{\Upsilon}_a), \\
q_a^{\text{tot}} &\approx \frac{2\rho^{\text{tot}}}{\rho} (q_a + 6
\bar{\cal Q}_a ).
\end{align}
Making the high-energy approximation $\rho \gg \lambda$ in
Eqs.~$\eqref{e:be1}$--$\eqref{e:be5}$, we obtain  
\begin{align}
\label{e:he2a}
9x^2 \Phi^{'}_k + 3x \Phi_k + (12+x^2) \sigma_k - x^2 {\cal Z}_k  &=
54 \left[ \frac{7 \bar{\cal P}_k}{x} - 3 \bar{\cal P}_k' \right]\;,  \\
\label{e:he2b}
3x^2 {\cal Z}^{'}_k + 3x {\cal Z}_k - 2x \sigma_k + 2x^2
\Phi_k + 12 \Delta_k &= -36 \bar{\cal P}_k\;, \\
\label{e:he2c}
3x \sigma^{'}_k + \sigma_k + x \Phi_k &= 18 \frac{\bar{{\cal P}}_k}{x}\;,
\\
\label{e:he2d}
\Delta_k^{'} - \frac{1}{3} q_k  + \frac{4}{9} {\cal Z}_k &=0\;, \\
\label{e:he2e}
q_k^{'} + \frac{1}{9} \Delta_k &=0\;.
\end{align}
The non-local quantities $\bar{\Upsilon}_k$ and $\bar{{\cal Q}}_k$ are
determined by the constraints
\begin{align}
\bar{\Upsilon}_k &= \frac{1}{18}x^2 \Phi_k + \frac{1}{18} x
({\cal Z}_k-\sigma_k) - \frac{1}{6} \Delta_k + \bar{{\cal P}}_k, \\
\bar{{\cal Q}}_k &= \frac{1}{54} x^2 ({\cal Z}_k - \sigma_k) - \frac{1}{6}
q_k.
\end{align}

We can manipulate Eqs.~$\eqref{e:he2a}$--$\eqref{e:he2e}$ to obtain
a fourth-order equation for the gravitational potential $\Phi_k$:
\begin{equation} \label{e:order4de}
729 x^2 \frac{\p^4 \Phi_k}{\p x^4} +3888 x \frac{\p^3 \Phi_k}{\p x^3} +
(1782+54x^2) \frac{\p^2\Phi_k}{\p x^2}
+144x \frac{\p\Phi_k}{\p x} + (90+x^2) \Phi_k = F_k(x)\;,
\end{equation}
where
\begin{equation}
\begin{split}
\label{e:p-de1}
F_k(x) &= -\frac{54}{x^4}\bigg( 243 x^4 \frac{\partial^4 \bar{\cal P}_k}{
\partial x^4} - 810 x^3 \frac{\partial^3 \bar{\cal P}_k}{\partial x^3}
+18x^2(135+2x^2) \frac{\partial^2 \bar{\cal P}_k}{\partial x^2}
-30 x(162+x^2) \frac{\partial \bar{\cal P}_k}{\partial x} \\
&\quad + [x^4 + 30(162+x^2)] \bar{\cal P}_k\bigg)\; .
\end{split}
\end{equation}
Since we do not have an evolution equation for $\bar{\cal P}_k$ we
adopt the strategy taken in the low-energy limit and look for solutions
of the homogeneous equations ($\bar{{\cal P}}_k=0$).
In principle one can use these solutions to construct formal solutions
of the inhomogeneous equations with Green's method.

To solve Eq.~(\ref{e:order4de}) with $\bar{{\cal P}}_k=0$ we
construct a Frobenius (or power) series solution for $\Phi_k(x)$:
\begin{equation} \label{e:frobenius}
\Phi_k(x) = x^m \sum^{\infty}_{n=0} a_n x^n\;,
\end{equation}
where $a_0 \neq 0$. The indicial equation for $m$ is
\begin{equation} \label{e:indicial1}
m (m-1) (3m+5) (3m-4)=0\;.
\end{equation}
For each value of $m$ we substitute into Eq.~(\ref{e:order4de}) and solve
the resulting recursion relations for the $\{a_n\}$. We then obtain the
other gauge-invariant variables by direct integration. The original set of
equations~(\ref{e:he2a})--(\ref{e:he2e}) has five degrees of freedom, so we
expect one additional solution with $\Phi_k=0$. This solution is the CDM
isocurvature solution discussed earlier, and has a finite series expansion:
\begin{align} 
\Phi_k &= 0\;, \\
\sigma_k  &= D x^{-\frac{1}{3}}\;, \\
{\cal Z}_k &= D x^{-\frac{7}{3}} (12 + x^2)\;, \\
\Delta_k &= 4 D x^{-\frac{4}{3}}\;, \\
q_k &= \frac{4}{3} D x^{-\frac{1}{3}}\;,
\end{align}
where $D$ is a constant. The non-local variables vanish.

The first two terms of the mode with $m=0$ are
\begin{align}
\Phi_k &= a_1 \left(1 - \frac{5}{198} x^2 \right)\;, \\
\sigma_k &= a_1 \left(-\frac{1}{4} x + \frac{1}{396} x^3 \right)\;,    \\
{\cal Z}_k &= a_1 \left(- \frac{3}{4} x + \frac{5}{864} x^3 \right)\;, \\
\Delta_k &= a_1 \left(\frac{1}{6} x^2 - \frac{1}{864} x^4 \right)\;,  \\
q_k &=  a_1 \left(-\frac{1}{162} x^3 + \frac{1}{38880} x^5 \right)\;, \\
\bar{\Upsilon}_k &= a_1 \left(-\frac{1}{972} x^4
+ \frac{1}{249480} x^6 \right) = \frac{a_1}{972} \l(-x^4+
\frac{3}{770} x^6 \r) \;, \\
\bar{{\cal Q}}_k &= a_1 \left(-\frac{2}{243} x^3 + \frac{1}{17820} x^5
\right) = \frac{2 a_1}{243} \l(-x^3 + \frac{3}{440} x^5 \r)\; ,
\end{align}
where $a_1$ is a constant.
The form of this solution is similar to the adiabatic growing mode of
general relativity.

The mode corresponding to $m=1$ is
\begin{align}
\Phi_k &= a_2 \left(x - \frac{13}{1890} x^3 \right)\;, \\
\sigma_k &= a_2 \left(-\frac{1}{7} x^2 + \frac{1}{1890} x^4 \right)\;, \\
{\cal Z}_k &= a_2 \left(\frac{72}{7} - \frac{12}{35} x^2 \right)\;,  \\
\Delta_k  &= a_2 \left(-\frac{18}{7} x + \frac{1}{15} x^3 \right)\;, \\
q_k  &= a_2  \left(6 + \frac{1}{7} x^2 \right)\;, \\
\bar{\Upsilon}_k &= a_2 \left(x + \frac{1}{30} x^3 \right)\;, \\
\bar{{\cal Q}}_k &= a_2 \left(-1 + \frac{1}{6} x^2  \right)\;,
\end{align}
with $a_2$ a constant. This mode has no analogue in general
relativity, except that $\Phi_k$ is a growing mode. As $t
\rightarrow 0$ there are non-zero but compensating contributions to
the effective peculiar velocity $q_a^{\text{tot}}  
/\rho^{\text{tot}}$ from the matter and the non-local energy
fluxes. The contributions of these components to the fractional total
density perturbation $a D_a \rho^{\text{tot}}/\rho^{\text{tot}}$
vanish as $t \rightarrow 0$.

The mode corresponding to $m=-\frac{5}{3}$ is singular as $t \rightarrow 0$
(it is a decaying mode with $a_3$ a constant):
\begin{align}
\Phi_k &= a_3 x^{-\frac{5}{3}} \left( 1 - \frac{5}{18} x^2 \right)\;, \\
\sigma_k &= a_3 x^{-\frac{2}{3}} \left( 1 + \frac{1}{18} x^2 \right)\;, \\
{\cal Z}_k &= a_3 \left(\frac{14}{99} x^{\frac{4}{3}} - \frac{1217}{1590435}
x^{\frac{10}{3}} \right)\;, \\
\Delta_k  &= a_3 \left(-\frac{8}{297} x^{\frac{7}{3}} + \frac{64}{433755}
x^{\frac{13}{3}} \right)\;, \\
q_k  &= a_3 \left(\frac{4}{4455} x^{\frac{10}{3}} - \frac{4}{1301265}
x^{\frac{16}{3}} \right)\; ,\\
\bar{\Upsilon}_k &= a_3 \left(-\frac{1}{162} x^{\frac{7}{3}} +
\frac{7}{43740} x^{\frac{13}{3}} \right) = \frac{a_3}{162} x^{\frac{7}{3}}
\l(-x + \frac{7}{270} x^2 \r) \;,  \\
\label{e:Qiso}
\bar{{\cal Q}}_k &= a_3 \left(-\frac{1}{54} x^{\frac{4}{3}} +
\frac{7}{4860} x^{\frac{10}{3}} \right) = \frac{a_3}{54}
x^{\frac{4}{3}} \l(- 1 + \frac{7}{90} x^2  \r) \;. 
\end{align}
A similar mode is found in general relativity but there the decay of $\Phi_k$
is more rapid ($\Phi_k \propto x^{-3}$) on large scales.  The solution
$\eqref{e:Qiso}$ describes an isocurvature velocity mode~\footnote{We note
that we have identified this mode incorrectly in \cite{leong1} since the
variable ${\cal Q}$ is rescaled by $x^{4/3}$.} where the early
time matter and non-local (Weyl) components have equal but opposite peculiar
velocities in the CDM frame. The existence of such isocurvature modes
was predicted in \cite{gordon} and \cite{langlois} for large-scale density
perturbations.  

Finally, for $m=\frac{4}{3}$, with $a_4$ a constant, we have
\begin{align}
\Phi_k &= a_4 x^{\frac{4}{3}} \left(1 - \frac{17}{3150} x^2 \right)\;, \\
\sigma_k &= a_4 x^{\frac{4}{3}} \left(- \frac{1}{8} x + \frac{17}{44100}
x^3 \right)\;, \\
{\cal Z}_k &= a_4 x^{\frac{1}{3}} \left(\frac{27}{2} - \frac{117}{392}
x^2 \right) \;, \\
\Delta_k  &= a_4 x^{\frac{4}{3}} \left(-\frac{9}{2} + \frac{3}{49} x^2
\right)\;, \\
q_k &= a_4 x^{\frac{4}{3}} \left(\frac{3}{14} x - \frac{1}{637} x^3
\right)\; ,\\
\label{e:Yiso}
\bar{\Upsilon} (x) &= a_4 x^{\frac{4}{3}} \left(\frac{3}{2} + \frac{1}{28}
x^2 \right) = \frac{3 a_4}{2} x^{\frac{4}{3}} \l( 1 + \frac{1}{42} x^2
\r) \;, \\
\bar{{\cal Q}} (x) &= a_4 x^{\frac{4}{3}} \left(\frac{3}{14} x -
\frac{29}{9828} x^3 \right) = \frac{3 a_4}{14} x^{\frac{4}{3}}  \l(x -
\frac{29}{2106} x^3 \r) \;.
\end{align}
In this mode the universe asymptotes to an FRW (brane) model in the past
as $t \rightarrow 0$. Note that this requires careful cancellation
between $a D_a \rho^{\text{tot}} / \rho^{\text{tot}}$ and $q_a^{\text{tot}}$
to avoid a singularity in the gravitational potential $\Phi_k$ (which
would diverge as $x^{-2/3}$ without such cancellation). Like the velocity
isocurvature mode ($m=1$) discussed above, this mode has no analogue in
general relativity. We identify the solution $\eqref{e:Yiso}$ to be
the non-local isocurvature density mode. The pair of non-local velocity and
density isocurvature modes $\eqref{e:Qiso}$ and $\eqref{e:Yiso}$ is
analogous to the neutrino density and velocity isocurvature modes (see
Eqs. $\eqref{e:neutrinoiso1}$ and $\eqref{e:neutrinoiso2}$). 

\subsection{The Energy Frame}
\label{sec:energyframe}

In this section, we present a complete set of evolution equations for the
total matter variables in the matter energy frame, $q_a = 0$. Note that
the four-velocity of the energy frame is not necessarily a timelike eigenvector
of the Einstein tensor in the presence of the non-local braneworld corrections
to the effective stress-energy tensor. We assume that the matter is
radiation dominated, the non-local energy density vanishes in the
background, and we ignore local anisotropic stresses. We also assume
that the baryons and CDM make a negligible contribution to the
fractional gradient in the total matter energy density and to the
energy flux, thus excluding the CDM and baryon isocurvature 
modes. We also give the evolution equations for the non-local density
gradient and energy flux in the matter energy frame. 

Denoting the variables in the energy frame by an tilde on the
quantity, for example, $\tilde{\Delta}_a$, the relevant equations for
scalar perturbations are 
\begin{align}
\dot{\tilde{\Delta}}_a &=\frac{1}{3}\Theta\tilde{\Delta}_a-\frac{4}{3}
\tilde{{\cal Z}}_a\;, \\
{\dot{\tilde{\cal Z}}}_a &= -\frac{2}{3}\Theta \tilde{\cal Z}_a -\frac{1}{4}
D^2 \tilde{\Delta}_a
- \frac{6 \rho}{\kappa^2 \lambda} \tilde{\Upsilon}_a
-\frac{1}{2} \kappa^2 \rho \tilde{\Delta}_a
\left(1+\frac{5\rho}{\lambda}\right)\;,\\
\dot{\tilde{\Upsilon}}_a &= -\frac{a}{\rho} D^2 {\tilde{\cal Q}}_a \;,\\
{\dot{\tilde{\cal Q}}}_a &= - \frac{4}{3}\Theta \tilde{\cal Q}_a - \frac{\rho}{3a}
\tilde{\Upsilon}_a - \frac{2 \kappa^4 \rho^2}{9 a}\tilde{\Delta}_a -
D^b \tilde{\cal P}_{ab}.
\end{align}

Solutions of these equations are related to those in the CDM frame
(\ref{sec:radiation}) by linearising the frame transformations given in
\cite{maartens4}. If the CDM projected velocity is $\tilde{v}^{(c)}_a$
in the energy frame, the variables in the CDM frame are given by
\begin{align}
\label{e:tranen1}
\Delta_a &= \tilde{\Delta}_a - \frac{4}{3} a \Theta \tilde{v}^{(c)}_a\; ,\\
\label{e:tranen2}
{\cal Z}_a &= \tilde{{\cal Z}}_a + \frac{1}{a} D_a D^b \tilde{v}^{(c)}_b
- \frac{2\kappa^2 \rho}{a}\left(1+\frac{\rho}{\lambda}\right)\tilde{v}^{(c)}_a
\;,\\
\label{e:tranen3}
\Upsilon_a &= \tilde{\Upsilon}_a \; ,\\
\label{e:tranen4}
{\cal Q}_a &= \tilde{{\cal Q}}_a \; , \\
\label{e:tranen5}
q_a &= - \frac{4}{3}\rho \tilde{v}_a^{(c)}\; ,
\end{align}
where we have used ${\cal U}=0$ in the background. Note that the
quantities $\Upsilon_a$ and ${\cal Q}_a$ are frame-invariant, hence we
can assume the same form for these quantities for the subsequent
calculations in the energy frame like the CDM frame.

The CDM peculiar velocity evolves in the energy frame according to
\begin{equation}
\dot{\tilde{v}}_a^{(c)} = - \frac{1}{3}\Theta \tilde{v}^{(c)}_a + \frac{1}{4a}
\tilde{\Delta}_a.
\end{equation}
%

\subsubsection{Energy frame solutions in the High-energy Regime}

In the high energy limit, $\rho/\lambda \gg 1$, using the variable $x$
defined in Eq. $\eqref{e:trans1}$, we obtain the following four equations:
\begin{align}
\label{e:energyhe1}
9x \tilde{\Delta}_k^{'} &= 3 \tilde{\Delta}_k  - 4x \tilde{\cal Z}_k \;, \\ 
\label{e:energyhe2}
3x^2 \tilde{\cal Z}_k^{'} &= - x \tilde{\cal Z}_k + \l(\frac{x^2}{4}-15 \r)
\tilde{\Delta}_k - 36 \bar{\Upsilon}_k \;, \\
\label{e:energyhe3}
3x \bar{\Upsilon}_k^{'} &=  4 \bar{\Upsilon}_k + x
\bar{\cal Q}_k  \;, \\
\label{e:energyhe4}
3x \bar{\cal Q}_k^{'} &= 4 \bar{\cal Q}_k - \frac{1}{3}  
\l(\bar{\Upsilon}_k + \frac{2}{3} \tilde{\Delta}_k - 2 \bar{\cal P}_k \r) \;.
\end{align} 

Setting $\bar{\cal P}_k=0$, we decouple the above equations to find a 4th
order differential equation, 
\begin{equation} \label{e:delta4eqn}
\begin{split}
&\frac{9}{16}x^4 \frac{\p^4  \tilde{\Delta}_k}{\p x^4} + \frac{3}{4} x^3
\frac{\p^3  \tilde{\Delta}_k}{\p x^3} + \l(\frac{x^4}{24} - \frac{5x^2}{4} \r)
\ppd{ \tilde{\Delta}_k}{x} + \l(3x + \frac{x^3}{36}\r) 
\pp{ \tilde{\Delta}_k}{x} \\
&+ \l(-\frac{7}{2} + \frac{x^2}{36} + \frac{x}{1296}  \r)  \tilde{\Delta}_k
= 0 \;.
\end{split}
\end{equation}
Using eq.~$\eqref{e:frobenius}$, we obtain the indical equation:
\begin{equation} \label{e:indical2}
(m-2)(3m-4)(3m-7)(1+m)=0 \;.
\end{equation}
We proceed to substitute each value of $m$ back in $\eqref{e:delta4eqn}$
and solve the recurrence relations for $\{ b_n \}$. The original set
of equations $\eqref{e:energyhe1}$--$\eqref{e:energyhe4}$ has four
degrees of freedom, and we note that the peculiar CDM isocurvature
mode vanishes in the energy frame. In addition, we compute the spatial
gradient of the 3-curvature, $\eta_k$ from eq. $\eqref{e:3curvature}$. 

The mode corresponding to $m=2$ are
\begin{align}
\label{e:ensol1}
 \tilde{\Delta}_k &= b_1 \l(x^2 - \frac{1}{150} x^4 \r) \;, \\
\label{e:ensol2}
\tilde{\cal Z}_k &= b_1 \l(-\frac{15}{4} x + \frac{11}{200} x^3 \r) \;, \\
\label{e:ensol3}
\bar{\Upsilon}_k &= \frac{b_1}{180} \l(- x^4 + \frac{3}{770} x^6
\r) \;, \\
\label{e:ensol4}
{\cal \bar{Q}}_k &= \frac{2 b_1}{45} \l(- x^3 + \frac{3}{440} x^5
\r) \;, \\
\label{e:etak1}
\eta_k &= b_1 \l( \frac{27}{4} - \frac{7}{40} x^2 \r)  \;.
\end{align}

The mode corresponding to $m=\frac{4}{3}$ are
\begin{align}
\label{e:ensol5}
\tilde{\Delta}_k &= b_2 x^{\frac{4}{3}} \l(1 - \frac{1}{78} x^2 \r)
\;, \\
\label{e:ensol6}
\tilde{\cal Z}_k &= b_2 \l(-\frac{9}{4} x^{\frac{1}{3}} + \frac{9}{104}
x^{\frac{7}{3}} \r) \;, \\
\label{e:ensol7}
\bar{\Upsilon}_k &= - \frac{7 b_2}{24} x^{\frac{4}{3}} \l(1  +
\frac{1}{42} x^2 \r) \;, \\
\label{e:ensol8}
{\cal \bar{Q}}_k &= -\frac{b_2}{24} x^{\frac{4}{3}} \l(x  -
\frac{29}{2106} x^3 \r) \; \\
\label{e:etak2}
\eta_k &= b_2 x^{\frac{4}{3}} \l( -\frac{1}{4} + \frac{19}{14040} x^2 \r) \;.
\end{align}

The mode corresponding to $m=\frac{7}{3}$ are
\begin{align}
\label{e:ensol9}
\tilde{\Delta}_k &= b_3 x^{\frac{7}{3}} \l(1 - \frac{1}{189} x^2\r) \;, \\
\label{e:ensol10}
\tilde{\cal Z}_k &= b_3 \l(-\frac{9}{2} x^{\frac{4}{3}} + \frac{1}{21}
x^{\frac{10}{3}} \r) \;,\\
\label{e:ensol11}
\bar{\Upsilon}_k &= -\frac{5 b_3}{24} \l(-x^{\frac{7}{3}} +
\frac{7}{270} x^{\frac{13}{3}} \r) \;, \\
\label{e:ensol12}
{\cal \bar{Q}}_k &= -\frac{5 b_3}{8}  \l(-x^{\frac{4}{3}} +
\frac{7}{90} x^{\frac{10}{3}} \r) \;, \\
\label{e:etak3}
\eta_k &= b_3 x^{\frac{1}{3}} \l( \frac{45}{4} - \frac{9}{56} x^2 \r) \;.
\end{align}

The mode corresponding to $m=-1$ are
\begin{align}
\label{e:ensol13}
 \tilde{\Delta}_k &= b_4 x^{-1} \l(1 + \frac{1}{6} x^2 \r) \;, \\
\label{e:ensol14}
\tilde{\cal Z}_k &= b_4 \l(\frac{3}{x^2} - \frac{1}{4} \r) \;, \\
\label{e:ensol15}
\bar{\Upsilon}_k &= -\frac{b_4}{18} \l(x + \frac{1}{30} x^3 \r) \;, \\
\label{e:ensol16}
{\cal \bar{Q}}_k &= -\frac{b_4}{18}  \l(-1 + \frac{1}{6} x^2 \r)
\;, \\
\label{e:etak4}
\eta_k &= b_4 \l(- \frac{1}{4x} - \frac{5}{72} x^2 \r) \;.
\end{align}
with $b_1$,$b_2$, $b_3$ and $b_4$ being constants.

From the solutions (Eqs. $\eqref{e:ensol1}$--$\eqref{e:ensol16}$)
in the high energy frame, we observe that powers and coefficients of
the solutions for $\bar{\Upsilon}$ and $\bar{\cal Q}$ match the ones
in the CDM frame with constants of integration (up to some
pre-factor). The solutions are checked by transforming them to the CDM
using Eqs. $\eqref{e:tranen1}$--$\eqref{e:tranen5}$. Eqs. $\eqref{e:ensol7}$
and $\eqref{e:ensol12}$ correspond to the pair of non-local
isocurvature density and velocity modes found earlier in the CDM
frame. From Eq. $\eqref{e:etak1}$, we find that the spatial gradient
of the 3-curvature for the adiabatic mode is a constant on large
scales as $x \to 0$. We have arrived at the same result found in
\cite{bruni2, gordon}. With a more detailed analysis, we can
characterize the $m=2$ and $m=-1$ modes as the growing and decaying
modes that are similar to the case in general relativity. We found
that the curvature is conserved for the GR-like modes for ${\cal
U}=0$, but not for the other two additional braneworld modes.     


\subsection{Discussion on the Scalar Modes} \label{sec:modes}

\begin{figure}[!bth]
\begin{center}
	\psfrag{A}{$\frac{\rho}{\lambda} \gg 1$}
     \psfrag{B}{$\frac{\rho}{\lambda} = 1$}
     \psfrag{C}{$\frac{\rho}{\lambda} \ll 1$}
\includegraphics{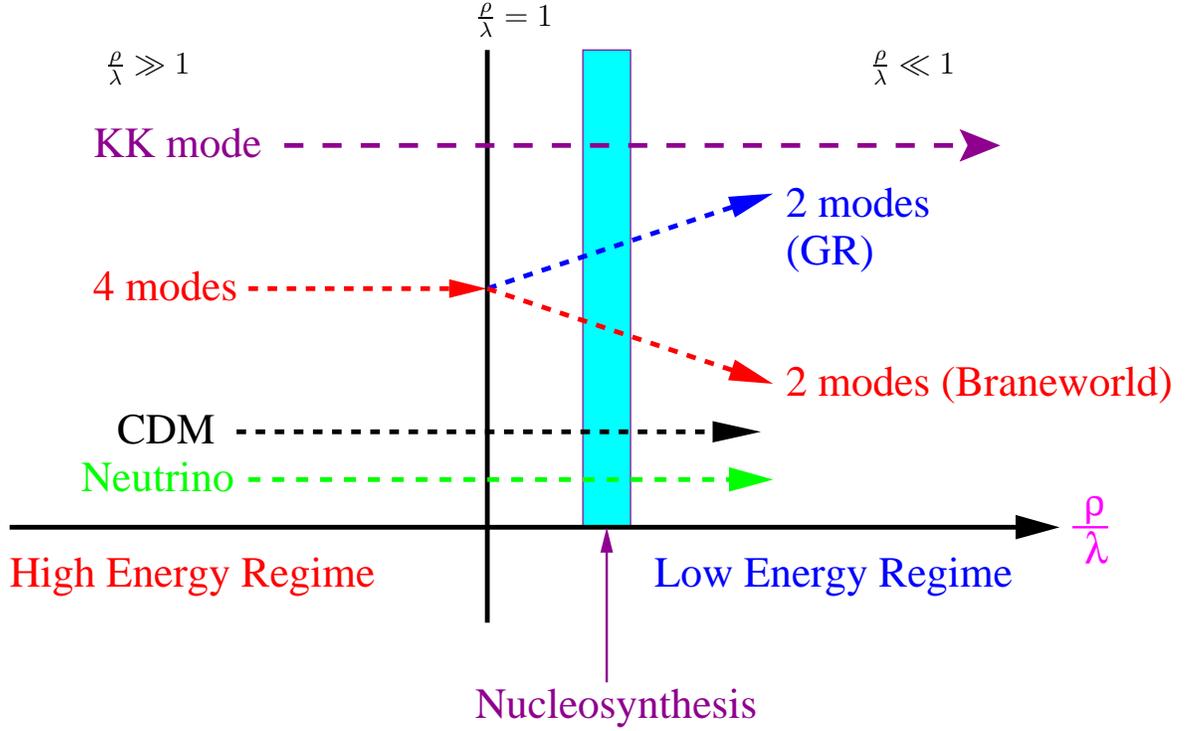}
\caption{A figure which depict history of the different modes for the
braneworlds. The KK mode, the CDM and the neutrino isocurvature modes
will trail from the high energy to the low energy regime, after
inflation. In the high energy regime, where $\rho/\lambda
\gg 1$, there are 4 degenerate modes in the case of the
braneworld. When  the instantaneous transition from high to low energy
regime ($\rho/\lambda \ll 1$) at $\rho/\lambda=1$, the modes will be
split up from four modes to two modes which correspond to the case in
general relativity and the other remaining two modes which are the
non-local isocurvature modes. The two remaining non-local isocurvature 
modes from the high energies are redshifted by approximately $10^{4}$
if there is no dark energy term. Hence the only effect
on the CMBR anisotropies is induced only by the KK mode.  
 \label{chapfig4-1}}
\end{center}
\end{figure}

Regarding the imprint of braneworld effects on the CMB, we note several
possible sources. Once the universe enters the low-energy regime the
dynamics of the perturbations are essentially general relativistic in the
absence of non-local anisotropic stress (see \ref{sec:lowenergy}).
If ${\cal P}_{ab}$ really were zero, the only imprints of the braneworld on
the CMB could arise from modifications to the power spectrum (and
cross correlations) between the various low-energy modes. Since there are two
additional isocurvature modes in the low-energy universe due to braneworld
effects, it need not be the case that adiabatic fluctuations produced during
high-energy (single-field) inflation give rise to a low-energy universe
dominated by the growing, adiabatic, general-relativistic mode (See
\ref{chapfig4-1} for a history of the scalar modes from the high to low
energy regime). Hence there exist the possibility of exciting the
low-energy isocurvature (brane) modes from plausible fluctuations in
the high-energy regime. 

However, if we consider linear scales at last scattering which project
onto angular scales corresponding to  $l \sim 200$
and $h =0.75$ in a $K=0$ and $\Lambda=0$ universe, assuming the
standard nucleosynthesis constraint [refer to eq.$\eqref{e:nucleo}$ in
the previous chapter] and ignoring the 
contributions of the dark radiation term, we find these modes enter
the Hubble radius at approximately
$10^{4}$. This large redshift would indicate that the modes are outside
the Hubble radius. If this is the case, the matching from the high energy
regime (brane-world) to the low energy regime (general relativity) would
render the modes to be super-Hubble. An immediate
consequence is that the CMB anisotropies would be insensitive to the
brane tension, $\lambda$, i.e. the high energy effects.  


Similarly in practice, we do not expect ${\cal P}_{ab}=0$. In this case
the non-local anisotropic stress provides additional driving terms to the
dynamics of the fluctuations, and we can expect a significant manifestation
of five-dimensional Kaluza-Klein effects on the CMB scalar
anisotropies. Barrow and Maartens have shown in \cite{barrow} that
the 5-dimensional KK graviton stresses can slow down the decay of the shear
anisotropy on the brane to observable levels. They found that with a
suitable approximation for ${\cal P}_{ab}$ that the initial shear to
the Hubble distortion of $\sim 10^{3} \Omega_0 H_0^2$ at 5D Planck
time would allow the large-angle CMB signal to be a relic of purely KK
effects. These possibilities may reveal new braneworld imprints that
would be more realistic for observational testing. 

\section{Braneworld Tensor Anisotropies in the CMB} \label{sec:tensor}

In this section, we investigate the dynamics of  tensor
perturbations in braneworlds. Recalling Eqs.~$\eqref{e:dotsigma0}$ and
$\eqref{e:dotE0}$ from Chapter \ref{chapter3}, we rewrite these
equations in conformal time:  
\begin{equation} \label{e:sigmadot} 
\frac{k}{a^2} \l(\sigma'_k + {\cal H} \sigma_k \r) +
\frac{k^2}{a^2} E_k - \frac{\kappa^2}{2} \rho \pi_k = \kappa^2
(2-3\gamma)\frac{\rho^2}{4\lambda}\pi_k  + \frac{\kappa^2}{2} \rho
\pi_k^{*}\,,
\end{equation}

\begin{equation}
\begin{split}
\label{e:Edot} 
&\frac{k^2}{a^2} \l(E'_k + {\cal
H} E_k \r) - k\l(\frac{k^2}{a^2} + \frac{3K}{a^2} -
 \frac{\kappa^2}{2} \g \rho  \r) \sigma_k +
\frac{\kappa^2}{2} \rho \pi'_k -\frac{\kappa^2}{2} (3 \g -1) {\cal
H}\rho \pi_k  \\
&= -\frac{\kappa^2}{4\lambda} \{ 2 k
\g \rho^2 \sigma_k - (3 \g-2) \rho^2 \pi'_k-[3 \g' - (3
\g-2)(6\g-1) {\cal H}] \rho^2 \pi_k \} \\
& \quad  - \frac{2}{3} k\kappa^2\rho^* \sigma_k  - \frac{\kappa^2}{2} \l[\rho
 \pi_k^{*'} + (1-3\g) {\cal H} \rho \pi_k^{*} \r]\,,
\end{split}
\end{equation}
where the non-local energy density is defined as in \cite{barrow}:
\begin{equation}
\rho^{*} = \frac{6}{\kappa^4 \lambda} {\cal U} \;.
\end{equation}
A prime denotes $d/d\tau$, ${\cal H}=a'/a$, and the
(non-constant) parameter $\gamma$ is defined in
Eq.~$\eqref{e:prho}$. Equations~$\eqref{e:sigmadot}$ and
$\eqref{e:Edot}$, which have all the braneworld terms on the right-hand
sides, determine the tensor anisotropies in the CMB, once $\pi_k$
and $\pi^*_k$ are given.
 
\subsection{A Local Approximation to the Nonlocal Anisotropic Stress} \label{sec:tensorapprox}
The general solution for the non-local anisotropic stress will be of the
form
\begin{equation}\label{e:soln}
\pi^*_k(\tau) \propto \int d \tilde\tau\,\,{\cal
G}(\tau,\tilde\tau) F[\pi_k,\sigma_k]\big|_{\tilde\tau} \,,
\end{equation}
where ${\cal G}$ is a retarded Green's function evaluated on the
brane. The functional $F$ is known in the case of a Minkowski
background~\cite{sasaki}, but not in the cosmological case. (An
equivalent integro-differential formulation of the problem is
given in \cite{mukohyama2}; see also \cite{soda}.) Once
${\cal G}$ and $F$ are determined, Eq.~$\eqref{e:soln}$ can in
principle be incorporated into a modified version of Boltzmann
codes such as CAMB~\cite{lewis} or CMBFAST~\cite{seljak}. It
remains a major task of braneworld cosmological perturbation
theory to find this solution, or its equivalent forms in other
formalisms. In the meanwhile, in order to make progress towards
understanding braneworld signatures on tensor CMB anisotropies, we
can consider approximations to the solution.

The non-local nature of $\pi^*_k$, as reflected in
Eq.~$\eqref{e:soln}$, is fundamental, but is also the source of
the great complexity of the problem. The lowest level
approximation to $\pi^*_k$ is local. Despite removing the key
aspect of the KK anisotropic stress, we can get a feel for its
influence on the CMB if we capture at least part of its
qualitative properties. The key qualitative feature is that
inhomogeneity and anisotropy on the brane are a source for KK
modes in the bulk which ``backreact"~\cite{maartens} or ``feed
back"~\cite{mukohyama}, onto the brane. The transverse traceless
part of inhomogeneity and anisotropy on the brane is given by the
transverse traceless anisotropic stresses in the geometry, i.e. by
the matter anisotropic stress $\pi_{ab}$ and the shear anisotropy
$\sigma_{ab}$. The radiation and neutrino anisotropic stresses are
in turn sourced by the shear to lowest order (neglecting the role
of the octupole and higher Legendre moments).

Thus the simplest local approximation which reflects the essential
qualitative feature of the spin-2 KK modes is
\begin{equation} \label{e:Pansatz1}
\kappa^2 \pi_{ab}^{*} = - \zeta H \sigma_{ab}\,,~~\zeta'=0\,,
\end{equation}
where $\zeta$ is a dimensionless KK parameter, with $\zeta=0$
corresponding to no KK effects on the brane, and
$\zeta=0=\lambda^{-1}$ giving the general relativity limit. [Note
that for tensor perturbations, where there is no freedom over the
choice of frame (i.e.\ $u^a$), there is no gauge ambiguity in
Eq.~(\ref{e:Pansatz1}). However, for scalar or vector
perturbations, this relation could only hold in one frame, since
$\pi_{ab}^{*}$ is frame-invariant in linear theory while
$\sigma_{ab}$ is not.]

The approximation in Eq.~(\ref{e:Pansatz1}) has the qualitative
form of a shear viscosity, which suggests that KK effects lead to
a damping of tensor anisotropies. This is indeed consistent with
the conversion of part of the zero-mode at Hubble re-entry into
massive KK modes~\cite{langlois2,gorbunov}. The conversion may be
understood equivalently as the emission of KK gravitons into the
bulk, and leads to a loss of energy in the 4D graviton modes on
the brane, i.e. to an effective damping. The approximation in
Eq.~(\ref{e:Pansatz1}) therefore also incorporates this key
feature qualitatively.

With this first approximation, we can close the system of
equations on the brane by adding the equation
\begin{equation} \label{e:Pansatz1a}
\kappa^2 \rho \pi_{k}^{*} = - \zeta {\cal H} \frac{k}{a^2}
\sigma_k\,.
\end{equation}
We will also assume $K=0=\rho^*$ in the background. The parameter
$\zeta$ (together with the brane tension $\lambda$) then controls
braneworld effects on the tensor CMB anisotropies in this simplest
approximation.

\subsubsection{Initial Conditions}
Ignoring the photon anisotropic stress (i.e. $\pi_{ab}=0$), 
we differentiate Eq. $\eqref{e:sigmadot}$ and decouple it with
Eq. $\eqref{e:Edot}$ and the modified Raychaudhuri equation
[see Eq. $\eqref{e:raychaudhuri}$ of Chapter \ref{chapter3}]. 
Using the variable $u_k \equiv a^{1+{\zeta}/{2}}$, the shear satisfies  
the following equation of motion 
\begin{equation} \label{e:ueqn2}
u_k'' + \l[k^2 +2 K - \frac{(a^{-1-
{\zeta}/{2}})''}{a^{-1-{\zeta}/{2}}} \r] u_k = 0\, ,
\end{equation}
where we have used Eq.~(\ref{e:Pansatz1a}). In flat models ($K=0$)
on large scales there is a decaying solution $\sigma_k \propto
a^{-(2+\zeta)} $. Since Eq.~(\ref{e:ueqn2}) contains no first
derivative term the Wronskian is conserved. On large scales we can
use the solution $\sigma_k \propto a^{-(2+\zeta)} $ to write the
conserved Wronskian as $W = \sigma_k' + (2+\zeta){\cal H} \sigma_k
$. (The Wronskian vanishes in the decaying mode.) Integrating
gives the following two independent solutions on large scales in
flat models:
\begin{equation}
\sigma_k=
\begin{cases}
A_k a^{-(2+\zeta)} \,,\\ B_k a^{-(2+\zeta)} \int^{\tau}
d\tilde\tau\, a(\tilde\tau)^{2+\zeta} \, ,
\end{cases}
\end{equation}
where $A_k$ and $B_k$ are constants of integration. If we let
$\zeta \to 0$, we recover the results in  ~\cite{lewis1} for
the general relativity case.

The conserved Wronskian is proportional to the metric perturbation
variable, $H_T$, characterising the amplitude of 4D gravitational
waves. In flat models $H_T$ is related to the covariant variables
quite generally by
\begin{equation} \label{e:HT1}
H_{Tk} = \frac{\sigma'_k}{k} + 2 E_k \, .
\end{equation}
Ignoring photon anisotropic stress, we can eliminate the electric
part of the Weyl tensor via the shear propagation
equation~(\ref{e:sigmadot}) to find $k H_{Tk} = - \sigma_k' -
(\zeta + 2) \mathcal{H} \sigma_k = -W$. The fact that $H_T$ is
conserved on large scales in flat models in the absence of photon
anisotropic stress can also be seen directly from its propagation
equation,
\begin{equation}
H_{Tk}'' + (2+\zeta) {\cal H} H_{Tk}' + k^2 H_{Tk} = 0\,.
\end{equation}

We can solve Eq.~(\ref{e:ueqn2}) on all scales in the high-energy
($\rho\gg \lambda$ and $a\propto \tau^{1/3}$) and low-energy
($\rho\ll\lambda$ and $a\propto \tau$) radiation-dominated
regimes, and during matter-domination ($a\propto \tau^2$). The
solutions are
\begin{align}
\label{e:highenergy1} u_k(\tau) &= \sqrt{k\tau} \left[ d_1
J_{\frac{1}{6}(5+\zeta)} (k \tau)  + d_2 Y_{\frac{1}{6}(5+\zeta)}
(k \tau)\right] & \quad &(\text{high energy radiation}), \\
\label{e:lowenergy1} u_k(\tau) &= \sqrt{k\tau} \left[d_3
J_{\frac{1}{2} (3+\zeta)} (k \tau) + d_4 Y_{\frac{1}{2}(3+\zeta)}
(k \tau)\right] & &(\text{low energy radiation}),\\
\label{e:matter} u_k(\tau) &= \sqrt{k\tau} \left[ d_5
J_{\frac{5}{2}+\zeta} (k \tau)  + d_6 Y_{\frac{5}{2}+\zeta} (k
\tau)\right] & &(\text{matter domination}),
\end{align}
where $d_i$ are integration constants. The solutions for the
electric part of the Weyl tensor can be found from
Eq.~(\ref{e:sigmadot}). For modes of cosmological interest the
wavelength is well outside the Hubble radius at the transition
from the high-energy regime to the low-energy. It follows that the
regular solution (labelled by $d_1$) in the high-energy regime
will only excite the regular solution ($d_3$) in the low-energy,
radiation-dominated era. Performing a series expansion, we arrive
at the appropriate initial conditions for large-scale modes in the
low-energy radiation era:
\begin{align}
\label{e:hijini} H_{Tk} &=  1 - \frac{(k \tau)^2}{2 (3 + \zeta)} +
\frac{(k \tau)^4}{8 (3+\zeta)(5+\zeta)}  + O[(k\tau)^6], \\
\sigma_k &= -\frac{k \tau}{3+\zeta} + \frac{k^3
\tau^3}{2(3+\zeta)(5+\zeta)} + O[(k\tau)^5], \label{e:shearini}\\
\label{e:elecini} E_k &=  \frac{(4 + \zeta)}{2(3+\zeta)} -
\frac{(k\tau)^2 (8+\zeta)}{4(3+\zeta)(5+ \zeta)} + O[(k\tau)^4].
\end{align}
In the limit $\zeta \to 0$, we recover the general relativity
results~\cite{challinor4}.

For modes that are super-Hubble at matter-radiation equality
(i.e.\ $k \tau_{\text{eq}} \ll 1$), the above solution joins
smoothly onto the regular solution labelled by $d_5$ in
Eq.~(\ref{e:matter}). For $k \tau_{\text{eq}} \gg 1$, the shear
during matter domination takes the form
\begin{equation}
\label{e:matterlong} \sigma_k = -
2^{\frac{3}{2}+\zeta}\Gamma(\tfrac{5}{2}+\zeta)
(k\tau)^{-(\frac{3}{2}+\zeta)} J_{\frac{5}{2}+\zeta}(k\tau).
\end{equation}
In the opposite limit, the wavelength is well inside the Hubble
radius at matter-radiation equality. The asymptotic form of the
shear in matter domination is then
\begin{equation}
\label{e:mattershort} \sigma_k \sim
\frac{\Gamma[\frac{1}{2}(3+\zeta)]}{\sqrt{\pi}} \left(
\frac{2\tau_{\text{eq}}}{\tau}\right)^{1+\zeta/2}
(k\tau)^{-(1+\zeta/2)} \sin(k\tau - \pi\zeta/4).
\end{equation}

\begin{figure}[!bth]
\begin{center}
\includegraphics[scale=0.8]{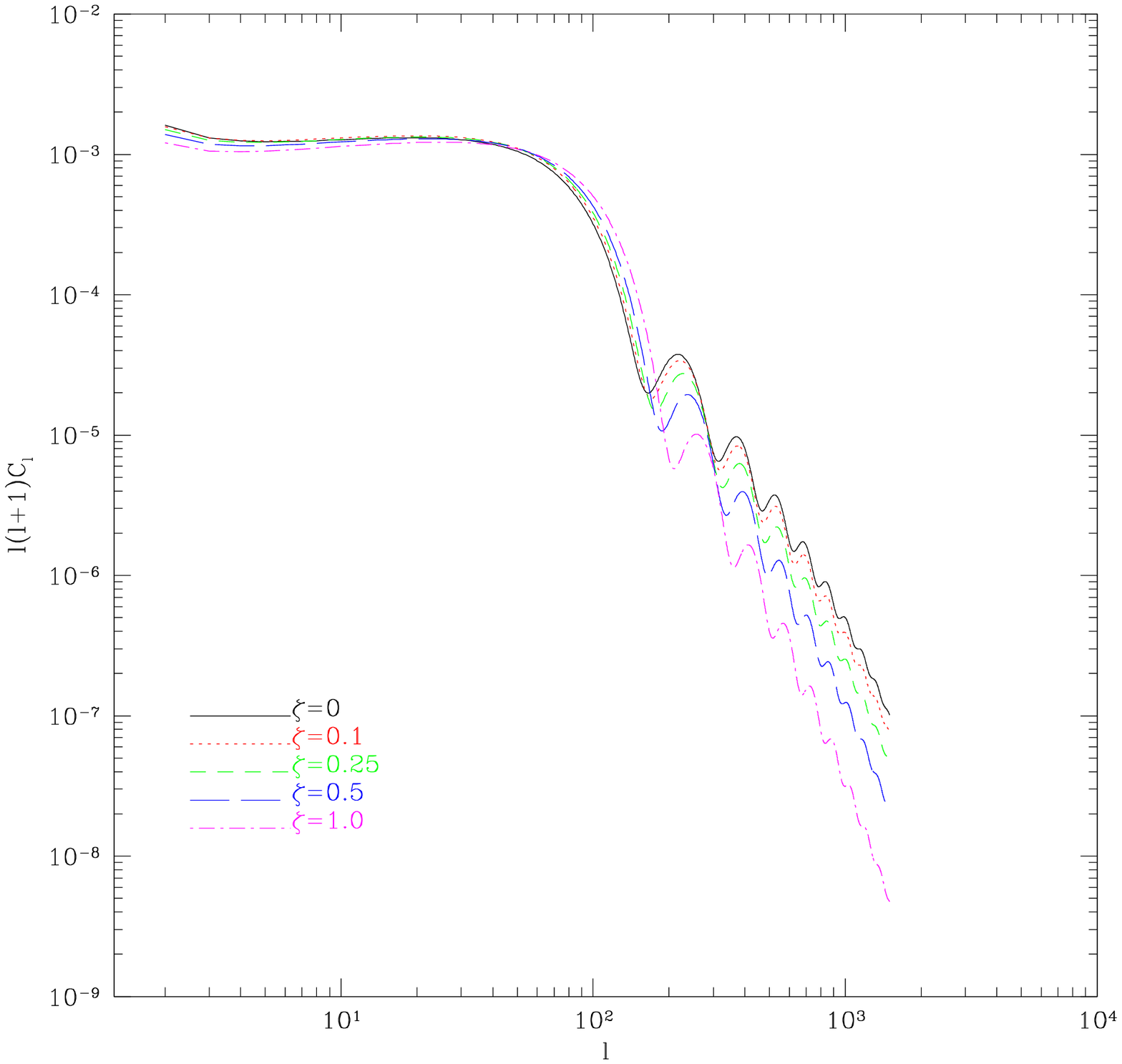}
\caption{The temperature power spectrum for tensor perturbations
in braneworld models using the approximation in
Eq.~(\ref{e:Pansatz1}), with $\zeta$ the dimensionless KK
parameter. Models are shown with $\zeta= 0.0$,  0.1, 0.25,  0.5
and 1.0. The initial tensor power spectrum is scale invariant and
we have adopted an absolute normalisation to the power in the
primordial gravity wave background. The background cosmology is
the spatially flat $\Lambda$CDM (concordance) model with density
parameters $\Omega_b=0.035$,  $\Omega_{c}=0.315$,
$\Omega_{\Lambda}=0.65$, no massive neutrinos, and the Hubble
constant $H_0=65\, \text{km~s}^{-1} \text{Mpc}^{-1}$.
\label{chapfig4-2}}
\end{center}
\end{figure}

\begin{figure}[!bth]
\begin{center}
\includegraphics[scale=0.8]{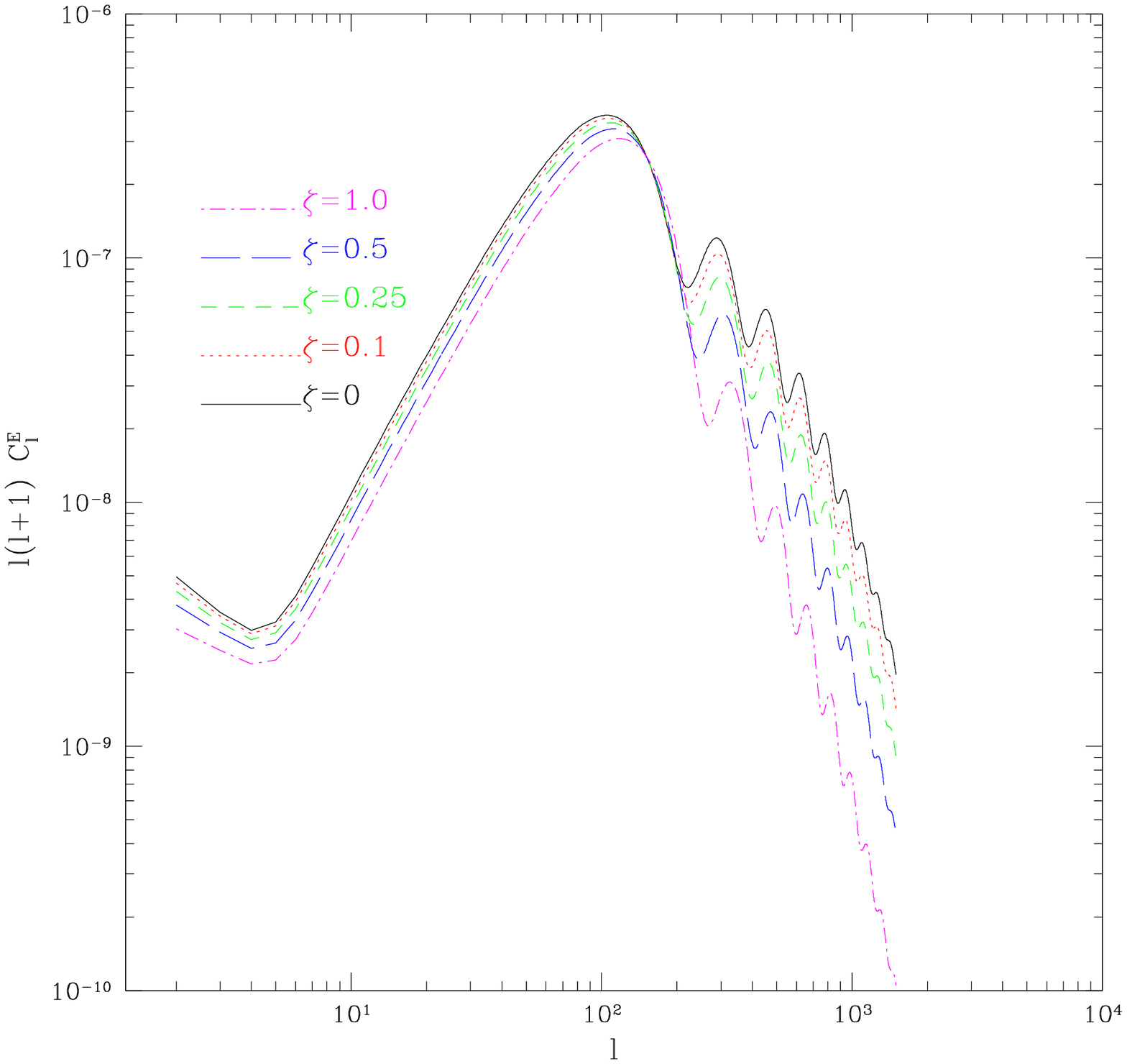}
\caption{ The electric polarization power spectrum for tensor
perturbations for the same braneworld models as in
Fig.~\ref{chapfig4-2}. \label{chapfig4-3}}
\end{center}
\end{figure}

We use the initial conditions,
Eqs.~(\ref{e:hijini})--(\ref{e:elecini}), in a modified version of
the CAMB code to obtain the tensor temperature and polarization
power spectra. The temperature and electric polarization spectra
are shown in Figs.~\ref{chapfig4-2} and \ref{chapfig4-3} for a
scale-invariant initial power spectrum. The normalisation is set
by the initial power in the gravity wave background. Figs
~\ref{chapfig4-2} and \ref{chapfig4-3}, together with
Eqs.~(\ref{e:highenergy1})--(\ref{e:elecini}), are the main result of
this work, and we now discuss the physical conclusions
following from these results.     

\begin{figure}[!bth]
\begin{center}
\includegraphics[scale=0.8]{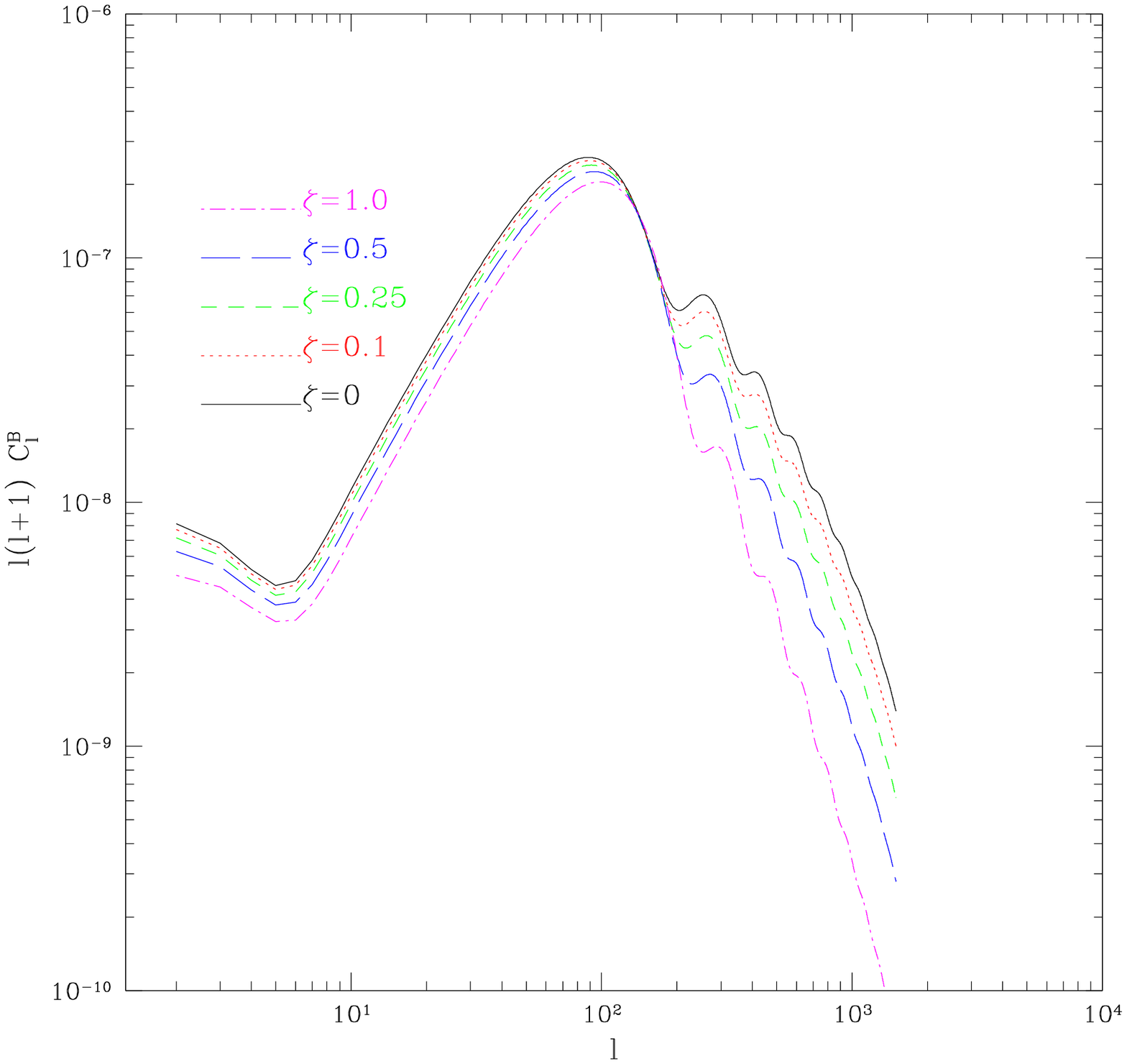}
\caption{ The magnetic polarization power spectrum for tensor
perturbations for the same braneworld models as in
Fig.~\ref{chapfig4-2}. \label{chapfig4-4}}
\end{center}
\end{figure}

\subsection{Discussion on the Tensor Power Spectra} \label{sec:tensorspectra}

As expected, we find that the power spectra are insensitive to high-energy
effects, i.e. effectively independent of the brane tension $\lambda$ : the
$\zeta=0$ curve in Fig.~1 is indistinguishable from that of the
general relativity model (both power spectra are identical 
at the resolution of the plot). For the computations, we have used the
lowest value of the brane tension $\lambda=(100~{\rm  GeV})^4$,
consistent with the tests of Newton's law. 

There are three notable effects visible in Figs.~1 and 2, from
our approximate model of the KK stress:
(i)~the power on large scales reduces with 
increasing KK parameter $\zeta$; (ii)~features in the spectrum
shift to smaller angular scales with increasing $\zeta$; and
(iii)~the power falls off more rapidly on small scales as $\zeta$
increases. Neglecting scattering effects, the shear is the only
source of linear tensor anisotropies (see e.g.~\cite{challinor4}).
For $1\ll l < 60$ the dominant modes to
contribute to the temperature $C_l$s are those whose wavelengths
subtend an angle $\sim 1 / l$ when the shear first peaks (around
the time of Hubble crossing). The small suppression in the $C_l$s
on large scales with increasing $\zeta$ arises from the reduction
in the peak amplitude of the shear at Hubble entry [see
Eq.~(\ref{e:matterlong})], qualitatively interpreted as the loss
of energy in the 4D graviton modes to 5D KK modes.

Increasing $\zeta$ also has the effect of adding a small positive
phase shift to the oscillations in the shear on sub-Hubble scales,
as shown e.g.\ by Eq.~(\ref{e:mattershort}). The delay in the time
at which the shear first peaks leads to a small increase in the
maximum $l$ for which $l(l+1)C_l$ is approximately constant, as is
apparent in Fig.~\ref{chapfig4-2}. The phase shift of the subsequent
peaks in the shear has the effect of shifting the peaks in the
tensor $C_l$s to the right. For $l > 60$ the main contribution to
the tensor anisotropies at a given scale is localized near last
scattering and comes from modes with wavenumber $k \sim l /
\tau_0$, where $\tau_0$ is the present conformal time. On these
scales the gravity waves have already entered the Hubble radius at
last scattering. Such modes are undergoing adiabatic damping by
the expansion and this results in the sharp decrease in the
anisotropies on small scales. Increasing the KK parameter $\zeta$
effectively produces more adiabatic damping and hence a sharper
fall off of power. The transition to a slower fall off in the
$C_l$s at $l \sim 200$ is due to the weaker dependence of the
amplitude of the shear on wavenumber at last scattering for modes
that have entered the Hubble radius during radiation
domination~\cite{starobinsky}. [The asymptotic expansion of
Eq.~(\ref{e:matterlong}) gives the shear amplitude $\propto k^{-(2
+ \zeta)}$ at fixed $\tau$, whereas for modes that were sub-Hubble
at matter-radiation equality Eq.~(\ref{e:mattershort}) gives the
amplitude $\propto k^{-(1 + \zeta/2)}$.]

Similar comments apply to the tensor electric polarization
$C^E_l$, shown in Fig.~\ref{chapfig4-3}. As with the temperature
anisotropies, we see the same shifting of features to the right
and increase in damping on small scales. Since polarization is
only generated at last scattering (except for the feature at very
low $l$ that arises from scattering at reionization, with an assumed 
optical depth $\tau_C=0.03$), the large-scale polarization is
suppressed, since the shear (and hence 
the temperature quadrupole at last scattering) is small for
super-Hubble modes. In matter domination the large-scale shear is
$\sigma_k = - k\tau / (5 + 2\zeta)$; the reduction in the
magnitude of the shear with increasing KK parameter $\zeta$ is
clearly visible in the large-angle polarization. The braneworld
modification to the tensor magnetic polarization $C^B_l$ has the
same qualitative features as in the electric case (see
Fig.~\ref{chapfig4-4}).


      \clearpage{\pagestyle{empty}\cleardoublepage}	
      \part{Changing Global Symmetry}
\chapter{Changing Global Symmetry} \label{chapter5}
\begin{verse} 
{\it Some say the world will end in fire \\
some say in ice. \\
From what I've tasted of desire  \\
I hold with those who favor fire.} \\
- Robert Frost 
\end{verse}

\begin{verse}
{\it This is the way the world ends, \\
Not with a bang but a whimper. }\\
- T.S. Eliot 
\end{verse}

{\it In this chapter, we examine the changing global symmetry
of the Einstein-de Sitter Universe and  see  how  it could  lead  to
a possible phase transition in  the future as in reference
\cite{leong3}. The  largest galaxy
cluster in a flat Einstein-de Sitter universe may grow indefinitely to
encompass most  of space  after an extremely  long time.  We  derive a
general  relativistic   metric  for  a  very   large,  bounded  nearly
isothermal cluster of  galaxies which is embedded in  such a universe.
The  embedding  is  done  by  means  of  a  "Schwarzschild  membrane".
Pressure  is   important,  unlike  previous   Tolman-Bondi  models  of
inhomogeneities.  The  cluster's expansion, represented  by a sequence
of models, alters the average  global symmetry of an increasing volume
of  space-time.  Initially  the metric  is homogeneous  and isotropic,
having translational  and rotational symmetry around  every point.  As
the  metric evolves,  it eventually  loses its  translational symmetry
throughout larger regions and  retains rotational symmetry around only
one  point:  the centre  of  the cluster  becomes  the  centre of  the
universe.  Our  sequence  of  hybrid  models  transforms  between  the
Einstein-de Sitter  and the isothermal cluster  limits and illustrates
how  a changing  equation  of state  of  matter can  alter the  global
symmetry of space-time.}

\section{Introduction} \label{chap5-1}

It is well  known in standard texts \cite{mtw, peebles,
weinberg1} that one can
have a changing equation of state in a single metric. For example, one
can study the  behaviour of the FRW metric  in the radiation-dominated
and the matter-dominated era  which correspond to different equations
of state.  In these  situations, the equation  of state does not
influence the global symmetry of the  metric,  and there  is  a smooth
transition from  the radiation-dominated  to the matter-dominated
era.  In  other previous inhomogeneous  models,  the  equation   of
state  has  zero  pressure \cite{bondi, einstein, kramer,
krasinski, lasenby}. Therefore, it
remains an interesting  fundamental question to ask whether such a
dynamical symmetry breaking process can 
be modelled  in general relativity.  To paraphrase this question  in a
more  concrete form,  can one  find a  metric to  describe  a changing
global symmetry, with a changing equation of state that will alter the
form of the metric?  Here, the basic effect of the pressure on the metric
arises not from  its contribution to the energy  density or to the
expansion, but from its support of growing inhomogeneity.

At first sight,  it may not seem possible to  solve the Einstein field
equations  directly for  a changing  equation of  state  which induces
symmetry changes in  the metric. Even the form of  such an equation of
state has  not been examined  previously.  Hence, to answer  the above
question, we need to adopt a different approach to find a more general
solution. Our method is to find a continuous hybrid metric which is
consistent with the known limit-solutions. The hybrid solution is
parameterised by two constants and applies only to the quasi-static
phases of the universe when it is dominated by one gigantic isothermal
cluster. As a second cluster of comparable size enters the horizon the
hybrid description breaks down. The clusters merge and eventually form
a new, larger isothermal sphere. After relaxation of this larger
system the hybrid solution is valid again with changed values  of its
characteristic constants. 

This allows us to separate the problem into two parts. First, we find 
the parameterised hybrid solution which describes those periods in the
evolution of the universe, which are dominated by a large isothermal
core. We then describe the temporal evolution as a discrete sequence
of such stages i.e. as a sequence of hybrid solutions with changed 
parameterisation.

Due  to  the  different equations  of  state, it  is  not  possible
to find  direct  matching conditions   for  the generalised isothermal 
metric and the FRW metric. The  reason   can  be   understood
physically.  Clustering induces  a change  from a  universe  with zero
pressure  to  a  universe  with  non-zero  cosmologically  significant
pressure. There  is a symmetry change from  the spherical, homogeneous
FRW  models, which  have  no  centre to  the  isothermal model,  which
singles out the point with highest density as its centre.

\section{The Isothermal Universe} \label{chap5-2}

\subsection{The Isothermal Metric}

We begin with a brief summary of the isothermal universe found by
Saslaw et al\cite{saslaw}. The starting point is to consider an
asymptotically static universe that satisfies the general static, 
spherically symmetric line element:  
\begin{equation} \label{e:lineelement1}
ds^2 = -e^{\nu(r)} dt^2 + e^{\lambda(r)} dr^2 + r^2 (d\theta^2 + \sin^2
\theta d\phi^2).
\end{equation}
The Einstein field equations are given by
\begin{equation} \label{e:einstein}
G_{ab} = R_{ab} - \frac{1}{2} R g_{ab}  = 8 \pi T_{ab},
\end{equation}
where the energy-momentum tensor of a perfect fluid is
\begin{equation} \label{e:em-perfectfluid}
T_{ab} = (\rho + P) u_a u_b + P g_{ab}.
\end{equation}
To derive  the solution  for the isothermal  universe, an  equation of
state is required to determine  the metric. The isothermal equation of
state  is characterized  by  a pressure  gradient  which balances  the
mutual self-gravity of its  constituent particles i.e. idealized point
galaxies.  The dispersion  of  the particles'  peculiar velocities  is
independent of position. Therefore the equation of state is:
\begin{equation} \label{e:eqs1}
P = \alpha \rho,
\end{equation}
where $\alpha$ is a constant for the whole spacetime and $0 \le \alpha
\le 2/3$. 

There are two assumptions in obtaining the solution of the isothermal
sphere. Firstly, we assume $e^{\lambda(r)}$ to be independent of
time. This has the geometrical interpretation that the radial coordinate of the metric
scales radially as a constant. Secondly, we assume the  density
distribution  of   a  finite  isothermal  sphere  (which realistically
would  be bounded  by  an  external  pressure or  tidal 
disruption)  is described  by the  relation $\rho  \propto  r^{-2}$ at
large radii, as was first shown by Emden~\cite{emden} for the Newtonian case
and    subsequently   extended   to    the   relativistic    case   by
Chandrasekhar~\cite{chandra2}.   To obtain the cosmological metric describing
an infinite isothermal universe one has to make the simplifying
approximation that $\rho \propto r^{-2}$ throughout the entire sphere. 
Solving   the   Einstein    field   equations $\eqref{e:einstein}$
using the line  element $\eqref{e:lineelement1}$ together   with Eqs. 
$\eqref{e:em-perfectfluid}$ and $\eqref{e:eqs1}$  and   the  density
distribution given above, gives the following relations: 
\begin{align}
e^{\nu} &= Ar^{\frac{4\alpha}{1+\alpha}}, \\
e^{\lambda} &= 1 + \frac{4\alpha}{(1+\alpha)^2},\\
\label{e:rho}
8\pi \rho &= \frac{4\alpha}{4\alpha+(1+\alpha)^2}~\frac{1}{r^2}.
\end{align}
The line element for the isothermal universe is
therefore stated as follows:
\begin{equation} \label{e:isothermal1}
ds^2 = -A r^{\frac{4 \alpha}{1+\alpha}} dt^2 + \l(1 + \frac{4
\alpha}{(1 + \alpha)^2} \r) dr^2 + r^2 (d \theta^2 + \sin^2
\theta d \phi^2),
\end{equation}
where $A$ is an arbitrary constant.

Here we find that the transformation
$\bar{r} = r^{U}$ casts the metric in Eq. $\eqref{e:isothermal1}$ into
its isotropic form  
\begin{equation} \label{e:isotropic1}
ds^2= - A \bar{r}^V dt^2 +
\bar{r}^{2W}  \l[d \bar{r}^2 + \bar{r}^2 d\Omega^2 \r],
\end{equation}
where the functions $U$, $V$ and $W$ are defined as
\begin{align} 
\label{e:termc}
U &= \l(1+ \frac {4\alpha}{(1+\alpha)^2} \r)^ \frac{1}{2}, \\
\label{e:termd}
V &= \frac{4\alpha}{U \l(1+\alpha \r)}, \\
\label{e:terme}
W &= \frac{1}{U} - 1.
\end{align}
For completeness, we also state the conformal form of Eq. $\eqref{e:isothermal1}$,
which relates to minimally curved spaces, as found by Dadhich~\cite{dadhich1}:  
\begin{equation}
ds^2 = r^{\frac{-2m}{1+m}} \l[-dt^2 +k_{1}^2dr^2 +r^2(d\theta^2 +
\sin^2\theta d\phi^2)\r]. 
\end{equation}
where $e^{\lambda}=k_{1}^2$ and $e^{\nu}=r^{-2m}$. Note that $k_{1}^2$ and
$m$ are constants relating to $\alpha$.

\subsection{Properties and Implications of the Isothermal solution with a Global
Phase Transition of the Universe}
Before we proceed further, we note a few interesting features with the
isothermal solution, $\eqref{e:isothermal1}$, found by Saslaw et al in
\cite{saslaw}. First of all, the isothermal universe in their case, is
a infinite isothermal sphere without a boundary. The solution is not a
realistic one which we could use to model the ultimate phase
transition of the universe. However, even with that limitation, from
examining the physical interpretation of the unbounded isothermal
sphere, we could infer why a global phase transition would eventually
happen. 

From an extensive study in the dynamics of galaxy clustering (which
was well discussed in the references \cite{saslaw2,saslaw5,saslaw4}),
there is a discussion on whether the many-body clustering of galaxies
in our expanding Universe would be represented as a phase transition. The
conclusion is reached in \cite{saslaw5} that galaxy clustering lacks
the near simultaneity of the transition over all scales, even though
it has some basic features of such a phase transition. In
Einstein-Friedmann cosmologies with $\Omega_0 < 1$, and in related
cosmologies with $\Omega_{\Lambda} >0$, the universe would eventually
expand so rapidly relative to the gravitational clustering timescales
such that the pattern of galaxies would freeze out on large scales, and
correlations cease to grow, and the attempt at a second order phase
transition would die away before it could ever happen. 

However, in the case of the flat $\Omega_0=1$, $\Omega_{\Lambda}=0$,
Einstein-de Sitter cosmology, there is a possibility for this phase
transition to occur. In present day observational cosmology, in the
viewpoint of the data from the observed fluctuations in the cosmic
microwave background anisotropies \cite{maxima1, boomerang1}, it
seems very likely that our Universe is nearly flat. However, whether there is a
cosmological constant, is still being debated, but if there is, the
consequence of a phase transition would have been different. 

One would ponder how the phase transition would enter in the
unbounded isothermal sphere scenario. The first thing is to recognise
that there is a change of fundamental symmetry between the
Einstein-de Sitter model and the isothermal model. We recall the form
of the Friedmann-Robertson-Walker metric (Eq.$\eqref{e:frw}$ in
Chapter \ref{prologue}). 

To match Eqs.$\eqref{e:frw}$ and $\eqref{e:isothermal1}$, we need to
show that the pressure must be discontinous at the junction across the
hypersurface where $r$ is a constant. Comparison of the two metrics
would show that for $\alpha >0$, there is no such hypersurface where
this match occurs. Therefore this change must be continuous. Since
there is a time dependence in the global expansion $a(t)$, and the
isothermal metric has no time dependence, it suggests that in the
limit that $t\to \infty$, the transition would take place
everywhere. Such an event would correspond to a phase transition. Of
course, a more complicated metric that allows growing
inhomogeneities would be required to follow this transition. 

This has set the scene and motivation for this chapter on the theme of
changing global symmetry in our thesis on alternative cosmologies. In
\cite{leong3}, we have found and studied a hybrid model which corresponds to
the global symmetry change mentioned above. In the next section, we
will subsequently derive  a generalized form  of their metric to include a
{\it bounded}  isothermal  sphere using  an  approach  by Tolman~\cite{tolman}. We
review and discuss the properties of the bounded isothermal
sphere. This is followed by a discussion of the approximations
involved in the general solution (\ref{chap5-3}) and the
thermodynamic consequences implied by the modified equation of state
(\ref{chap5-4}).  In \ref{chap5-5},  we  first match  the
bounded isothermal sphere to a Schwarzschild background, and then embed the
combined  metric  into   the  expanding  Einstein-de Sitter  universe,
adapting earlier techniques (McVittie \cite{mcvittie, mcvittie2} and
Hogan \cite{hogan}). In   
\ref{chap5-6}, we describe the evolution of the universe as 
a sequence of static stages. Each stage is described by a hybrid
metric with characteristic parameterisation. Finally, in 
\ref{chap5-7}, we discuss the implications of such a cosmological
solution, linking the dynamical breaking of symmetry to a phase
transition.  In the same light, we  mention an additional role for the
cosmological constant which counteracts the  formation of the
isothermal sphere. 

\subsection{Generalization of the Isothermal Metric to Finite Spheres}

The basic problem in the search for a solution by direct matching of the
isothermal metric and the FRW metric, lies in the finite pressure-mismatch
between the FRW models, which are based on the assumption of zero
pressure, whereas our FRW fitting uses zero pressure, but the
isothermal sphere could have a nonzero uniform pressure. To circumvent this
difficulty, one has to generalise the infinite isothermal sphere in
\cite{saslaw} to a finite sphere, where the pressure drops to zero at
the boundary. Then it is possible to match this finite sphere solution
to the expanding space surrounding it. 

To generalise the isothermal solution we employed the technique used
by Tolman \cite{tolman}. By starting with the general static, spherically
symmetric line element described by Eq. $\eqref{e:lineelement1}$ one
arrives at the following field equations for the perfect fluid case: 
\begin{equation}
\label{e:tolmanefe}
\frac{d}{dr} \l(\frac{e^{-\lambda}-1}{r^2}\r) + \frac{d}{dr}
\l(\frac{e^{-\lambda}\nu'}{2r}\r)+ e^{-\lambda-\nu}\frac{d}{dr}
\l(\frac{e^{\nu}\nu'}{2r}\r) = 0 ,
\end{equation}
\begin{align}
8\pi \rho &= e^{-\lambda}\l( \frac{\nu'}{r}+\frac{1}{r^2}\r) -
\frac{1}{r^2}, \\
8\pi P &= e^{-\lambda}\l( \frac{\lambda'}{r}-\frac{1}{r^2}\r) +
\frac{1}{r^2}.
\end{align}
We assume $e^{-\lambda}=$ constant in order to obtain a simple
generalization of the isothermal equation of state, so that a very
large central cluster can have a finite boundary.
This allows us to integrate Eq. $\eqref{e:tolmanefe}$, and leads to
the corresponding solution:  
\begin{align}
e^{\nu} &= \l( Cr^{1-n} - Dr^{1+n}\r)^2, \\
e^{\lambda} &= 2-n^2,  \\
8\pi \rho &= \frac{1-n^2}{2-n^2} ~\frac{1}{r^2}, \\
8\pi P &= \frac{1}{2-n^2}~\frac{1}{r^2}~\frac{(1-n)^2 C -(1+n)^2 Dr^
{2n}}{C-Dr^{2n}},
\end{align}
where $n$, $C$ and $D$ are arbitrary constants of integration. 

To relate Tolman's solution to the isothermal metric, one makes the
following identification
\begin{equation}
e^{\lambda} \equiv 1+ \frac{4\alpha}{(1+\alpha)^2} = 2-n^2,
\end{equation}
where
\begin{equation}
n = \pm \frac{1-\alpha}{1+\alpha}.
\end{equation}
By taking the positive root\footnote{Choosing the negative root gives
the same solutions with $C$ and $D$ interchanged. This reflects an
implicit symmetry of the Tolman solution.} of the above relation 
the isothermal metric is generalised to the form 
\begin{align}
e^{\nu} &= \l( Cr^{\frac{2\alpha}{1+\alpha}} -
Dr^{\frac{2}{1+\alpha}}\r)^2, \\
e^{\lambda} &= 1 + \frac{4\alpha}{(1+\alpha)^2},\\
8\pi \rho &= \frac{4\alpha}{4\alpha+(1+\alpha)^2} ~\frac{1}{r^2},\\
8\pi P &= \frac{4\alpha^2}{4\alpha+(1+\alpha)^2}~\frac{1}{r^2}~\frac{C
-\frac{1}{\alpha^2} Dr^{\frac{2(1-\alpha)}{1+\alpha}}} {C-Dr^{
\frac{2(1-\alpha)}{1+\alpha}}}. 
\end{align}
$D$ is interpreted as a deviation term for the isothermal sphere. 
The original isothermal metric without a boundary is included in this
form in the limit $D \to 0$. 

The specific case for $n=1/2$ was discussed by Tolman~\cite{tolman}. In the
limit $D/C \to 0$, the ratio of pressure to density ($\alpha$)
approaches one-third throughout and the sphere grows without limit. 
Physically, it represents the blackbody radiation
solution in the Oppenheimer-Volkoff analysis. In the limit of large
$\rho$, the equation of state for the sphere goes over to
the approximate form, $\rho-3P \propto \rho^{1/2}$, as for a highly
compressed Fermi gas.  

In the generalised form of the bounded
isothermal metric, we still require the fixed density distribution
$\rho \propto r^{-2}$, but allow the pressure-density relation to
deviate from the perfect isothermal equation of state.
Hence, we deduce the following generalised isothermal equation of
state:
\begin{equation} \label{e:eqs2}
\frac{P}{\rho}=\alpha ~\frac{C-\frac{1}{\alpha^2}
Dr^{\frac{2(1-\alpha)}{1+\alpha}}} {C-Dr^{
\frac{2(1-\alpha)}{1+\alpha}}}, 
\end{equation}  
where $\alpha$ is defined as the constant of proportionality between
pressure and density {\it at the centre} of the sphere. 
Note that the above equation of state allows the pressure to drop to
zero at a finite radial distance, $r_b$. Hence, our metric incorporates
the feature of describing a finite sphere, each shell of
which is approximately isothermal. It is the requirement that the isothermal
solution has a boundary, which introduces deviations
from the perfect isothermal equation of state. A related class of
spherically symmetric solutions of Einstein's equation which consist
of a perfect fluid with uniform density but non-uniform pressure, was
found by Wesson and Ponce de Leon~\cite{wesson1, wesson2}. In their
solution, the equation of state is similar to Eq. $\eqref{e:eqs2}$ except
that their equation of state for the metric is both dependent on
time, $t$ and radial position, $r$. These solutions are recognised to
be scale free and self-similar and were discussed in detail by
Carr~\cite{carr1, carr2}. The nature of equation $\eqref{e:eqs2}$ is
that at any given radius and for  
any given value of $\alpha$, the pressure $P$ is proportional to $\rho$,
but the constant of proportionality varies with position (see \ref{chap5-4} for
a discussion). We denote this as a generalised isothermal sphere. Formally,
this is a result of the finite boundary. Physically, this could result
from incomplete dynamical relaxation in the outer regions  of the
cluster where the densities are lower and the relaxation timescales
longer, but a detailed model of these dynamics is beyond the scope of
this chapter. The pressure decreases to zero at the boundary and it
remains zero for larger radii.  This means that infall or expansion
around the cluster is radial, moving mainly with the Hubble flow, and
has negligible random velocity. 

The boundary of the generalised isothermal sphere is physically
defined by $P(r=r_b) = 0$:
\begin{equation}\label{e:rb}
r_b=\l(\frac{\alpha^2 C}{D}\r)^{\frac{1+\alpha}{2(1-\alpha)}}.
\end{equation} 
The pressure parameter at the centre, $\alpha$, is
directly related to the scale of the generalised isothermal
sphere. For vanishing $\alpha$, Eq. $\eqref{e:rb}$ agrees with the fact
that in this limit there is no isothermal sphere. Larger values of the
constant $\alpha$ correspond to a sequence of static solutions for
larger spheres which have grown through the merger of clusters at
later stages. We find that $C$ and $D$ can be used to normalise the
scale of the generalised isothermal  sphere. We will also discover
that $C$ and $D$ are not independent, since the matching to the
Schwarzschild exterior provides a relation between them.  

The mass of the generalised isothermal sphere can be determined by
integrating its density up to the boundary:
\begin{equation}\label{e:mass}
M_{iso} = \int_{0}^{r_{b}} \rho(r) 4\pi r^2 dr
=\frac{2\alpha}{4\alpha+(1+\alpha)^2} ~r_{b}.
\end{equation}
 
Given the functional form of the isothermal boundary, $r_b$, in terms of
$C$, $D$ and $\alpha$ we can rewrite the generalised isothermal metric
in the following form which closely resembles the original isothermal
metric, but with a radially dependent deviation term. 
\begin{align} \label{e:isothermal3}
e^{\nu} &= Ar^{\frac{4\alpha}{1+\alpha}} \l[1- \alpha^2
\l(\frac{r}{r_b}\r)^{\frac{2(1-\alpha)}{1+\alpha}}\r]^2, \\ 
e^{\lambda} &= 1 + \frac{4\alpha}{(1+\alpha)^2},\\
8\pi \rho &= \frac{4\alpha}{4\alpha+(1+\alpha)^2} ~\frac{1}{r^2},\\
P &= \alpha\rho
~\frac{1-\l(\frac{r}{r_b}\r)^{\frac{2(1-\alpha)}{1+\alpha}}}{1-\alpha^2\l(\frac{r}{r_b}\r)^{\frac{2(1-\alpha)}{1+\alpha}}},  
\end{align}
where we have set $C^2 \equiv A$. 

Hence we have found the interior metric for a generalised isothermal
sphere of finite radius:
\begin{equation}\label{e:interior}
ds^2= -Ar^{\frac{4\alpha}{1+\alpha}}\l[1-\alpha^2
  \l(\frac{r}{r_b}\r)^{\frac{2(1-\alpha)}{1+\alpha}}\r]^2dt^2
  +\l(1+\frac{4\alpha}{(1+\alpha)^2}\r)dr^2
  +r^2(d\theta^2+\sin^2\theta d\phi^2)
\end{equation}
for $r < r_b$.

\section{The Isothermal Approximation} \label{chap5-3}
Our generalization of the isothermal universe involves two
approximations which we now want to consider in turn.
First we review the approximation of the generalised metric to the
ideal isothermal counterpart. For the generalised isothermal metric
to be a good approximation to Eq. $\eqref{e:isothermal1}$,
Eq. $\eqref{e:isothermal3}$ requires the condition 
\begin{equation}
j \equiv \alpha^2 \l(\frac{r}{r_b}\r)^{\frac{2(1-\alpha)}{1+\alpha}} \ll 1,
\end{equation}
where $0 \le \alpha \le \frac{2}{3}$. This condition on $\alpha$
incorporates the possibilities of relativistic pressure ($\alpha=1/3$)
and of a perfect monatomic gas ($\alpha=2/3$). 

\begin{figure}[!bth]
\psfrag{A}{$\alpha$}
\psfrag{X}{$x$}
\begin{center}
\includegraphics{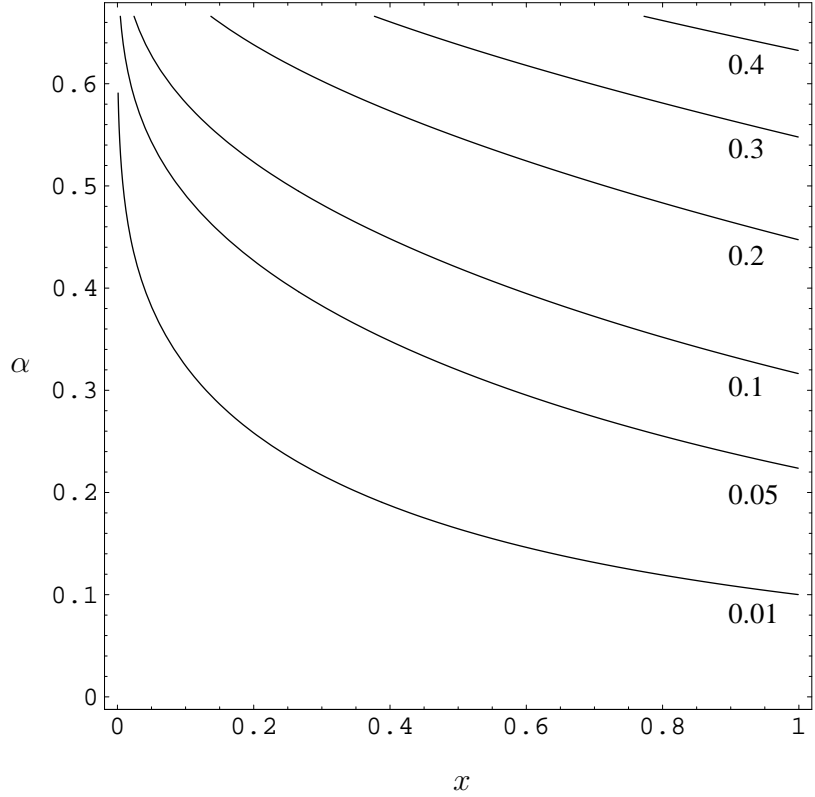}
\caption{Contour plot of $j$. Here $x \equiv (r/r_b)$.
     For $j \ll 1$, which characterizes almost all combinations of
     $\alpha$ and $x$, the deviation of the generalised
     metric from the ideal isothermal form is neglegible. \label{chapfig5-1}
}
\end{center}
\end{figure}

A contour plot of $j\l(r/r_b, \alpha\r)$ is given in Fig.
\ref{chapfig5-1}. It shows that $j < 1$ for all $\alpha$ and $(r/r_b)$-ratios. 
The generalised metric therefore resembles the perfect isothermal
metric very closely. Deviations from the perfect isothermal metric
express the feedback  of the finite boundary of the isothermal sphere
on the metric.  

The generalised equation of state can be rewritten in the form 
$P = \alpha \rho \l[1-\Delta\r]$ where we define the deviation term 
\begin{equation}
\Delta \equiv \frac{(1-\alpha^2)\l(\frac{r}{r_b}\r)^{\frac{2(1-\alpha)}{1+\alpha}}}{1-\alpha^2 \l(\frac{r}{r_b}\r)^{\frac{2(1-\alpha)}{1+\alpha}}}.
\end{equation}
Next we define the transformation $x \equiv (r/r_b)$ to obtain
the normalised radial gradient of the deviation: 
\begin{equation}
\pp{\Delta}{x}=\frac{2(1-\alpha)^2
x}{\l(x^{\frac{2\alpha}{1+\alpha}}-\alpha^2 x^{\frac{2}{1+\alpha}}\r)^2}.
\end{equation}
For a good approximation to the isothermal equation of state $p=
\alpha \rho$, we suggest two necessary limits: (a) $\Delta \to 0$
which scales the pressure-density dependence and (b)
$|d\Delta/dx| \ll 1$, which expresses the 
fact that in an ideal isothermal sphere, the form of the equation of
state is independent of position.  

\begin{figure}[!bth]
     \psfrag{A}{$\alpha$}
     \psfrag{X}{$x_c$}
\begin{center}
\includegraphics[scale=0.8]{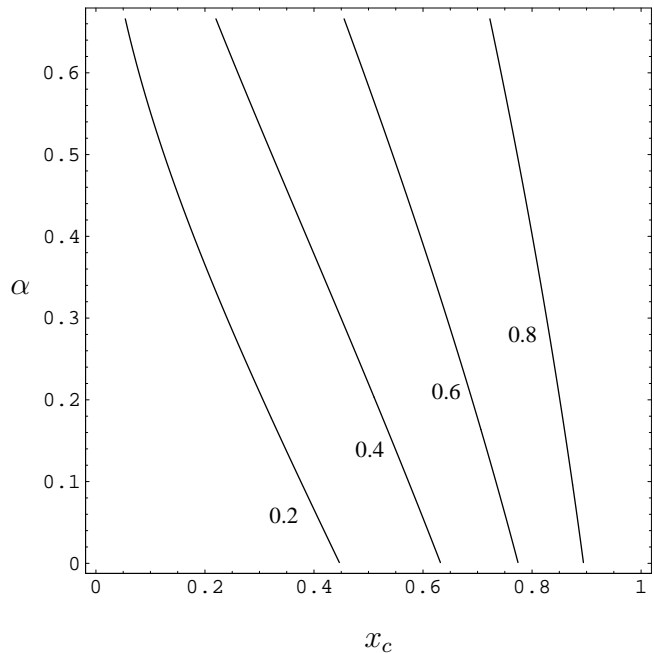}
\caption{Contour plot of the deviation term $\Delta$. 
     The plot gives the  characteristic core region where the
     deviation from the perfect isothermal equation of state is less
     than a particular contour value.
     Note that although the ratio of the core radius to the boundary radius
     $x_c=(r_c/r_b)$ decreases with increasing $\alpha$ (and hence with
     step number), both $r_c$ and $r_b$ are monotonically increasing with
     each step. The decreasing $x_c$ merely reflects their relative rates
     of increase. \label{chapfig5-2}}
\end{center}
\end{figure}

Fig. \ref{chapfig5-2} shows a contour plot of the deviation term
$\Delta$. We see that for a particular value of the constant $\alpha$
the approximation gets
worse with increasing distance from the center, as the radius-dependent
perturbation becomes more significant. The plot gives a characteristic
core region where the deviation from the perfect isothermal equation
of state is less than a particular contour value. Note that although
the ratio of the core radius to the boundary radius
$x_c=(r_c/r_b)$ decreases with increasing $\alpha$, at later
evolutionary steps, 
both $r_c$ and $r_b$ are monotonically increasing with each step. The
decreasing $x_c$ merely reflects their relative rates of increase.

\begin{figure}[!bth]
     \psfrag{C}{$\dd{\Delta}{x}$}
     \psfrag{XAXIS}{$x$}
     \psfrag{YAXIS}{$\alpha$}
     \psfrag{ZAXIS}{$\frac{d\Delta}{dx}$}
\begin{center}
\includegraphics[scale=0.75]{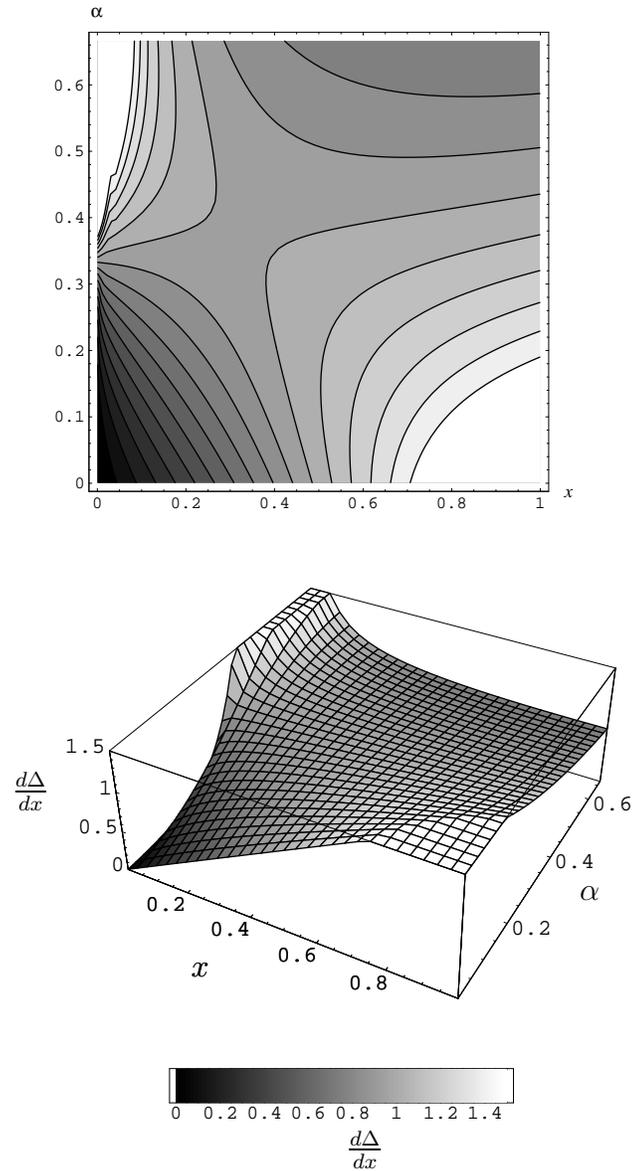}
\caption{Contour plot of the derivative of the deviation term
     $(d \Delta /dx)$. The plots gives the core region (small
     $\alpha$, $r \ll r_b$) where the derivative is less than a
     particular value and the ratio of pressure to density
 is therefore nearly independent
     of position. \label{chapfig5-3}}
\end{center}
\end{figure}

Fig. \ref{chapfig5-3} shows diagrams of the radial derivative of the
deviation term $d\Delta/dx$.
This derivative becomes very small for both small $\alpha$ and small
$r/r_b$. For large $\alpha$ and $r/r_b$ there is a significant variation
of the equation  of state with position. The derivative $d\Delta/dx$ shows the
complex functional form expected for a finite sphere. 	

The contour plots of Figs. \ref{chapfig5-1} - \ref{chapfig5-3} indicate that
there exists a core region within our generalised sphere for which the
metric and the equation of state are close to the ideal isothermal limit. 
There are several alternative ways of defining this core region. All
of these arbitrary definitions, however, have in common that they
describe the same qualitative picture. In that picture our metric
describes a sphere which tends to the perfect isothermal limit at the
centre, has a core region where it is close to the isothermal state
and then a transition region approaching the boundary where the
isothermal approximation breaks down. 

For small $\alpha$, a good approximation to the ideal isothermal
equation of state requires
\begin{equation} \label{e:approx0}
\Delta \equiv k = \l(\frac{r}{r_b}\r)^{\frac{2(1-\alpha)}{1+\alpha}}\ll 1.
\end{equation}
We may ask for the range of $r < r_{c}$ (relative to
$r_{b}$) that satisfies a fixed constraint, for example, $k_{c} \equiv const
=(1/3)$. This gives a core sphere, $r_{c}$, within the isothermal
sphere for which the 
approximation does not deviate by more than the set limit. This critical
core radius is given by
\begin{equation} \label{e:approx}
r_{c} = k_{c}^{\frac{1+\alpha}{2(1-\alpha)}}~r_b.
\end{equation}
As $r \to 0$, Eq. $\eqref{e:approx0}$ implies that $\Delta \to 0$,
and the approximation to the isothermal equation of state goes to
its ideal limit. For larger $r$ the approximation gets continuously
worse at a rate described by Eq. $\eqref{e:approx0}$.
Both the outer radius $r_b$ and the core radius $r_c$ grow
with each step as $\alpha$ and $A$ change. Moreover, from equation
$\eqref{e:approx}$, we see that $r_b$
expands faster than $r_c$ and the `wobble' region between the
isothermal core and the boundary radius grows.

\section{Thermodynamic Consequences implied by the Generalised
Equation of State} \label{chap5-4}
Although an isothermal equation of state is well-known to be a
reasonable zeroth-order approximation for rich clusters of galaxies
(like the Coma cluster) it is also known not to be realistic,
especially in the outer regions, for clusters with finite
boundaries. From a physical point of view, this is because the longer
dynamical relaxation timescales at the lower densities in the outer
regions do not allow sufficient energy exchange for
thermalization. From a mathematical point of view, the departure from
isothermality is forced by the outer boundary at a finite radius. 
  
The generalised equation of state can be written as
\begin{equation}\label{e:generaleqs}
P= \alpha_{\text{eff}}(r)\cdot\rho
\end{equation}
where
\begin{equation}\label{e:fraceqs}
\alpha_{\text{eff}}(\alpha, r) \equiv \alpha \cdot
\frac{1-(\frac{r}{r_b})^{\frac{2(1-\alpha)}{1+\alpha}}}{1-\alpha^2(\frac{r}{r_b})^{\frac{2(1-\alpha)}{1+\alpha}}}  
\end{equation} 
We determine the limits: (i) $r=0$, $\alpha_{\text{eff}} \to \alpha$,   
(ii) $(r/r_b) \ll 1$, $\alpha_{\text{eff}} \approx \alpha$, and
(iii) $(r/r_b) \to 1$, $\alpha_{\text{eff}} \to 0$. Note that we have
implicitly introduced a new definition of $\alpha$: 
\begin{equation}
\alpha \equiv \frac{P(r=0)}{\rho(r=0)}
\end{equation}
i.e. $\alpha$ characterizes the constant of proportionality between
pressure and density {\it at the centre} of the sphere. 

Eq. $\eqref{e:generaleqs}$ can be regarded as an 
isothermal equation of state with a temperature that depends on $r$ according
to the fractional term in $\eqref{e:fraceqs}$.  This type of
temperature behavior is quite common in problems involving boundary
conditions. In our case, the pressure is zero at the boundary since
the temperature is zero there for a good physical reason, namely the
galaxies there have not relaxed and randomised their peculiar
velocities significantly since the density of 
galaxies is low and there are fewer gravitational scatterings.  The peculiar
motions of galaxies near the boundary are mainly streaming motions, probably
mostly infall into the approximately isothermal cluster.

Note that for any constant $r$ (or given temperature), Eq. $\eqref{e:generaleqs}$ 
shows that  $dP/d \rho$ is positive so this equation of state has a positive
isothermal compressibility, as needed for mechanical stability.  Because
Eq. $\eqref{e:generaleqs}$ is close to an isothermal equation of state,
and equivalent to it at any fixed radius, it should have reasonable
thermodynamic properties as long as the number density of galaxies is
large enough that one can define the thermodynamic properties in a
volume smaller than that of the range of radii one is considering.
This is similar (though not exactly the same) as defining the local
temperature in a gas which has a temperature gradient. 

 All that the equation of state is saying is that the temperature is a
   function of position.  This is similar to the case of polytropes in stars,
   where it is sometimes generalised to include inhomogeneous polytropes
   \cite{chandra2}.  In real stars the temperature
   dependence of the equation of state on position occurs because of the
   nuclear reactions and energy transport processes.  In our largest cluster
   it occurs because the relaxation timescales are dependent on the local
   density.  Calculating this is a separate problem.  For the equation of state
   to be physically acceptable, it must satisfy the first and second laws of
   thermodynamics locally.  An isothermal equation of state does this, although
   our case is complicated by the long range nature of gravity in the cluster.
   We can regard the isothermal case with variable temperature as an
   approximation, just as it is in stellar structure.

\section{A Solution of an Expanding, Centrally \\ Isothermal Universe}
\label{chap5-5} 
In  this section,  we derive  a hybrid  metric which  encompasses the
behavior  of  both  an  expanding  universe and  a  static  nearly isothermal
core. Physically, we consider the evolution of such a hybrid metric to
occur through three stages. First,  the universe follows the usual 
globally  homogeneous   Einstein-de  Sitter  expansion,   which  tends
asymptotically   to  stationarity,   i.e.   the  expansion   timescale,
$a/\dot{a}$, of the universe tends  to infinity at late times. Second,
the  condensation stage  (which  is attributed  to galaxy  clustering)
gradually  produces a significant  cosmological pressure  which alters
the initial  global homogeneity and supports  a radially inhomogeneous
global  distribution of  matter. Consequently,  the equation  of state
changes  with time, resulting  in a  dynamical  feedback on  the
metric. Third, the postulated asymptotic final state is the
generalised isothermal sphere  with a  boundary characterized  by  the
equation  of state  in Eq. $\eqref{e:eqs2}$.
Our solution only applies to stages one and three. It only describes
the relaxed states of the universe with a static approximately isothermal core. The
temporal evolution will be a sequence of such relaxed, static
stages. We do not have a continuous description in time, but only
discrete steps between quasi-static periods with constant $\alpha$. 
Our solution breaks down in between.
We note that the sequence of hybrid metrics (with different values of
the constants $\alpha$ and $A$) is constrained by various boundary conditions
in both space and time, namely, (a) at early times (for $t=0$) the metric
is FRW for all $r$, and (b) at late times (for $t \to \infty$) the
universe would have a bounded approximately isothermal metric with a
horizon expanding proportional to $t$, and be FRW on large scales ($r
\to \infty$) which increase at $t^{(2/3)}$.  

A possible counter argument to this proposal would be the following:
the horizon of the Einstein-de Sitter can expand rapidly enough that
the matter inside is not consolidated into a single bound system, no
matter how long one waits. The only loophole is if the primordial
perturbation spectrum tilts towards long wavelengths. However it would
seem at first sight that the scenario we propose is not valid. 

However, the key point against the skeptic's argument is that we are
considering the very non-linear case of structure formation after very long
times when the clustering parameter $b \to 1$ and nearly all the
matter is in a few dominant
massive clusters. The metric still expands at the Einstein-de Sitter rate which
is proportional to $t^{(2/3)}$ since this is determined to a first
approximation by an average over all horizons, so the approximate
expansion timescale, $t_{expansion}$, is given by the usual
\begin{equation}
t_{expansion} = \frac{2a}{3 \dot{a}} = \frac{1}{\sqrt{ 6 \pi G
\bar{\rho}_{crit} }}~.
\end{equation}
We wish to compare this expansion with the timescale, $t_{merge}$, for
the clusters to merge. We are interested in the {\bf non-linear}
merging of smaller galaxy clusters with a larger cluster in the
expanding Einstein-de Sitter Universe. In the simplest approximation,
as an example to estimate the basic merging timescale, $t_{merge}$, for
collapse into a single cluster, consider a massive cluster surrounded
by a roughly spherical distribution of less massive clusters, where
the whole configuration has an average large initial overdensity,
$\bar{\rho}_{i}$, compared to the initial critical average density
$\bar{\rho}_{ci}$ in that volume. For simplicity, suppose the clusters
have negligible initial comoving velocity. We know that this region
which has gradually become overdense will subsequently expand more
slowly than the universe, turn around and collapse leading to a merger
of the cluster. 

To estimate the collapse (merger) timescale, we use the illustrative
example in \cite{saslaw3} (See Eqs. 
(37.10)--(37.19) in the reference). For $\bar{\rho}_i \gg \bar{\rho}_{ci}$
(and thus $E \approx 1$ where $E = a_{\text{max}}^{-1}$), the time
scale to maximum expansion is  
\begin{equation}
\frac{\pi}{2} = t_{max}\sqrt{\frac{8 \pi G \bar{\rho}_i}{3}}
\end{equation}
where $\bar{\rho}_{i}$ is the initial average density in a volume
containing the clusters, and
\begin{equation}
t_{max}^2 = \frac{\pi^2}{4} \frac{3}{8 \pi G \bar{\rho}_i} = \frac{3
\pi}{32 G \bar{\rho}_i}~.
\end{equation}
Thus
\begin{equation}
\frac{t^2_{max}}{t_{expansion}^2} \approx  \frac{3
\pi}{32 G \bar{\rho}_i} \times (6 \pi G \rho_{crit}) = \frac{9
\pi^2}{16} \frac{\bar{\rho}_{crit}}{\bar{\rho}_i} \approx 5.65
\frac{\bar{\rho}_{crit}}{\bar{\rho}_i} 
\end{equation}
and we obtain
\begin{equation}
t_{max} \approx 2.36
\left(\sqrt{\frac{\bar{\rho}_{crit}}{\bar{\rho}_{i}}}\right)
t_{expansion}~.
\end{equation}
If the clusters are a factor of $\sim 2$ closer than the average separation,
$\bar{\rho}_i \approx 8 \rho_{crit}$ and $t_{max} \approx 0.8
t_{expansion}$. The timescale for the clusters to merge is less than
about $2 t_{max}$. If the clusters have initial velocities (which they will
probably acquire while they form) that are mainly toward the most
massive cluster, then this merging
timescale is shorter. Therefore, any clusters that have less than about half
of the average separation of clusters (i.e. where
$\bar{\rho}_i \gtrsim 8 \rho_{crit}$) are likely to merge within one
expansion timescale. Clusters with larger initial separations will
merge after longer times. On these longer timescales, the horizon
scale will double on a timescale equal to $t_{expansion}$, which
becomes infinite. This may or may not be the timescale to introduce
new clusters into a particular horizon,  depending on where 
they are located in the adjacent horizons. In the asymptotic limit, $t
\to \infty$, which we are interested for this chapter, an arbitrary large
horizon can contain an arbitrary large dominant cluster, separated
from such clusters in other horizons by arbitrarily large
distances. 

What is the effect of the matter that enters the horizons?  At any given
time within a horizon centred on a dominant supercluster, the galaxies not
in the supercluster will be accreted by it on a timescale less than the 
expansion timescale.  If the accretion is nonlinear, e.g. if the supercluster
dominates the local gravitational field, this accretion timescale is shorter
than the average expansion timescale by a factor of about $\sqrt{2}$.  At any
finite time, the infinite E-deS universe will contain an infinite number of 
these expanding horizons.  The positions of their central superclusters
throughout the whole universe may have a Poisson distribution since they have
not had time to influence each other.  However, they could, have long
wavelength primordial 
perturbations which would speed up their clustering.  Some of the
horizons at any given time will be dominated by one supercluster,
others may contain more, or may contain merging superclusters.
Outside any of these horizons, the other clusters may be averaged over
an increasingly large scale to give the effective average density of
the E-deS universe. 

     In the limit of infinite future time (which is what we consider), for a
Poisson distribution of superclusters initially outside each other's horizons,
they come within each other's horizons and can merge into our isothermal
sphere solution which grows to ultimately incorporate all the matter of the
universe in the limit of infinite time.  If long wavelength initial 
perturbations are important, this could happen sooner.  In either case, the
average separation of superclusters increases more slowly than the horizon
expands, which is what our model requires.

\subsection{The derivation of the hybrid metric}

Bounded isothermal universes with non-zero pressure (but with
zero-pressure and non-zero density at the boundary at $r=r_b$) cannot be matched
directly to the Einstein-de Sitter model across a specific
$r=constant$ hypersurface. However, one can 
achieve a hybrid solution by first matching the isothermal metric to a
Schwarzschild exterior. This is physically justified by Birkhoff's
theorem. A formal mathematical treatment is given by
Fayos et al~\cite{fayos} for the general matching of two spherically symmetric
metrics for various pressure and density conditions. 
One can perceive the Schwarzschild metric as a kind of
membrane (equivalent to the requirement of a closed trapped surface 
discussed by Fayos et al~\cite{fayos} for the spherically symmetric
solution, such as the Vaidya metric, to match onto the FRW metric)
which facilitates the matching between the {\it bounded static
isothermal sphere} and the {\it expanding Einstein-de 
Sitter universe}. Such a procedure would be analogous to the
astrophysical case for an isothermal star whose external
Schwarzschild metric can be embedded in its small region of the universe. 

The continuity condition of the metric coefficients, requires
the following relations to be satisfied:
\begin{align}
e^{\nu(r=r_{b})} &=
\l(Cr_b^{\frac{2\alpha}{1+\alpha}}-Dr_b^{\frac{2}{1+\alpha}}\r)^2= 1 -
\frac{2m}{r_{b}}, \\
e^{\lambda(r=r_{b})} &= 1+\frac{4\alpha}{(1+\alpha)^2}= \l( 1 -
\frac{2m}{r_{b}} \r)^{-1}. 
\end{align}
where $r_b$ is defined by Eq. $\eqref{e:rb}$.
These provide the matching relation between $C$ and $D$:
\begin{equation}
D(\alpha)= E(\alpha)C^{\frac{1}{\alpha}},
\end{equation}
where
\begin{equation}\label{e:edef}
E(\alpha)\equiv \l[1+\frac{4\alpha}{(1+\alpha)^2}\r]^{\frac{1-\alpha}{2\alpha}}\alpha^2\l[1-\alpha^2\r]^{\frac{1-\alpha}{\alpha}}.
\end{equation}
Substituting the matching condition into Eq. $\eqref{e:rb}$ we find 
\begin{equation}\label{e:rb2}
r_{b}(\alpha) = F(\alpha)A^{-\frac{1+\alpha}{4\alpha}}
\end{equation}
where $A \equiv C^2$ and 
\begin{equation}\label{e:fdef}
F(\alpha) \equiv \l(\frac{\alpha^2}{E(\alpha)}\r)^{\frac{1+\alpha}{2(1-\alpha)}} 
\end{equation}
with $E(\alpha)$ defined in Eq. $\eqref{e:edef}$. The matching eliminates
the constant $D$ and therefore shows that the boundary radius $r_b$ is
only dependent on the parameters $\alpha$ and $A$. Hence, we also
deduce immediately that the generalised isothermal metric $\eqref{e:interior}$
is parameterised by the two constants $\alpha$ and $A$.

This procedure also determines a Schwarzschild mass from the matching
condition:
\begin{equation}\label{e:malpha}
m(\alpha) = \l(\frac{2\alpha}{4\alpha + (1+\alpha)^2}\r) r_b = M_{iso}
\end{equation}
Hence, the Schwarzschild mass $m(\alpha)$ is equal to the mass of the
bounded generalised isothermal sphere, $M_{iso}$, as found
independently by direct integration of the density. 
The equality of the results indicates consistency in our approach.
The corresponding Schwarzschild radius is
\begin{equation}
r_s = 2m = \l(\frac{4\alpha}{4\alpha + (1+\alpha)^2}\r) r_b, 
\end{equation}
which indicates that $r_s < r_b$, i.e. the Schwarzschild radius is 
always inside the bounded isothermal sphere (as is the case for a star).

To complete the picture, we adapted the method in McVittie \cite{mcvittie} to
enable us to embed the finite generalised isothermal sphere with given
properties in the Einstein-de Sitter model. A sequence of embeddings
with different properties is later used to describe the time evolution
of the universe. McVittie's solution applies to a mass particle in an
expanding universe and has been applied by Noerdlinger and Petrosian
in \cite{noerdlinger} to investigate the motion of galaxies in galaxy
clusters under the influence of cosmic expansion. In \cite{mcvittie},
he found a way of embedding a Schwarzschild 
geometry in a FRW spacetime. Having matched the generalised isothermal
metric to a Schwarzschild exterior, we use McVittie's approach to
embed this exterior spacetime in an expanding Einstein-de Sitter
cosmological background. The most general form of the McVittie metric was
summarised in a later paper by McVittie in \cite{mcvittie2} which
incorporates a larger class of such solutions. 

Hence, we find 
\begin{equation}\label{e:exterior}
ds^2= -\l(
 \frac{1-\frac{m(\alpha)}{2a\bar{r}}}{1+\frac{m(\alpha)}{2a\bar{r}}}\r)^2
dt^2 + \l(1+\frac{m(\alpha)}{2a\bar{r}}\r)^4 a^2(t)
\l[ d\bar{r}^2+\bar{r}^2(d\theta^2+\sin^2\theta d\phi^2) \r]  
\end{equation}    
for the exterior ($r > r_b$) matched solution of a generalised
isothermal sphere of mass
$m(\alpha)$ in an expanding FRW universe. Note that the solution is
stated in isotropic coordinates. Here $a(t)$ is the scale
factor of the standard cosmological solution and $m(\alpha)$ is
defined by Eq. $\eqref{e:malpha}$. 

If one makes the substitution $\bar{r}_1=a\cdot \bar{r}$, the 
line element in Eq. $\eqref{e:exterior}$, {\it at any particular instant
$t_1$}, reduces to the Schwarzschild line element for small $\bar{r}$
(isotropic coordinate). McVittie also showed that the transformation
of the time-dependent metric in Eq. $\eqref{e:exterior}$ into the static
Schwarzschild field is independent of the moment in time at which the
transformation is made in \cite{raychaudhuri}. This
proves that the embedding is valid for all times $t$.

To relate the interior generalised isothermal metric and the exterior
 metric, we give the transformation between the spherical
polar radial coordinate $r$ used in Eq. $\eqref{e:interior}$ and the
isotropic coordinate $\bar{r}_1$ in Eq. $\eqref{e:exterior}$:
\begin{align}
\label{e:isotropicr}
r &\equiv \bar{r}_1\l(1+\frac{m}{2\bar{r}_1}\r)^2,\\
\bar{r}_1&\equiv \frac{1}{2}\l[\sqrt{r^2-2mr}+(r-m)\r].
\end{align}
It can then be shown that the interior and exterior metrics match at
$r=r_b$, for a particular instant $t_1$ and that the matching is valid
for all $t$ in \cite{nolan}.  For a given value of
$\alpha=\text{constant}$, the derivatives of the metric
coefficients also match according to the junction conditions described
by Bonnor and Vickers \cite{bonnor}. In the next section, we will
discuss the matching of the metrics in a more rigourous manner. 

At the beginning of this section we discussed the limits in space and
time that the sequence of hybrid solutions has to satisfy.  
We now compare these constraints to the
corresponding limits of Eqs. $\eqref{e:interior}$ and
$\eqref{e:exterior}$. For vanishing $\alpha$, the mass and the boundary radius
of the isothermal sphere tend to zero and the metric is FRW on all
scales. For small $\alpha$ (i.e. early evolutionary steps) the
applicable hybrid metric has
the generalised isothermal form for small $r$ and is FRW for large
$r$. For relatively large $\alpha$ (i.e. late evolutionary steps) the
metric tends  to the isothermal form on much larger scales. We therefore see that
all limits of our hybrid solution in both time (discrete steps between
static periods) and space agree with
the limits required by physical arguments.  

\subsection{The validity of the matching of the metric}
We may ask more generally whether using the generalised isothermal
sphere parameters in the Schwarzschild interior metric could lead to
a problem in matching the FRW exterior. Mars\cite{mars} has shown that
{\it the only static, spatially compact, vacuum region in a 
Friedman-Lemaitre spacetime must be a 2-sphere comoving with the
cosmological flow and with the Schwarzschild-exterior geometry.}
In our model, at the boundary between the Schwarzschild metric and
the bounded isothermal sphere, there are no problems that arise from the
energy momentum tensor. This is because at that boundary between the
bounded isothermal sphere and the Schwarzschild exterior, the pressure
$P$ is  zero. Therefore, there is no problem embedding the matched
system of a bounded isothermal sphere and the Schwarzschild metric
into the FRW metric, if the solution of the combined system is
expanding. 

Since the energy momentum tensor is not an issue in the matching, 
we can work out the intrinsic and extrinsic curvatures of the metric
by matching on a hypersurface $\Sigma$, at a constant comoving radius, $r$.
The smooth matching of the 2 spacetimes across a hypersurface $\Sigma$
is guaranteed if the Darmois junction conditions  are satisfied, i.e. if
the first fundamental (intrinsic metrics) forms and second fundamental
(extrinsic metrics) forms are calculated in terms of the coordinates
on the hypersurface, and they are identical on both sides of the
hypersurface. If both the first and second fundamental forms are
continuous on $\Sigma$, then the matching is not piecewise i.e. it
will match for all times rather than just at  single instants. The
Darmois junction conditions allow us to use different coordinate
systems on both sides of the hypersurface. We adopt the procedure in
\cite{dyer} and have  calculated explicitly the interior and the
exterior metrics and they match on a 3-dimensional timelike tube and
exhibit no curvature discontinuity for all times. In the process, we
applied the procedure in \cite{dyer} to check the conditions for matching the 
Schwarzschild-Tolman system and the FRW metric on a hypersurface
$\Sigma$, where $r=r_0$ for the radial constant comoving coordinate in
the FRW metric. 

We recall the Schwarzschild-Tolman metric (bounded isothermal sphere
interior with a Schwarzschild exterior), which is written in
coordinates $(T,~\rho,~\theta,~\phi)$, 
\begin{equation} \label{e:stmetric}
ds^2 = -\l(1- \frac{2M(\alpha)}{\rho} \r) dT^2 + \l(1-
\frac{2M(\alpha)}{\rho} \r)^{-1} d \rho^2 + \rho^2 d \theta^2 + \rho^2
\sin^2 \theta d \phi^2 \;,
\end{equation}
where 
\begin{equation} 
M(\alpha) = \frac{2 \alpha}{4 \alpha +(1+\alpha)^2} r_b \;,
\end{equation}
and $r_b$ is a constant. The general FRW metric is given in Eq.
$\eqref{e:frw}$. 

The first fundamental form is the metric which the hypersurface
$\Sigma$ inherits from its spacetime in which it is imbedded:
\begin{equation} \label{e:firstform}
\g_{ij} = g_{\mu \nu} \pp{x^{\mu}}{u^i} \pp{x^{\nu}}{u^j} \;,
\end{equation}
where $u^i=(u^1, u^2, u^3)=(u,v,w)$ is the coordinate system defined
on the hypersurface. Note that $i,j$ denotes only the coordinates of
the hypersurface of 3-dimensions while $\mu, \nu$ denotes the
coordinates of the 4-dimensional spacetime. The second fundamental
form is defined by 
\begin{equation} \label{e:secondform}
\Omega_{ij} = (\Gamma^{\mu}_{\nu \lambda} n_{\mu} - \p_{\lambda}
n_{\nu}) \pp{x^{\nu}}{u^i} \pp{x^{\lambda}}{u^j} \;,
\end{equation}
where $n_{\mu}$ is the unit normal to the hypersurface,
$\Sigma$. Finally if $\Sigma$ is given by the function
$f[x^{\mu}(u^i)]=0$, then $n_{\mu}$ can be calculated by: 
\begin{equation} \label{e:normal}
n_{\mu} = - \frac{\p_{\mu} f}{\sqrt{g^{\alpha \beta} \p_{\alpha} f
\p_{\beta} f}}
\end{equation}
We will denote the FRW and Schwarzschild-Tolman (ST) frames by the
subscripts or brackets with F and S in our quantities. 

We consider a spherical hypersurface $\Sigma$ given by the function
$f_F (x^{\mu}_F)=r-r_0=0$, where $r_0$ is constant. 

For the FRW frame
\begin{equation}
x^t_F = t = u, \quad x^r_F = r = r_0, \quad x^{\theta}_F=\theta=v,
\quad x^{\phi}_F = \phi=w \;,
\end{equation}
and similarly for the ST frame,
\begin{equation}
x^t_S = T = T(u), \quad x^{\rho}_S = \rho= \rho(u), \quad
x^{\theta}_S=\theta=v, \quad x^{\phi}_S = \phi=w \;. 
\end{equation}
The first matching condition $\g_{ij} (F) = \g_{ij} (S)$ gives
\begin{align}
\label{e:eqn1}
\l(1- \frac{2M(\alpha)}{\rho} \r) \l(\dd{T}{u}\r)^2 -\l(1-
\frac{2M(\alpha)}{\rho} \r)^{-1} \l(\dd{\rho}{u} \r)^2 &= 1 \;,\\ 
\label{e:eqn2}
a(t)^2 r_0^2 &= \rho^2.    
\end{align}
Next we calculate the second fundamental forms. The (outward
pointing) unit normal in the FRW frame follows from
$\eqref{e:normal}$, and since $f_F (x^{\mu}_F)=r-r_0=0$, we get
$n_{\mu} (F) = \delta^{r}{}_{\mu} n_{r} (F)$. 

For the unit normal in the Schwarzschild case, we cannot derive the
result directly from Eq. $\eqref{e:normal}$, but we use the result that
$n_\mu (S)$ must satisfy the following conditions:
\begin{align}
\label{e:cond1}
n^{\mu} (F) n_{\mu}(F) &= n^{\mu} (S) n_{\mu}(S) = 1 \;,\\
\label{e:cond2}
n_{\mu} (S) \pp{x^{\mu}_S}{u^i} &= 0 \;,
\end{align}
where the second condition comes from the fact that partial
differentiation of $f_S [x^i_{S} (u^i)]=0$ with respect to $u^i$. With 
these conditions we get the following equations in component form:
\begin{align}
n_{\theta} (S) = n_{\phi}(S) &= 0 \;, \\
n_T (S) \dd{T}{u} + n_{\rho} (S) \dd{\rho}{u} &=0 \\
\l(1- \frac{2M(\alpha)}{\rho} \r)^{-1} n_T^2 (S)  -\l(1-
\frac{2M(\alpha)}{\rho} \r) n_{\rho}^2 (S) &= -1 \;.
\end{align}
Comparing the last equation with Eq. $\eqref{e:eqn1}$, we derive
$n_{\mu}(S)$ to be
\begin{equation}
n_{\mu} (S) = \l(- \dd{\rho}{u},~ - \dd{T}{u},~0,~0 \r) \;.
\end{equation}

We can compute $\Omega$ from Eq. $\eqref{e:secondform}$ to be
\begin{equation}
\Omega_{ij} = -\frac{1}{2} \frac{1}{\sqrt{g_{rr}} } \p_r g_{ij} \;,
\end{equation}
and this equation is valid for any coordinate hypersurface
$r=r_0=$constant, in an orthogonal coordinate system and parametrised
by $x^{i}=u^i$. 

From Eq. $\eqref{e:cond2}$, we obtain the following equation by
differentiating with respect to $u^i$, 
\begin{equation}
\p_{\mu} n_{\nu}(S) \pp{x^{\mu}_S}{u^i} \pp{x^{\nu}_S}{u^j} = -
n_{\mu} \frac{\p^2 x^{\mu}_s}{\p u^i \p u^j} \;,
\end{equation}
and we get 
\begin{equation}
\Omega_{ij} (S) = \Gamma^{\mu}_{\nu \lambda} n_{\mu} \pp{x^{\nu}}{u^i}
\pp{x^{\lambda}}{u^j} + n_{\mu} \frac{\p^2 x^{\mu}_s}{\p u^i \p u^j}.
\end{equation}

We know that $\Omega_{ij} (S)=\Omega_{ij}(F)=0$ for $i \neq j$, and for
the remaining diagonal components, we have $\Omega_{ij}
(S)=\Omega_{ij}(F)$ on $\Sigma$, and we get the following three
differential equations:
\begin{align}
\label{e:de1}
\Omega_{uu} (S) &\equiv  \Omega_{uu}(F) = \Gamma^{\rho}_{TT} n_{\rho}
\l(\dd{T}{u}\r)^2 + \Gamma^{\rho}_{\rho \rho} n_{\rho} \l(\dd{\rho}{u}
\r)^2 + 2 \Gamma^{T}_{T \rho} n_{T} \dd{T}{u} \dd{\rho}{u} + n_T
\ddp{T}{u} + n_{\rho} \ddp{\rho}{u} = 0 \;, \\
\label{e:de2}
\Omega_{vv} (S) &\equiv \Omega_{vv} (F) = \zeta |a(t)| r_0 =
\Gamma^{\rho}_{\theta \theta}(S) n_{\rho} \;, \\
\label{e:de3}
\Omega_{ww} (S) &\equiv \Omega_{ww} (F)= \zeta |a(t)| r_0 \sin^2 \theta=
\Gamma^{\rho}_{\phi \phi}(S) n_{\rho} \;,
\end{align}
where in the Schwarzschild frame, the Christoffel connections are
\begin{align}
\Gamma^{T}_{T \rho} &= -\Gamma^{\rho}_{\rho \rho} =
\frac{M(\alpha)}{\rho(\rho-2 M(\alpha))} \;, \\
\Gamma^{\rho}_{TT} &= \frac{(\rho-2M(\alpha)) M(\alpha)}{\rho^2} \;,\\
\Gamma^{\rho}_{\theta \theta} &= - (\rho - 2M) \;,\\
\Gamma^{\rho}_{\phi \phi} &= \Gamma^{\rho}_{\theta \theta} \sin^2
\theta \;,
\end{align}
and $\zeta \equiv \sqrt{1- k r_0^2}$. Note that this will also prove
that the matching is continuous for the cases where $k \neq 0$,
i.e. open and closed cases as well. In retrospect, if we follow
the procedure of Hogan\cite{hogan}, we will also obtain the generalization of
the Schwarzschild-Tolman metric with the open and closed FRW
universes, but their forms will be complicated. By the same physical
reasons mentioned in the discussion of the infinite isothermal sphere
solution, the open and closed FRW universes would deter the phase
transition from happening.   

Using Eqs. $\eqref{e:de2}$, $\eqref{e:eqn2}$ and
$\eqref{e:eqn1}$, we get
\begin{align}
\label{e:dTdu}
\dd{T}{u} &= \frac{\zeta \rho}{\rho - 2M(\alpha)} \\
\label{e:drhodu}
\l(\dd{\rho}{u} \r)^2 &= \zeta^2 - \l(\frac{\rho-2M(\alpha)}{\rho} \r).
\end{align}
Differentiating Eqs. $\eqref{e:dTdu}$ and $\eqref{e:drhodu}$ gives
\begin{align}
\ddp{T}{u} &= - \zeta \frac{2M(\alpha)}{(\rho - 2M(\alpha))^2}
\dd{\rho}{u} \\
\ddp{\rho}{u} &= - \frac{M(\alpha)}{\rho}.
\end{align}
Because $M(\alpha)=$constant for values of $\alpha$, we can substitute
Eqs. $\eqref{e:dTdu}$ and $\eqref{e:drhodu}$ into
$\eqref{e:de1}$ to show that Eqs. $\eqref{e:de1},
\eqref{e:de2}$ and $\eqref{e:de3}$ are consistent. From these
solutions, one would realise that both the first and second
fundamental forms are continuous on $\Sigma$, hence the matching is
not piecewise, i.e. it will match for all times rather than just at
single instants.  

\section{Evolution of the Isothermal Universe - A Sequence of Hybrid
Solutions} \label{chap5-6}

Thus far, we have derived the hybrid metric to describe a generalised
isothermal sphere embedded via a ``Schwarzschild membrane'' in the expanding
Einstein-de Sitter universe at any particular cosmic time. Now we
consider briefly how such a hybrid model might form and evolve. 

Greater than average densities in any volume of an Einstein-de Sitter
universe produce local gravitational clustering timescales shorter
than the current expansion timescale. This facilitates the growth and
merging of a small number of large clusters within an apparent or
particle horizon. Collective relaxation \cite{saslaw3} could
thermalize the velocities of the galaxies on a timescale $\sim
(G\bar{\rho})^{-\frac{1}{2}}$ where $\bar{\rho}$ is the average density
in a volume containing the merging clusters. Eventually, this would
lead to an approximately isothermal virialized cluster. This process
would continue indefinitely as other great clusters appear within
the horizon and merge.

\begin{figure}[!bth]
     \psfrag{Xaxis}{$\alpha$}
     \psfrag{Yaxis}{$\log(A)$}
     \psfrag{Zaxis}{$\log(r_b)$}
\begin{center}
\includegraphics{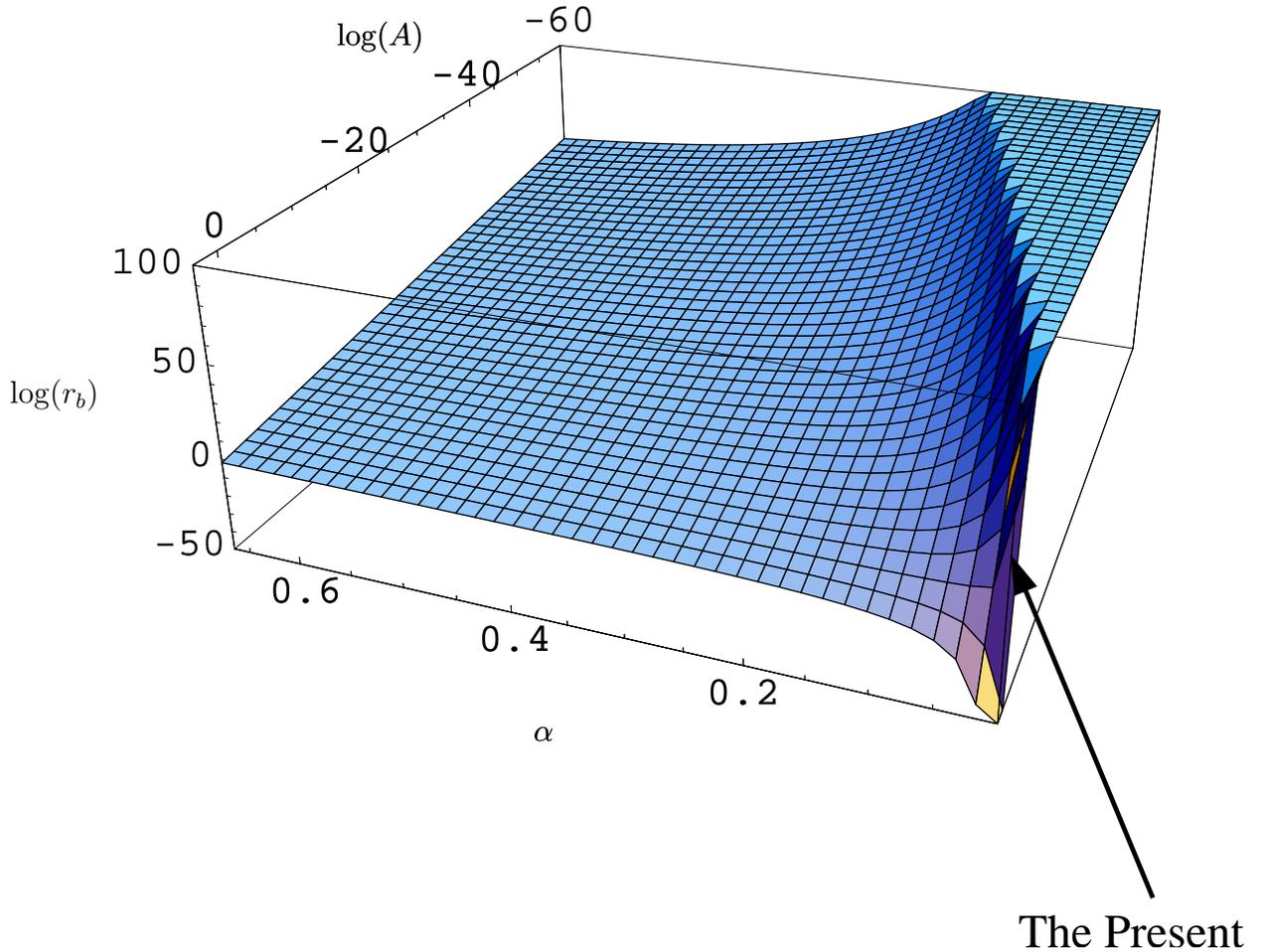}
\caption{Surface plot of $r_b$ -- the model's future laid out before us: It
     is interpreted as a map of the future scale of the generalised
isothermal universe for particular values of $\alpha$ and $A$. The
radius of the generalised isothermal sphere, $r_b$, is in units of
centimeters, $A$ is measured in $[cm^{-\frac{4\alpha}{1+\alpha}}]$ and
$\alpha$ is dimensionless. $r_b$ and $A$ are plotted on a logarithmic
scale. The hypothetical present size of the isothermal sphere is taken
to be $r_b \sim 10^{25}~\text{cm}$ (the size of the Coma
cluster).\label{chapfig5-4}}
\end{center} 
\end{figure}

During periods when one massive isothermal cluster dominates, our
idealised hybrid metric would apply. More generally, a much more
complicated metric would be needed to describe the periods when two or
more great clusters dominate a horizon. However, we may envisage
our hybrid metric applying at intervals, for particular values of
$\alpha$ and $A$ when one cluster dominates. ``Snapshots'' of the
long term evolution would then be characterized by a discrete sequence
of increasing values of $\alpha$ as virialization progresses, and
consequently by decreasing values of $A$, as seen in Fig.
\ref{chapfig5-4}.

How long will it take an isothermal universe to form? We can estimate
this timescale and compare it to timescales for several possible
competing effects. The timescale for the formation of the isothermal
sphere can be estimated from the evolution of $N$-body galaxy clustering.
For clustering in an adiabatically expanding universe this gives 
\cite{saslaw2}
\begin{equation}
a(t)=a_{0} \frac {b^{\frac{1}{8}}}{(1-b)^{\frac{7}{8}}},
\end{equation}
where $a$ is the time-dependent scale factor of the universe and $b$
is defined as the average ratio of the gravitational correlation
energy, $W$, to twice the kinetic energy, $K$, of peculiar velocities
in an ensemble of galaxies.
Inverting this gives the time-dependence of $b(t)$ within the limits
$b(0)=0$ (no interaction, perfect gas) and $b(\infty)=1$ (complete
virialization, isothermal).
The transition to the isothermal state therefore corresponds to a $b$
close to unity, where the universe is dominated by voids and a small
number of very large clusters.
We define $\tau$ as the time after which  the  void probability  of
the  universe  becomes greater  than $0.5$.   This  describes  the
timescale  required   for an approximately isothermal cluster to grow to
cosmologically significant size. From the galaxy distribution function in
\cite{saslaw2}, we get the condition
\begin{equation}
e^{-\bar{N} (1-b)} \ge 0.5,
\end{equation}
where $\bar{N}$ is the total number of average mass galaxies in the
observable universe. If we assume $\bar{N}\approx 10^{10}$, 
this allows us to  define the condition for transition to the
isothermal state as $b_{c}=1-10^{-10}$. Hence, we find the scale
factor $a_c(\tau)= 10^{8}a_{0}$, where $a_0$ is the present Hubble radius.
For an Einstein-de Sitter expansion with $a(t)=t^{\frac{2}{3}}$ this
corresponds to 
\begin{equation}
t_{exp} \approx 10^{12} t_0=3 \times 10^{29}~\text{seconds},
\end{equation} 
where we used an estimate for the present Hubble time $t_0=10^{17}$ seconds.
Hence, the universe would take approximately $10^{29}$ seconds to form
the isothermal sphere.

The solution permits a sequence of generalised isothermal universes
whose radii grow with each step according to
\begin{equation}\label{e:radius1}
r_{b}=\l(\frac{\alpha^2 C}{D}\r)^{\frac{1+\alpha}{2(1-\alpha)}}=
F(\alpha)A^{-\frac{1+\alpha}{4\alpha}},      
\end{equation}
where
\begin{equation}
F(\alpha)\equiv
\l(\frac{\alpha^2}{E(\alpha)}\r)^{\frac{1+\alpha}{2(1-\alpha)}}
\end{equation}
and $E(\alpha)$ is defined by Eq. $\eqref{e:edef}$. The values of $r_b$
 increase as $\alpha$ and $A$ {\it change their values with time}
 (discrete steps: for each stage $\alpha$ and $A$ are constant; see 
 Fig. \ref{chapfig5-4}). We explore next how this constrains the {\it 
time-dependence} through discrete steps of the sequence of
$\alpha$ and $A= C^2$ values.

We can normalize our metric by finding approximate limits on the
sequence of $A$.
These limits can be found by the following approximate argument:
$r_b=0$ at early times requires $\lim_{t \to 0}A_t\ge 1$ (see Fig.
\ref{chapfig5-4}). We next consider the limit for late times.
From the previous discussion we can set $r_b \sim a_c = 10^8a_0=
O(10^{36})\text{cm}$ as an approximate scale of the isothermal universe at
late times ($t=t_c$). 
  
This allows us to use Eq. $\eqref{e:radius1}$ and find $A_{t_c}$ at
$\alpha=(2/3)$, 
\begin{equation}
A_{t_c}\sim O(10^{-58})~\text{cm}^{-\frac{8}{5}},
\end{equation}
which gives an approximate limit for the time evolution of the
sequence of the constants $A_t$.
The proposed time evolution of $A_t$ therefore goes as follows:
$A$ starts off at $t=0$ with $A_0\ge 1$.   At $t_c$
(corresponding to $\alpha(t_c)=2/3$), $A_{t_c} \sim
O(10^{-58})~\text{cm}^{-\frac{8}{5}}$. For $t>t_c$, $A_t$ decreases further,
so that $\lim_{t \to \infty} A_t\to 0$ and $\lim_{t \to \infty} r_b(t)
\to \infty$, meaning that the isothermal sphere grows indefinitely.
Fig. \ref{chapfig5-4} should therefore be interpreted as a map of the future
scale of the isothermal universe for changes in $\alpha$ and $A$.
Note, however, that our present state of the universe is still very
close to $\alpha=0$, $r_b=0$. As the universe evolves in the far
future, changes of $\alpha$ and $A$ correspond to an increasing scale
$r_b$ of the isothermal part of the universe. This evolution as a
sequence of quasi-static steps can be characterized by discrete points
on the surface of Fig. \ref{chapfig5-4}.
 
There are competing  effects, which  could inhibit  the
formation  of the generalised isothermal state,  namely, (a) the
conversion of all the  matter into gravitational radiation, (b)
Hawking radiation and (c) proton decay and (d) a positive cosmological
constant. 

Next we estimate the amount of gravitational radiation produced by the
interaction between a pair of galaxies.  Using the standard quadrupole
formula  given by Landau and Lifschitz~\cite{landau} and  making
reasonable  estimates for masses and  mean distances  of galaxy
binaries  we find that  it would require about $10^{51}$ seconds for the
universe to convert all its mass into gravitational radiation. This is
$10^{22} t_{exp}$.  The mass which could be evaporated  by Hawking
radiation of  small black holes  in the time needed  to reach  the
isothermal phase  transition  is calculated  as 
$10^{28}$ grams which is  equal to the mass of the  Earth and can therefore
safely be  neglected. The isothermal  state is therefore  reached long
before Hawking radiation could have influenced the mass content of our
universe.  Some grand unified  theories suggest the instability of the
proton.  Experimentally,  however, it was  found that the  lower limit
for the half-life of the proton must be greater than $10^{33}$ seconds
(if not infinity).   That leaves us with only  one more possibility, a
positive    non-zero   cosmological    constant.     If   we    assume
$\Omega_{\Lambda}=0.7$   in   standard   big   bang   cosmology,   the
cosmological constant  would become dominant  after the radius  of the
universe increases  to about  three times its  current value.  This is
significantly  faster  than the  timescale  of  the  formation of  the
isothermal sphere.  It suggests  a possible role  for $\Lambda$,  as a
buffer  in preventing our  universe from  undergoing a  possible phase
transition.

\section{Discussion} \label{chap5-7}
We have presented a relativistic approach to describe the evolution of
the universe from an Einstein-de Sitter expansion with no centre to an
isothermal  sphere with  a centre.  We  have shown  how this  changing
symmetry   of  the   universe   can  be   incorporated  into   general
relativity.  This  demonstrates  that  general  relativity  permits  a
changing symmetry on a global scale, a nontrivial result.

Our solution required a Schwarzschild metric to act as an intermediate
background to match the static  generalised isothermal metric to the
FRW metric of the   expanding  universe.  
The   necessity  for   this  Schwarzschild
intermediary  can  be understood  both  intuitively  and  on a  formal
mathematical  basis.  The  Schwarzschild  metric  acts as  a  kind  of
membrane  linking   a  static   space-time  region  to   an  expanding
space-time. In \cite{fayos, mars, matravers}, the general matching 
of  two  spherically symmetric  spacetimes was investigated in
detail. Their  results provide   a  formal   basis  for
understanding  the   role   of  the ``Schwarzschild
membrane''. Applying  the  general matching  conditions derived in
their paper to the expanding, centrally isothermal universe proves
that a  direct matching without the ``Schwarzschild membrane'' 
between  the generalised isothermal and the  FRW spacetimes  is  not
possible. An  extension  of  this   idea  of  a ``Schwarzschild
membrane''  to general spherically  symmetric spacetimes other  than
the  isothermal  metric  might  be   possible  and  have interesting
implications. 

In our hybrid model the field equations are solved for constant
pressure/density ratio, $\alpha$ and, as stated, the solution
therefore does not incorporate a time-dependent $\alpha$.
We therefore introduced a ``sequence of quasi-static stages''.
We envisage the universe evolving through a sequence of models with
increasing $\alpha$ as its central cluster grows in size and the
isothermal approximation becomes better for larger regions. The main
physical argument for this is that the local clustering timescale in a nonlinear 
positive density inhomogeneity in an Einstein-de Sitter 
model is shorter than the expansion timescale since the average local 
density is greater than the average global density.
Our metric only describes those phases of the evolving universe when
it is dominated by one relaxed, static isothermal cluster. As a second
large cluster enters the horizon our description breaks down and only
applies again after the system has relaxed to a larger static cluster
with new values of $\alpha$ and the boundary radius of the
cluster. This leads to a sequence of discrete changes
which characterize the evolving model. Our solution therefore describes only
discrete steps between static intervals and not continuous changes in
time. Finding a continuous description remains a problem for the future.

      \clearpage{\pagestyle{empty}\cleardoublepage}
      \part{Epilogue}	
      \clearpage{\pagestyle{empty}\cleardoublepage}
\chapter{Discussion} \label{epilogue}

\begin{quote} 
{\it
"After a lifetime of crabwise thinking,  I have gradually become aware
of the towering  intellectual structure of  the world, One article  of
faith I have about it is that, whatever the end may be for each of us,
it cannot be a bad one."}
\par{\bf - Fred Hoyle}
\end{quote}

\begin{quote}
{\it "I do not know what I may appear to the world. But to myself I seem to
 have been like a boy playing on the seashore, and enjoying myself in
 now and then finding a smoother peeble, or a prettier shell than
 ordinary, while the great ocean of truth lay all undiscovered before
 me." }
\par{\bf - Sir Issac Newton}
\end{quote}

\section*{From the Beginning to the End}

In the epilogue, we summarize the main results of this
thesis and discuss possible future directions. We began this thesis
with a modest aim to study possible and viable alternatives to the
standard inflationary big-bang cosmology, in the hope that we can
understand these alternative cosmologies better and search for
phenomenological effects if possible. As such, we have explored three
interesting alternative themes, namely, spin-torsion theories, extra
dimensions and changing global symmetry. 


Beginning from chapter \ref{chapter2} for spin-torsion theories, we
found new massive, cosmological solutions for
the Dirac field coupled self-consistently to gravity with a
cosmological constant in a $K=0$ (flat) universe.
In addition to the generalized particle and anti-particle
solutions found in \cite{challinor} with a cosmological constant, we
discovered another anti-particle solution which has interesting
implications. All these solutions have the presence of a particle
horizon but they do not possess an inflationary phase. Although these
spin-torsion solutions are interesting, they are shown not to be
viable alternatives to the standard scalar field inflation.  It
has been shown in \cite{ale, ochs} that the vacuum polarization terms
in the framework of Einstein-Dirac equations would provide an
alternative to standard inflationary models. Hence the next
step is to reconsider the model with vacuum polarisation terms.


Next, we come to the theme of theories with large extra
dimensions. We concentrated on a sector of the braneworld cosmology in
chapters \ref{chapter3} and \ref{chapter4}, particularly on the study
of linear perturbations using the (1+3)-covariant approach and the
possible effects on the CMB from the extra dimensions. In chapter
\ref{chapter3}, we set up the (1+3)-covariant formalism for the study
of cosmological perturbations, and provided a complete set of
frame-independent equations for the total matter variables, 
and a partial set of equations for the non-local variables which arise
from the projection of the Weyl tensor in the bulk. We also derived the
covariant form of the line of sight solution for the CMB temperature
anisotropies in braneworld cosmologies. However, our setup still lacks
the propagation equation for the non-local anisotropic stress, which
appears to be an important ingredient in the study of CMB
anisotropies.

In the following chapter, we studied the scalar and tensor
perturbations and their implications on the
CMB anisotropies. For the scalar perturbations, we provide the
equations for the total matter variables with equations for the
independent constituents in a cold dark matter cosmology, and supplement
solutions in the high and low-energy radiation-dominated phase under
the assumption that both the dark energy and non-local anisotropic
stress vanish. These solutions reveal an additional pair of
non-local isocurvature velocity and density modes which emerge from
the two additional non-local degrees
of freedom. These mode solutions would be useful in setting up initial
conditions for numerical codes (for example, CAMB~\cite{lewis} and
CMBFAST~\cite{seljak}) aimed at exploring the effect of braneworld
corrections on the cosmic microwave background (CMB) power
spectrum. For the tensor perturbations, we set out
the framework of a program to compute the tensor anisotropies in the
CMB that are generated in braneworld models. In the simplest
approximation, we show the braneworld imprint as a correction to the
power spectra for standard temperature and polarization anisotropies
and similarly show that the tensor anisotropies are also insenitive to
the high energy effects.

In principle, future observations can constrain the KK parameter $\zeta$,
which controls the generation of 5D modes within our simplified
local approximation, Eq.~(\ref{e:Pansatz1}). The other braneworld
parameter $\lambda$, the brane tension, is not constrained within
our approximation. In practice, the tensor power spectra have not been
measured, and the prospect of useful data is still some way off.
What is more important is the theoretical task of improving on the
simplified local approximation we have introduced. This
approximation has allowed us to encode aspects of the qualitative
features of braneworld tensor anisotropies, which we expect to
survive in modified form within more realistic approximations.
However, a proper understanding of braneworld effects must
incorporate the nonlocal nature of the KK graviton modes, as
reflected in the general form of Eq.~$\eqref{e:soln}$. A possible
direction is to investigate the tensor perturbations in the bulk and
determine how it would actually affect the tensor power spectrum. 

It is also necessary to investigate the scalar anisotropies, which have a
dominant contribution to the measured power spectra. There are a few
possible directions for the scalar anisotropies. First, there is
the possibility of exciting these non-local velocity and density
isocurvature modes. Another possibility comes from the fact that the
dark radiation, ${\cal U}$ is not necessarily zero. It has been shown by
Langlois et al \cite{langlois5} that that the dark radiation, ${\cal U}_0$ is
asymptotically a constant upon reaching the low energy regime, while
it is time-dependent as ${\cal U} \sim a^4$ at high energies. Hence it
would be natural to extend the mode solutions we found for cases
${\cal U}\neq 0$ and work out the scalar anisotropies. All these
possibilities for both scalar and tensor anisotropies may reveal new
braneworld imprints that are more amenable to observational testing.


Finally we come to the last theme of changing global symmetry. In
chapter \ref{chapter5}, we constructed a possible relativistic description of
how an Einstein-de Sitter expansion with no centre would evolve to an
isothermal sphere with  a centre. We  have shown how this  changing
symmetry   of  the   universe  might possibly lead to a phase
transition in the future. This construction  demonstrates  that
general  relativity  permits  a changing symmetry on a global scale, a
nontrivial result.  

The  largest galaxy cluster in a flat Einstein-de Sitter universe may
 grow indefinitely to  encompass most  of space  after an extremely
 long time.  We  derive a
general  relativistic   metric  for  a  very   large,  bounded  nearly
isothermal cluster of  galaxies which is embedded in  such a universe.
The  embedding  is  done  by  means  of  a  ``Schwarzschild  membrane''.
Pressure  is   important,  unlike  previous   Tolman-Bondi  models  of
inhomogeneities.  The  cluster's expansion, represented  by a sequence
of models, alters the average  global symmetry of an increasing volume
of  space-time.  Initially  the metric  is homogeneous  and isotropic,
having translational  and rotational symmetry around  every point.  As
the  metric evolves,  it eventually  loses its  translational symmetry
throughout larger regions and  retains rotational symmetry around only
one  point:  the centre  of  the cluster  becomes  the  centre of  the
universe.  Our  sequence  of  hybrid  models  transforms  between  the
Einstein-de Sitter  and the isothermal cluster  limits and illustrates
how  a changing  equation  of state  of  matter can  alter the  global
symmetry of space-time.

Gravitational   clustering   creates   a  cosmologically   significant
pressure,  which can feed back  into  the spacetime  structure  of  the
universe. This  induces a dynamical symmetry breaking  effected by the
redistribution of matter. Is this transition between distinct symmetry
states of the universe an  actual phase transition? The answer depends
on  the timescale and the definition of a phase transition. The experimental
study  of  numerous condensed  matter  systems  has demonstrated  that
transitions,   which   show   a   macroscopic   discontinuity,   might
fundamentally  be  continuous when  examined  on  a microscopic  scale
(e.g. a  ferromagnetic phase transition).  This may be analogous  to the
transition   in  the  isothermal   universe  in   which  gravitational
clustering provides the  continuous, ``microscopic'' description of what
is ``macroscopically'' a discontinuous phase transition when viewed on a
long enough  timescale. The different symmetries of the  universe
simply arise as different limits of the sequence of hybrid
solutions. It has also suggested a new role for the cosmological
constant, that is to deter the Universe from going into a phase
transition in the future. 

This possible end of our Universe is conditional on the
extremely long term stability of the matter. If
the matter decays into radiation before the isothermal metric develops,
the expansion can continue and possibly accelerate. Since matter can
be converted into radiation in the isothermal universe, the equation
of state would change and produce another possible phase
transition. 


Of course, it is speculative to conceive the end of the
simple Einstein-de Sitter Universe without a cosmological constant,
just as it is difficult to conceive the beginning of our
Universe that might be endowed with spin-torsion or extra-dimensions. 
It would be appropriate to conclude this thesis along the same thought as 
\cite{saslaw4} that perhaps T. S. Eliot had some intuition of this idea
in East Coker, where he said,{\it ``In my beginning is my end.''}

      \clearpage{\pagestyle{empty}\cleardoublepage} 
      \appendix

\part*{Appendix}

\chapter{Conventions, Units, Sign Manifesto \\ and Table of Symbols} \label{appendix1}

In this thesis, we would like to follow the spirit in \cite{kolb} and
\cite{liddle} to state the conventions used and provide (i) the units for the
conversion factors and fundamental constants and (ii) cosmological
parameters, the sign manifesto for the covariant approach (which is
useful for the modifying the CAMB code) and the
table of symbols adopted throughout this thesis.  

\section{Conventions}

We would like to follow the conventions of the metric signature
$(-,~+,~+,~+)$ for Chapters 1,3,4 and 5 and $(+,~-,~-,~-)$ for Chapter
2, and we would adopt the choice of natural units,
i.e. we set the fundamental constants $\hbar=c=k_B=1$, and there is
one fundamental dimension, energy, $[E]$ such that
\begin{equation}
[E] = [M] = [T_E] = [L] = [T]^{-1}
\end{equation}
where $[M]$, $[T_E]$, $[L]$ and $[T]$ are the dimensions of mass,
temperature, length and time respectively. \\

The unit of energy is taken to be a GeV$=~10^3$ MeV$~=~$10$^6$
keV$~=~$10$^9$ eV. For the case of braneworlds, the more popular unit
for energy is in the TeV scale i.e. TeV= 10$^3$ GeV.   

\section{Units}

\subsection{Conversion Factors}

\begin{center}
\begin{tabular}{p{4 cm} p{12cm}}
Temperature: & 1 GeV = $1.1605 \times 10^{13}$K  \\
Mass:        & 1 GeV = $1.7827 \times 10^{-27}$ kg  \\
Length:      & 1 GeV$^{-1}$ = $1.9733 \times 10^{-16}$ m \\
Time:        & 1 GeV$^{-1}$ = $6.5822 \times 10^{-25}$ s  \\
Megaparsec:  & 1 Mpc = $10^{6}$ pc = $3.0856 \times 10^{22}$ m = $1.5637
\times 10^{38}$ GeV$^{-1}$
\end{tabular}
\end{center}

\subsection{Fundamental Constants}

\begin{center}
\begin{tabular}{p{4 cm} p{12cm}}
Planck's constant: & $\hbar~$=$~6.6261 \times 10^{-34}$m$^2$ kg s$^{-1}$ \\
Speed of light: & $c~$=~$29972458$ m s$^{-1}$ \\
Boltzmann's constant: & $k_B$ = $1.3807 \times 10^{-23}$ J K$^{-1}$ \\
Newton's constant: & $G~$=$~6.672 \times 10^{-11}$ kg m$^3$ s$^{-2}$
$~\equiv~m_{Pl}^{-2}$ \\
Planck energy: & $m_{Pl}~\equiv~\sqrt{\frac{\hbar c^5}{G}}~=~1.1211
\times 10^{19}$ GeV \\
Planck mass: & $m_{Pl}~\equiv~\sqrt{\frac{\hbar c}{G}}~=~2.1768 \times
10^{-8}$ kg \\
Planck time: & $t_{Pl}~\equiv~\sqrt{\frac{\hbar G}{c^5}}~=~5.3904
\times 10^{-44}$ s \\ 
Planck length: & $l_{Pl}~\equiv~\sqrt{\frac{\hbar G}{c^3}} =1.6160
\times 10^{-35}$ m \\
Electron mass: & $m_e~=~0.5110$ MeV \\
Neutron mass: & $m_n~=~939.566$ MeV \\
Proton mass: & $m_p~=~938.272$ MeV \\
Thomas cross section: & $\sigma_T~=~6.652 \times 10^{-29}$ m$^2$ \\ 
Solar mass & $M_{\odot}~=~2 \times 10^{30}$ kg
\end{tabular}
\end{center}

\section{Physical Parameters}

\subsection{Cosmological parameters}

\begin{center}
\begin{tabular}{p{5 cm} p{10cm}}
Hubble constant: & $H_0$ = 100 $h$ km s$^{-1}$ Mpc$^{-1}$ \\
Present Hubble distance: & $c H_0^{-1}$= 2998 $h^{-1}$ Mpc\\
Present Hubble time: & $H_0^{-1}$ = $9.78 \times 10^{6} h^{-1}$ yr \\
Present critical density: & $\rho_c,0$ = $1.88 h^2 \times 10^{-32}$ kg
m$^{-3}$  
\end{tabular}
\end{center}

\section{Sign Manifesto}

In order to incorporate the equations in the CAMB code, we need to
re-write the equations in the convention $(+,~-,~-,~-)$. To do this,
we adopt the convention for the change of signs in \cite{challinor4}
and expand the range of quantities in this thesis, for the following
quantities: 
\begin{equation}
\begin{split}
g_{ab} &\to -g_{ab}, \quad \n_a \to \n_a \quad R_{abcd} \to R_{abcd} 
\quad R_{ab} \to R_{ab}, \\
R &\to -R, \quad h_{ab} \to -h_{ab}, \quad {\cal R} \to - {\cal R}, 
\quad D_a \to -D_a,  \\
u_a &\to -u_a, \quad q_a \to -q_a, \quad A_a \to - A_a, \quad \sigma_{ab} \to -
\sigma_{ab} \\
\rho &\to \rho, \quad P \to P, \quad \Theta \to \Theta, \quad {\cal U}
\to {\cal U}, \\ 
\Delta_a &\to - \Delta_a, \quad {\cal Z}_a \to -{\cal Z}_a, \quad \eta_a \to
\eta_a, \quad \pi_{ab} \to \pi_{ab},  \\
\Upsilon_a &\to -\Upsilon_a, \quad {\cal Q}_a \to - {\cal Q}_{a},
\quad {\cal P}_{ab} \to {\cal P}_{ab}.  
\end{split}
\end{equation}

The above sign manifesto would help to translate equations from the
signature of $(-,~+,~+,~+)$ to the other signature $(+,~-,~-,~-)$. 

\section{Table of Symbols}

\section*{For all chapters using General Relativity}
\begin{center}
\begin{longtable}{|p{1.8cm}|p{12cm}|}
\hline
{\bf Symbol} & {\bf Meaning of Symbol} \\ \hline \hline
$a(t)$ & Scale Factor of the Universe with respect to cosmic time, $t$ \\
\hline 
$a(\tau)$    & Scale Factor of the Universe with respect to conformal
time, $\tau$ \\ \hline 
$H$          & Hubble parameter with respect to cosmic time, $t$  \\ \hline
${\cal H}$   & Hubble parameter with respect to conformal time, $\tau$ \\ \hline
$u^a$        & 4-velocity/tangent vector \\ \hline 
$\n_a$       & Covariant derivative \\ \hline
$D_a$        & Projected (spatial) covariant derivative \\ \hline
$\Lambda$    & Cosmological Constant \\ \hline
$\lambda$    & Tension of the brane \\ \hline
$R_{abcd}$   & Riemann tensor  \\ \hline
$R_{ab}$     & Ricci tensor \\ \hline
$R$          & Ricci scalar  \\ \hline
${\cal R}$   & 3-curvature scalar in general relativity \\ \hline
$K$          & Constant of curvature in general relativity \\ \hline
$G_{ab}$     & Einstein tensor \\ \hline 
$T_{ab}$     & Energy-momentum tensor \\ \hline
$\rho$       & Density of a fluid (or generally energy density) \\ \hline
$\Delta_a$   & Spatial gradient in the density (or energy density) of
the fluid \\ \hline
$P$          & Pressure of a fluid \\ \hline
$q_a$	     & Energy flux of a fluid \\ \hline
$\pi_{ab}$   & Local anisotropic stress/pressure \\ \hline
$\eta_a$     & Spatial gradient of the 3-curvature \\ \hline
$E_{ab}$ & Gravito-electric part of the Weyl tensor on the brane\\ \hline
$H_{ab}$ & Gravito-magnetic part of the Weyl tensor on the brane\\ \hline
$\omega_{ab}$ & Vorticity tensor \\ \hline
$\sigma_{ab}$ & Shear \\ \hline
$\Theta$      & Comoving expansion \\ \hline
${\cal Z}_{a}$ & Spatial gradient of the comoving expansion\\ \hline
$A_{a}$       & Acceleration \\  \hline
$\Phi$        & Gravitational Potential \\ \hline 
$c_s$         & Adiabatic sound speed \\ \hline
$n_e$         & Electron density \\ \hline
$\sigma_T$    & Thomson cross section  \\ \hline
\end{longtable}
\end{center}

\subsection*{Brane Worlds} 
\begin{center}
\begin{longtable}{|p{1.8cm}|p{12cm}|}
\hline
{\bf Symbol} & {\bf Meaning of Symbol} \\ \hline \hline
$S_{ab}$     & Symmetric tensor which contains the local quadratic
energy corrections from the bulk to brane \\ \hline
${\cal E}_{ab}$ & Symmetric projection of the bulk (5D)-Weyl tensor \\ \hline
${\cal U}$   & Non-local energy density projected from bulk to brane \\ \hline
$\Upsilon_a$ & Spatial gradient in non-local energy from bulk to brane
\\ \hline 
${\cal Q}_a$   & Non-local energy flux projected from bulk to brane \\
\hline
${\cal P}_{ab}$   & Non-local anisotropic stress projected from bulk to
brane \\ \hline
\end{longtable}
\end{center}

\section*{For the chapter using Gauge Theory of Gravity}
\begin{center}
\begin{longtable}{|p{1.8cm}|p{12cm}|}
\hline
{\bf Symbol} & {\bf Meaning of Symbol} \\ \hline \hline
$\g_{\mu}$ & Dirac vectors in Spacetime Algebra (STA) \\ \hline
$\ps$      & Pseudoscalar \\ \hline
$A_r$      & Grade-$r$ multivector \\ \hline
$\n$       & Vector derivative \\ \hline
$\bar{h} (a)$ & Position gauge field \\ \hline
$\Omega(a)$ & Rotation gauge field \\ \hline
${\cal D}$ & Covariant derivative with respect to a multivector \\
\hline
$\psi$ & Spinor in GTG\footnote{Gauge Theory of Gravity} \\ \hline
${\cal R}(a \w b)$ & Covariant Riemann tensor in GTG \\ \hline
${\cal R} (a)$ & Covariant Ricci tensor in GTG \\ \hline
${\cal G} (a)$ & Covariant Einstein tensor in GTG \\ \hline
${\cal T} (a)$ & Covariant energy momentum tensor in GTG \\ \hline
${\cal S} (a)$ & Covariant spin torsion trivector \\ \hline
\end{longtable}
\end{center}

      \clearpage{\pagestyle{empty}\cleardoublepage} 
      
\chapter{Identities in Geometric Algebra} \label{appendix2}

In this appendix, we would like to summarize some of the important
identities in geometric algebra. A more detailed exposition of the
geometric algbera formalism is given in Hestenes et al\cite{hestenes3, hestenes4,
hestenes1, hestenes2} and Lasenby, Doran and Gull in \cite{doran1}.

\section{Multivectors}

Let $A$ and $B$ be multivectors of grade $r$ and $s$, $\cdot$ be the
interior product, $\w$ be the exterior product and $\times$ be the
commutator product, where these operations are
defined in Eqs. $\eqref{e:ga1}$, $\eqref{e:ga2}$ and $\eqref{e:ga3}$
respectively in chapter \ref{chapter2}, then the following identities hold:  
\begin{align}
\label{e:b1}
&A = \langle A \rangle_0 + \langle A \rangle_1 + ... = \sum_r \langle
A \rangle_r \;, \\
\label{e:b2}
&A_r B_s = \langle A_r B_s \rangle_{|r-s|} + \langle A_r B_s
\rangle_{|r-s|+2} + ..... + \langle A_r B_s \rangle_{|r+s|} \;, \\
\label{e:b3}
&A_r \dt \lambda = 0 \;, \\
\label{e:b4}
&\langle AB \rangle = \langle BA \rangle \;, \\
\label{e:b5}
&A_r \dt (B_s \dt C_t) = (A_r \w B_s) \dt C_t \quad (\text{for}~r+s
\leq t~\text{and}~r,s>0) \;,\\
\label{e:b6}
&A_r \dt (B_s \dt C_t) = (A_r \dt B_s) \dt C_t \quad
(\text{for}~r+t\leq s) \;, \\
& A \times  (B \times C) + C \times (A \times B) + B \times (C \times
A) = 0 \;.
\end{align}
With the pseudoscalar, $\ps$, we have the following identities,
\begin{align}
\label{e:b16}
& A_r \dt (B_s \ps) = A_r \w B_s \ps \quad (r+s \leq n) \;,\\
\label{e:b17}
& A_r \w (B_s \ps) = A_r \dt B_s \ps \quad (r \leq s) \;.
\end{align}

We now specialize to the case for the vector and bivector cases. 
Let $\{ a,b,c \}$ be a set of vectors and $\{ A, B,C \}$ be a set of
bivectors, and the following identities will hold:

\begin{align}
\label{e:b7}
&a \dt (b \w c) = a \dt b c - a \dt c b \;, \\
\label{e:b8}
&(a \w b) \dt (c \w d) = a \dt d b \dt c - a \dt c b \dt d \;,\\
\label{e:b9}
&a \dt (b \dt B) = (a \w b) \dt B \;,\\
\label{e:b10}
&(a \w b) \times (c \w d) = b \dt c a \w d - a \dt c b \w d + a \dt d b
\w c - b \dt d a \w c \;, \\
\label{e:b11}
& (a \w b) \times B = (a \dt B) \w b + a \w (b \dt B) \;, \\
\label{e:b12}
& BC = B \dt C + B \times C + B \w C \;, \\
\label{e:b13}
& CB = B \dt C - B \times C + B \w C \;, \\
\label{e:b14}
&B \times A_r = \langle B \times A_r \rangle_r \;, \\
\label{e:b15}
&A \times (BC) = A \times B~C + B~ A \times C  \;.
\end{align}

\section{Geometric Calculus}

Let $F$ be an arbitrary function of some multivector argument $X$,
such that $F=F(X)$. The derivative of $F$ with respect to $X$ in the
$A$-direction is defined to be 
\begin{equation} \label{e:b34}
A * \p_X F(X) = \lim_{x \to 0} \frac{F(X+ x A) - F(X)}{x}
\end{equation}
where we define the multivector derivative $\p_X$ to be
\begin{equation} \label{e:b35}
\p_X \equiv \sum_{i < \dots < j} e^{i} \w \dots \w e^{j} (e_j \w \dots
\w e_i) * \p_X
\end{equation}
Most of the properties of the multivector derivative follow from the
result 
\begin{equation} \label{e:b36}
\p_X \langle XA \rangle = P_X (A)
\end{equation}
where $P_X(A)$ is the projection of the multivector $A$ onto the
graded element contained in
$X$. The multivector derivative acts on the next object to its right
unless brackets are present. Denoting the multivector which the
derivative acts, we can write Leibniz rule in the form:
\begin{equation} \label{e:b37}
\p_X (AB) = \dot{\p}_X \dot{A} B + \dot{\p}_X A \dot{B}
\end{equation}

The derivative with respect to the vector $a$, $\p_a$ is used to
perform linear alegbra operations. The following identities are
obtained as follows:
\begin{align}
\label{e:b38}
& \p_a \dot b = b \;, \\
\label{e:b39}
& \p_a \dot a = n \;, \\
\label{e:b40}
& \p_a \w a = 0 \;,\\
\label{e:b41}
& \p_a a = n \;, \\
\label{e:b42}
& \p_a a^2 = 2a \;,\\
\label{e:18}
&(B \dt \p_a ) \w a = 2 B \;, \\
\label{e:b19}
&\p_a \w \p_b (b \w a) \dt B = 2 B \;, \\
\label{e:b20}
&\p_a A_r a = (-1)^r (n-2r) A_r \;,\\
\label{e:b21}
&\p_a a \dt A_r = r A_r  \;,
\end{align}

\section{Linear Algebra}

Let $a,b,c$ be vectors, $\underline{f}(A)$ be the linear function with
the multivector $A$ in its argument and $\bar{f}(A)$ be its
adjoint, from the definition in Eq. $\eqref{e:ga6}$, it follows that
\begin{equation} \label{e:b31}
\bar{f} (a) = \p_b \langle \underline{h} (b) a \rangle
\end{equation}
Eq. $\eqref{e:b31}$ is frequently employed perform algebraic
manipulations of linear functions. The advantage is that all
manipulations are frame-frame. We note that the $\p_a$ and $a$ vectors
can be replaced by the sum over a set of constant frame vectors and
their reciprocals, if required.

Linear functions extend to act on multivectors via 
\begin{equation} \label{e:b32}
\underline{f} (a \w b \w \dots \w c) \equiv \underline{f} (a) \w
\underline{f} (b) \w \dots \w \underline{f} (c)
\end{equation}
so that $\underline{f}$ is a grade-preserve linear function mapping
multivectors to vectors.

In particular, since the pseudoscalar $\ps$ is unique to a scale
factor, we define 
\begin{equation} \label{e:b33}
\text{det} (\underline{f}) = \underline{f}(\ps) \ps^{-1}
\end{equation}

From the Eq. $\eqref{e:b32}$ and $\eqref{e:b33}$, the following
identities hold:
\begin{align}
\label{e:b22}
& f(a\w b\w ...\w c) = f(a) \w f(b) \w ... \w f(c) \;, \\ 
\label{e:b23}
& f(\ps ) = \text{det} (f) \ps \;,\\
\label{e:b24} 
& \bar{f} (b) = \p_a f(a) \dt b \;,\\
\label{e:b25}
& f(A_r) \dt B_s = f [ A_r \dt \bar{f} (B_s) ] \quad (r \geq s) \;,\\
\label{e:b26}
& A_r \dt \bar{f} (B_s) = \bar{f} [f(A_r) \dt B_s] \quad (r \leq s) \;, \\
\label{e:b27}
&f^{-1} (A) = \text{det}(f)^{-1} \ps \bar{f} (\ps^{-1}A) \;, \\
\label{e:b28} 
&\bar{f}^{-1} (A) = \text{det}(f)^{-1} \ps f (\ps^{-1}A) \;, \\
\label{e:b29}
&\p_{\bar{h}(a)} \text{det} \underline{h}^{-1} = - \text{det}
\underline{h}^{-1} \underline{h}^{-1} (a) \;, \\
\label{e:b30}
&\p_{\bar{h}(a)} \underline{h}^{-1} (b) \dt v = - \underline{h}^{-1}
(a) \dt v \underline{h}^{-1} (b) \;. 
\end{align}

      \clearpage{\pagestyle{empty}\cleardoublepage} 
\chapter{Identities in Cosmological Perturbation Theory} \label{appendix3}
\section{Differential Identities in the PSTF Approach}

We will summarize the following covariant linearized identities on a
Friedmann background used in \cite{challinor4, maartens8,
maartens, maartens6,  maartens9}, which are useful for the derivation of the
propagation and evolution equations in chapter \ref{chapter3}.  

The following linearized identities with the operation of curl and
div, which is valid for any arbitrary first order PSTF vector, $V_a$,
tensor, $S_{ab}$, and arbitrary scalar $f$ with first order projected
gradients are summarized as follows: 
\begin{align}
\label{e:c32}
\text{curl} D_a f &= - 2 \dot{f} \omega_a \;,\\ 
\label{e:c33}
D^2 (D_a f) &= D_a (D^2 f) + \frac{2K}{a^2} D_a f + 2 \dot{f}
\text{curl} \omega_a \;,\\
\label{e:c34}
(D_a f)^{\cdot} &= D_a \dot{f} - H D_a f + \dot{f} A_a \;,\\
\label{e:c35}
(a D_a V_b \dots)^{\cdot} &= a D_a \dot{V}_b \dots \;,\\
\label{e:c36}
(D^2 f)^{\cdot} &= D^2 \dot{f} - 2 H D^2 f + \dot{f} D^a A_a \;, \\
\label{e:c37}
D_{[a} D_{b]} V_c &= \frac{K}{a^2} V_{[a} h_{b]c} \;,\\
\label{e:c38}
D_{[a} D_{b]} S^{cd} &= \frac{2K}{a^2} S_{[a}{}^{(c} h_{b]}{}^{d)} \;,
\\
\label{e:c39}
D^a \text{curl} V_a &=0  \;, \\
\label{e:c40}
\text{curl} \text{curl} V_a &= D_a (D^b V_b) - D^2 V_a +
\frac{2K}{a^2} V_a \;, \\
\label{e:c26}
(\text{curl} S_{ab})^{\cdot} &= \text{curl} \dot{S}_{ab} - \frac{1}{3}
\Theta \text{curl} S_{ab} \;, \\
\label{e:c27}
(D^a S_{ab})^{\cdot} &= D^{a} \dot{S}_{ab} - \frac{1}{3} \Theta D^a
S_{ab} \;, \\
\label{e:c28}
D^a (\text{curl} S_{ab}) &= \frac{1}{2} \text{curl} (D^a S_{ab}) \;, \\
\label{e:c29}
\text{curl} (\text{curl} S_{ab}) &= - D^2 S_{ab} + \frac{3}{2}
D_{\langle a} D^c S_{b \rangle c} + \frac{3K}{a^2} S_{ab} \;,
\end{align} 
where the vectors and tensors vanish in the background, i.e. $S_{ab} =
S_{\langle ab \rangle}$ and all the identities except $\eqref{e:c32}$
are linearized. 

The identities in \cite{challinor4} are also useful for the
evaluation of the $\text{curl}$ quantities for parity tensor harmonics are
also summarized here:
\begin{align}
\label{e:c30}
\text{curl} Q_{ab} &= \frac{k}{a} \l(1 + \frac{3K}{k^2} \r)^{1/2}
\bar{Q}_{ab}^{(k)} \;, \\
\label{e:c31}
\text{curl} \bar{Q}_{ab} &= \frac{k}{a} \l(1 + \frac{3K}{k^2} \r)^{1/2}
Q_{ab} \;.
\end{align}

\section{The identitites for the harmonics}

\subsection{Scalar Harmonics}

As in \cite{dunsby, hawking, kodama2}, we define the scalar harmonics as being
eigenfunctions of the covariantly Laplace-Beltrami operator. 
\begin{equation} \label{e:c1}
D^2 Q \equiv D_a D^a Q = - \l(\frac{k}{a} \r)^2 Q \;,
\end{equation}
where we have dropped the index $k$ from the $Q$s. One can use the
scalar harmonic to expand scalars, and with this we define the
following vector which correspond to the definition in \cite{kodama2}. 
\begin{equation} \label{e:c2}
Q_a = -\frac{a}{k} D_a Q \;,
\end{equation}
and a trace-free symmetric tensor,
\begin{equation} \label{e:c3}
Q_{ab} = \l(\frac{a}{k} \r)^2 D_b D_a Q + \frac{1}{3} h_{ab} Q \;.
\end{equation}
These harmonics are defined to have the following feature:
\begin{equation} \label{e:c4}
\dot{Q} = \dot{Q}_a = \dot{Q}_{ab} =0 \;,
\end{equation}
and by the above property, we can apply the commutations in \cite{ellis1} to
get
\begin{align}
\label{e:c5} D_{[a} D_{b]} Q &= - \omega_{ab} \dot{Q} = 0 \;, \\ 
\label{e:c6} D_{[a} D_{b]} Q_c &= -\frac{K}{2a} (h_{ac} Q_b - h_{bc} Q_a) \;,\\
\label{e:c7} D_{[a} D_{b]} Q_{cd} &= \frac{K}{2a^2} [(h_{ac} Q_{bd} -
h_{bc} Q_{ad}) + (h_{ad} Q_{bc} - h_{bd} Q_{ac}) ] \;.
\end{align}

We derive the properties of the scalar harmonics, by making use of
the projected (spatial) covariant derivative, $D^a$, 
\begin{align}
\label{e:c8}
D_a Q^a &= \l(\frac{k}{a} \r) Q \;, \\
\label{e:c9}
D^2 Q_a &= - \l(\frac{k^2 -2 K}{a^2} \r) Q_a \;, \\
\label{e:c10}
D_b Q_a &= -\frac{k}{a} \l(Q_{ab} - \frac{1}{3} h_{ab} Q \r) \;, \\
Q^{a}{}_{a} &=0 \;, \\
\label{e:c11}
D^b Q_{ab} &= \frac{2}{3} \l(\frac{1}{ka} \r) (k^2 - 3K) Q_a \;,\\
\label{e:c12}
D^a D^b Q_{ab} &= \frac{2}{3} \l(\frac{1}{a^2} \r) (k^2 - 3K) Q \;, \\
\label{e:c13}
D_b D^c Q_{ac} &= \frac{2}{3} \l(\frac{1}{a^2} \r) (3K - k^2)
\l(Q_{ab} - \frac{1}{3} h_{ab} Q \r) \;, \\
\label{e:c14}
D^2 Q_{ab} &=  \l(\frac{6K - k^2}{a^2} \r) Q_{ab} \;.
\end{align}

\subsection{Vector Harmonics}

These harmonics are defined as eigenfunctions of the Helmholtz
equation:
\begin{equation} \label{e:c15}
D^2 Q_{a} = - \l(\frac{k}{a} \r)^2 Q_a \;,
\end{equation}
with $Q_a$ being the solenoidal vector, i.e. 
\begin{equation} \label{e:c16}
D_a Q^a = 0 \;.
\end{equation}
From this, we can construct a symmetric trace-free tensor,
\begin{equation} \label{e:c17}
Q_{ab} = -\frac{a}{2k} (D_b Q_a + D_a Q_b)  \;,
\end{equation}
and we note that both these harmonics are covariantly constant along
the flow lines
\begin{equation} \label{e:c18}
\dot{Q}_a = \dot{Q}_{ab} = 0 \;.
\end{equation}

The following relations hold for the defined tensor $Q_{ab}$
(equation $\eqref{e:c17}$) from $Q_a$ in this
section: 
\begin{align}
\label{e:c19}
Q^{a}{}_{a} &= 0 \;, \\
\label{e:c20}
D^b Q_{ab} &= \frac{1}{2} \l(\frac{1}{ka} \r) (k^2 - 2K) Q_a \;, \\
\label{e:c21}
D_b D^c Q_{ac} + D_a D^c Q_{bc} &= - \l(\frac{k^2-2K}{a^2} \r) Q_{ab} \;,
\\
\label{e:c22}
D^2 Q_{ab} &= - \l(\frac{k^2 - 4K}{a^2} \r) \;.
\end{align} 
Note that $Q_{a}$ and $Q_{ab}$ also satisfy relations $\eqref{e:c6}$
and $\eqref{e:c7}$ respectively.

\subsection{Tensor Harmonics}

For tensor harmonics, we define again
\begin{equation}  \label{e:c23}
D^2 Q_{ab} =  - \l(\frac{k}{a} \r)^2 Q_{ab} \;,
\end{equation}
and $\dot{Q}_{ab}=0$. The tensor is also trace-free and
divergenceless, 
\begin{align}
\label{e:c24} Q^{a}{}_{a} &= 0  \;, \\
\label{e:c25} D^b Q_{ab} &=0 \;.
\end{align}

      \clearpage{\pagestyle{empty}\cleardoublepage}    
   \backmatter
      \bibliographystyle{authordate3}
      \bibliography{thesis}

\end{document}